\documentclass[fleqn,usenatbib]{mnras}

\usepackage{newtxtext,newtxmath}

\usepackage[T1]{fontenc}

\DeclareRobustCommand{\VAN}[3]{#2}
\let\VANthebibliography\thebibliography
\def\thebibliography{\DeclareRobustCommand{\VAN}[3]{##3}\VANthebibliography}

\newcommand{\OIII}{$[\text{OIII}]$}
\newcommand{\Ha}{H$\alpha$}
\newcommand{\disp}{\sigma_0}
\newcommand{\reff}{$r_{e}$}
\newcommand{\rotsupp}{\text{v}/\sigma_0} 

\newcommand{\Mstar}{$M_{\star}$}

\newcommand{\Mdyn}{$M_{\text{dyn}}$}
\newcommand{\PAmorph}{$\text{PA}_{\text{morph}}$}
\newcommand{\PAkin}{$\text{PA}_{\text{kin}}$}
\newcommand{\vre}{$\text{v}_{\text{re}}$}

\newcommand{\geko}{\textsc{geko}}

\usepackage{graphicx}	
\usepackage{amsmath}	
\usepackage{hyperref}
\usepackage{natbib}
\usepackage{caption}
\usepackage{multirow}
\usepackage{geometry}
\usepackage{booktabs}
\usepackage[table]{xcolor}
\usepackage{array}

\usepackage{adjustbox}  

\newcommand\orcid[1]{\href{http://orcid.org/#1}{\adjustbox{trim={-.15\width} {0\height} {-.15\width} {0\height},clip}{\includegraphics[height=10pt]{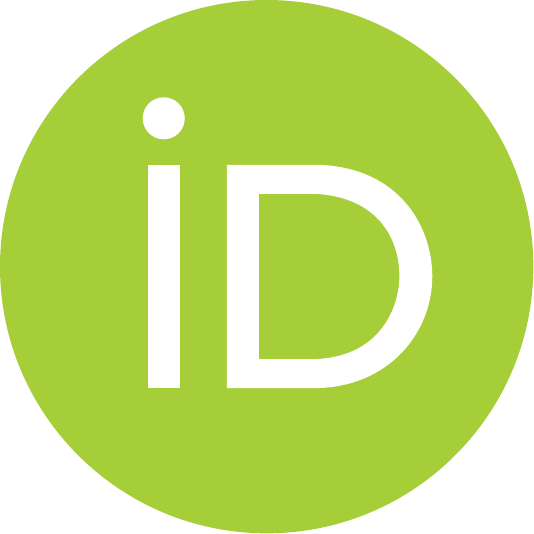}}}}

\title[Ionised gas kinematics of high redshift galaxies]{ The dawn of disks: unveiling the turbulent ionised gas kinematics of the galaxy population at $z\sim4-6$ with JWST/NIRCam grism spectroscopy}

\author[Danhaive et al.]{A. Lola Danhaive\orcid{0000-0002-9708-9958}$^{1,2}$\thanks{ald66@cam.ac.uk},
Sandro Tacchella\orcid{0000-0002-8224-4505}$^{1,2}$, Hannah Übler\orcid{0000-0003-4891-0794}$^{3}$, Anna de Graaff\orcid{0000-0002-2380-9801}$^{4}$, Eiichi Egami\orcid{0000-0003-1344-9475}$^{5}$, 
\newauthor Benjamin D. Johnson\orcid{0000-0002-9280-7594}$^{6}$, Fengwu Sun\orcid{0000-0002-4622-6617}$^{5}$, Santiago Arribas\orcid{0000-0001-7997-1640}$^{7}$, Andrew J. Bunker\orcid{0000-0002-8651-9879}$^{8}$, Stefano Carniani\orcid{0000-0002-6719-380X}$^{9}$, 
\newauthor Gareth C. Jones\orcid{0000-0002-0267-9024}$^{1,2}$, Roberto Maiolino\orcid{0000-0002-4985-3819}$^{1,2}$, William McClymont\orcid{0009-0009-5565-3790}$^{1,2}$, Eleonora Parlanti\orcid{0000-0002-7392-7814}$^{3,9}$,  
\newauthor Charlotte Simmonds\orcid{0000-0003-4770-7516}$^{1,2}$, Natalia C. Villanueva\orcid{0000-0001-6917-4656}$^{10}$, William M. Baker\orcid{0000-0003-0215-1104}$^{11}$, Daniel T. Jaffe$^{10}$, 
\newauthor Daniel Eisenstein\orcid{0000-0002-2929-3121}$^{6}$,   Kevin Hainline\orcid{0000-0003-4565-8239}$^{5}$, Jakob M. Helton\orcid{0000-0003-4337-6211}$^{5}$, Zhiyuan Ji\orcid{0000-0001-7673-2257}$^{5}$, Xiaojing Lin\orcid{0000-0001-6052-4234}$^{12}$, 
\newauthor Dávid Puskás\orcid{0000-0001-8630-2031}$^{1,2}$, Marcia Rieke\orcid{0000-0002-7893-6170}$^{5}$,  Pierluigi Rinaldi\orcid{0000-0002-5104-8245}$^{5}$, Brant Robertson\orcid{0000-0002-4271-0364}$^{13}$, Jan Scholz\orcid{0000-0001-6010-6809}$^{1,2}$, 
\newauthor Christina C. Williams \orcid{0000-0003-2919-7495}$^{14}$, Christopher N. A. Willmer\orcid{0000-0001-9262-9997}$^{5}$
\\
\\
\emph{\normalsize Affiliations are listed at the end of the paper.}}

\date{Accepted XXX. Received YYY; in original form ZZZ}

\pubyear{2025}

\begin{document}
\label{firstpage}
\pagerange{\pageref{firstpage}--\pageref{lastpage}}
\maketitle

\begin{abstract}
Recent studies of gas kinematics at high redshift have reported disky systems which appear to challenge models of galaxy formation, but it is unclear whether they are representative of the underlying galaxy population. We present the first statistical sample of spatially resolved ionised gas kinematics at high redshift, comprised of $272$ H$\alpha$ emitters in GOODS-S and GOODS-N at redshifts $z\approx3.9-6.5$, observed with JWST/NIRCam slitless spectroscopy and imaging from JADES, FRESCO and CONGRESS. The sample probes two orders of magnitude in stellar mass ($\log (M_{\star}[\mathrm{M}_{\odot}])\approx8-10$) and star formation rate ($\text{SFR}\approx0.3-100\thinspace M_{\odot}/$yr), and is representative down to $\log(M_{\star}[\mathrm{M}_{\odot}])\approx 9$. Using a novel inference tool, \textsc{geko}, we model the grism data to measure morphological and kinematic properties of the ionised gas, as probed by H$\alpha$. Our results are consistent with a decrease of the rotational support $v/\sigma_0$\ and increase of the velocity dispersion $\sigma_0$ with redshift, with $\sigma_0\approx100$ km/s and $v/\sigma_0\approx1-2$ at $z\approx3.9-6.5$. We study the relations between $\sigma_0$, and $v/\sigma_0$, and different star formation tracers and find a large scatter and diversity, with the strongest correlations between $\sigma_0$ and SFR and SFR surface density. The fraction of rotationally supported systems ($v/\sigma_0>1$) slightly increases with cosmic time, from $(36\pm6)\%$ to $(41\pm6)\%$ from $z\sim 5.5$ to $z\sim 4.5$, for galaxies with masses $9<\log(M_{\star}[\mathrm{M}_{\odot}])<10$. Overall, disks do not dominate the turbulent high-redshift galaxy population in the mass range probed by this work. When placed in the context of studies up to cosmic noon, our results are consistent with a significant increase of disk-like systems with cosmic time. 
\end{abstract}




\section{Introduction}

In the $\Lambda $CDM framework, galaxy evolution is driven by four key processes: the accretion of cold gas, the formation of stars from the collapse of the gas, the feedback from both stars and black holes, and galaxy-galaxy mergers. The morphology and kinematics of a galaxy trace these processes and are hence deeply connected to the galaxy's evolutionary stage and mass assembly history. In this work, we focus on constraining the kinematics and stellar populations of galaxies in the first 2 billion years of the Universe's history (redshifts $z=3.9-6.5$) to shed new light onto the interplay between these different physical processes in the assembly of galaxies.

In the simple analytical models of galaxy formation, gas flows into the centre of dark matter halos and cools to form stars. The gas conserves most of its angular momentum as it falls into the halo, approximately reflecting the size of the halo at the time of accretion. This means that gas cooling at later times will have higher angular momentum and settle further out into an extended star-forming disk \citep{Fall:1980aa,Mo:1998aa}.  This formalism implies that galaxies are expected to grow from the inside-out, with new stars forming in the disk. 

Beyond this simple model, feedback from stars and black holes also plays an important role in regulating star formation. Stars inject momentum into the interstellar medium through supernovae, stellar winds, and radiation pressure \citep{Ostriker:2011aa, Newman:2012aa, Shetty:2012aa,Rathjen:2023aa}. These processes heat and disrupt surrounding gas, and sufficiently strong feedback can drive strong outflows which eject gas from the galaxy entirely \citep{Oppenheimer:2008aa,Oppenheimer:2010aa,Brook:2012aa,Ubler:2014aa}. Such processes re-distribute gas and affect the kinematics of cold ($\approx 100 \rm ~K$, e.g. CO and [CII]) and, in particular, warm ionised ($\approx 10^4 \rm ~K$, e.g. \Ha and [OIII]) gas. Similarly, the accretion of baryons onto a nuclear supermassive black hole, hence triggering an active galactic nucleus (AGN), can generate powerful outflows and heavily disrupt the star formation and dynamics of its host galaxy \citep{Silk:1998aa,Cattaneo:2009aa,Kormendy:2013aa,Reines:2015aa,Harrison:2017aa,Florian:2020aa,DEugenio:2024ab}. Furthermore, complex processes such as the formation of dust \citep{Inoue:2011aa,Marshall:2022aa,Matharu:2023uj,Le-Bail:2024aa,Li:2024aa,Schneider:2024aa} and metals \citep{Mannucci:2010um,Wuyts:2016ab, Simons:2021wp,Sharda:2021aa,Boardman:2023wb,Venturi:2024aa}, galaxy-galaxy interactions and mergers \citep{Wright:2009ub,Qu:2010aa,Duan:2024ab,Nakazato:2024aa,Puskas:2025aa}, and violent gas inflows \citep{Dekel:2009aa,Klessen:2010aa,Gabor:2014aa,Forbes:2023aa} can affect the morphology and kinematics of the galaxy by introducing clumpy star formation and turbulence \citep{van-den-Bosch:2001aa, Bournaud:2007aa, Mandelker:2017aa,Oklopcic:2017aa}. These effects are especially important at early cosmic times when the gas fractions and overall densities are higher, causing more intense star formation, and galaxies have lower stellar masses, and are therefore less resistant to feedback processes due to their shallow potential wells \citep{El-Badry:2016aa, Hopkins:2023aa}. Such effects are thought to be the cause of the highly variable, or ``bursty'', star formation which has been identified at high redshift \citep{Tacchella:2023aa,Endsley:2024aa,Looser:2024aa,Baker:2025aa,McClymont:2025aa,Witten:2025aa}. The direct impact of bursty star formation on the kinematics has not been directly studied, however, it has been associated with massive gas inflows and outflows on short timescales \citep{Tacchella:2020aa,Zhu:2024ab,McClymont:2025aa,Saldana-Lopez:2025aa} and rapid morphological evolution, including the fluctuations of galaxy sizes \citep{El-Badry:2016aa,McClymont:2025ab}, implying that the impact on kinematics is likely important.

Observational studies of the local Universe paint a picture of a variety of galaxy dynamical states and morphologies, which are outcomes of their complex mass assembly history \citep{Sales:2012aa,Dubois:2016aa, Tacchella:2016aa, Tacchella:2019aa}. Galaxies are often separated into two broad categories -- star-forming spiral disk galaxies with younger stellar populations and bulge-dominated quiescent galaxies with older stellar populations \citep[e.g.,][]{Hubble:1926aa,Kormendy:2004aa,Simard:2011aa}. However, various dynamical studies have shown that the picture is more complex, with the kinematics and morphology not always tracing the state of star formation. This can be seen through the structural and dynamical variety of early-type galaxies, which are made up of both fast and slow rotators \citep[e.g. review by][]{Cappellari:2016aa}. 

The theoretical framework outlined above allow us to reproduce well the large population of $M_{*}=10^{9-11}\thinspace M_{\odot}$ disk galaxies found in the local Universe \citep{van-der-Wel:2014aa, Zhu:2018aa, Wang:2020aa,Rigamonti:2024aa}. Studies of both ionised and molecular gas kinematics find that these star-forming disks are characterized by high rotational support and low velocity dispersions of the order of $10-20$ km/s \citep{Dib:2006aa, Walter:2008aa, Tacconi:2013aa, Yu:2019aa}. Studies of stellar kinematics, the key tracer of mass assembly history, have found strong correlations between shape (i.e., ellipticity), age, and rotational support. This supports the idea that the young stellar populations in star-forming disks have low stellar velocity dispersions compared to the older bulge-dominated elliptical systems \citep{Cappellari:2007aa, Emsellem:2007aa,Cappellari:2011ab,Krajnovic:2011aa,Naab:2014aa,van-de-Sande:2018aa}, although quiescent galaxies have been found in a variety of kinematic states \citep{Bezanson:2018aa,Newman:2018aa,Ji:2024aa}. Studies of stellar kinematics at high redshift are difficult due to detection limits of current observational facilities, with current measurements reaching $z\approx 2-3$ \citep{Newman:2015aa,DEugenio:2024ab}.

The studies from the local Universe out to cosmic noon (at $z\sim 1-3$) agree that star-forming disks grow inside out at these epochs \citep{ Munoz-Mateos:2007te, Wuyts:2011vn, van-der-Wel:2014aa, Nelson:2016wo, Ji:2023ab}, which aligns with predictions from theoretical models and cosmological simulations \citep{Mo:1998aa,Hasheminia:2024aa}. However, the picture becomes more complex at cosmic noon, where the bulk of the cosmic star formation occurs \citep{Madau:2014aa}. Star-forming disks exist at these epochs but they are less massive, more turbulent, thicker, and more gas-rich than in the local Universe \citep{Genzel:2017aa, Forster-Schreiber:2020aa, Tacconi:2020aa}, and kinematic studies have shown that they are characterized by a range of dark matter profiles \citep{Genzel:2020aa, Bouche:2022aa, Nestor-Shachar:2023aa}. Furthermore, these star-forming galaxies have significant bulges in their cores and their specific star-formation rate (sSFR) radial profiles are roughly flat, indicating little inside-out growth on average \citep{Tacchella:2015aa,Mosleh:2020aa,Jain:2024aa}.

Spatially resolved studies of ionised gas kinematics have shown that the median rotational support $\rotsupp$\footnote{We discuss our definitions of $\rotsupp$ and $\disp$ in Sec. \ref{sec:model-der-params}} of galaxies decreases with redshift, while the intrinsic velocity dispersion $\disp$ increases, from $\sigma_0\sim 10-20$ km/s to average values of $\sigma_0\sim 50$ km/s at $z\sim 2.5$ \citep{Wisnioski:2015vx,Price:2016uv,Simons:2017aa,Turner:2017aa,Ubler:2019vg, Wisnioski:2019tg,Price:2020wf} with some systems going up to $\sigma_0\sim 100$ km/s. Some works find that the higher turbulence is mainly driven by gravitational instabilities in the disk, 
with a less important role of stellar feedback \citep{Krumholz:2018aa, Ubler:2019vg}, while others report that stellar feedback alone is sufficient to explain the measured turbulence in cold gas kinematics across redshifts \citep{Rizzo:2024aa}.  Many works find a dependence of $\disp$ on global star-formation rate (SFR) \citep{Wisnioski:2015vx,Arribas:2014aa,Simons:2016aa,Yu:2019aa, Price:2020wf}, but some studies highlight no local dependence on SFR surface density $\Sigma_{\text{SFR}}$ \citep{Ubler:2019vg}, whose detection would support a scenario of stellar feedback-induced turbulence. However, most works agree on the large scatter in the $\disp-z$ relation, which could be caused by the non-disk systems probing different origins of turbulence (including mergers) from their rotationally-supported disk counterparts. Some of the increase in the measured  $\disp$ could also be explained by an increased contribution of diffuse ionized gas, which typically has higher $\disp$ than gas emitted from HII regions \citep{Della-Bruna:2020aa}, to the \Ha\ flux at high-redshift \citep{Ejdetjarn:2024aa}.

All of these works highlight the importance of linking resolved studies of star formation to the stellar and gas kinematics of galaxies. Studying the dependence of $\disp$\ and $\rotsupp$\ on both redshift and physical galaxy properties such as SFR and its evolution with cosmic time beyond $z>3$ is crucial to understand when and how galaxies settle into rotationally-supported systems.

Many galaxy simulations predict that, on average, disk galaxies only become prominent at $z<4$ \citep{Zolotov:2015aa,Pillepich:2019aa, Lapiner:2023aa, Hopkins:2023aa, Semenov:2024ac}, but observations in the last decade, firstly with the Atacama Large Millimeter Array (ALMA), and now with the James Webb Space Telescope (JWST) have unveiled a different picture. Studies of molecular gas kinematics with ALMA have found dynamically cold (typically $\rotsupp\gtrsim4$) disks out to $z\sim 6$ \citep{Rizzo:2020aa,Lelli:2021aa, Pope:2023aa}, with a $\rotsupp\sim 11$ system found as early as $z = 7.31$ \citep{Rowland:2024aa}. Studies of ionised gas kinematics are now possible out to similar redshifts, and down to lower masses ($\log (M_{\star} [\rm M_{\odot}]) < 9$) and SFR regimes, with JWST, both with the NIRSpec integral field unit (IFU) and slit spectrograph \citep[e.g.,][]{Arribas:2024aa,de-Graaff:2024ab, Ubler:2024aa}, and with NIRCam slitless spectroscopy \citep{Li:2023aa,Nelson:2023ab}. Taken at face value, these studies imply that galaxies are able to settle into cold disks much earlier than many models are able to explain. However, it is unclear whether these studies probe a representative sample of star-forming galaxies.

Nonetheless, recent theoretical work has explored mechanisms to generate disks at high redshift, and some simulations have been successful in promoting the formation of disks beyond $z\sim4$ through smooth accretion of cold gas from the cosmic web filaments \citep{Feng:2015aa,Kohandel:2024aa} and high resolution on-the-fly modelling of the cold interstellar medium (ISM) \citep{Semenov:2024aa, Semenov:2024ab}. While such efforts are encouraging, it is now crucial to understand the kinematics of high-redshift galaxies with representative statistics in order to inform the next generation of models. While purely morphological studies support this idea, with disk-like morphologies found out to $z\sim 7$ \citep{Ferreira:2022aa,Ferreira:2023aa,Robertson:2023aa,Tohill:2024aa,Baker:2025ab}, there has thus far been no systemic study of ionised gas kinematics on a large galaxy sample beyond $z=4$.

The NIRCam grism mode offers a unique opportunity to address the questions arising from these recent results in a statistically significant way. Its $R\sim 1600$ ($\Delta v = 80$ km/s at $\lambda = 3-5 \thinspace \mu$m) spectral resolution allows for the study of kinematic signatures in the resolved 2D spectra. This mode is only available in the long-wavelength filters, yielding a spatial resolution of $\approx 0.12-0.16''$ ($\approx0.9$ kpc at $z\approx5$). One of the main benefits of slitless spectroscopy is that each exposure contains spectra for all of the galaxies in the instrument's field of view, resulting in a (relatively) unbiased large sample of galaxies to analyse. This is in contrast with slit spectrographs such as the NIRSpec micro-shutter assembly (MSA), which can benefit from greater sensitivity and higher spectral resolution, with point source resolution up to $R\sim 4000-5000$ \citep{de-Graaff:2024ab}, and lower sky background noise, but covers fewer galaxies, which are also usually specifically selected, introducing a bias that is hard to quantify. Also, the slit typically only covers part of a galaxy. However, recent efforts in using slit-stepping strategies with the NIRSpec MOS have been shown to produce integral field spectrograph (IFS) data with a significant gain in efficiency over the NIRSpec IFU \citep{Barisic:2024aa,Ju:2025aa}. The latter benefits from high spatial resolution ($\approx 0.1-0.2''$) pixel-by-pixel spectra that provide resolved constraints on the kinematics of the galaxy, ideal for detailed studies of interesting systems \citep[e.g.,][]{Arribas:2024aa,Jones:2024aa,Parlanti:2024aa,Ubler:2024aa}. However, the IFU’s limited field of view generally implies that these studies are restricted to single objects, or relatively small samples. In order to push our understanding of if and how disks form at high redshift, the large samples offered by the NIRCam grism are crucial in order to place single/few object observations with NIRSpec MOS and IFU into a broader picture.

In this paper, we present a systematic study of $\sim 250$ \Ha\ emitters at $z=3.9-6.5$ found in the First Reionization Epoch Spectroscopically Complete Observations survey \citep[FRESCO,][ PI: Oesch, PID: 1895]{Oesch:2023aa}, in both GOODS fields \citep{Giavalisco:2004aa}, and the COmplete Nircam Grism REdShift Survey (CONGRESS, PIs: Egami, Sun, PID: 3577), in the GOODS-N field. We use a new tool, the Grism Emission-line Kinematics tOol (\textsc{geko}; Danhaive et al. in prep), to combine NIRCam F444W grism measurements with NIRCam multi-band imaging from the JWST Advanced Deep Extragalactic Survey  \citep[JADES;][PIs: Rieke and Lützgendorf]{Bunker:2020aa,Rieke:2020aa,Eisenstein:2023aa} in order to infer morphological and kinematic constraints on the \Ha\ emission for each galaxy in the sample.

This paper is organised as follows. The data and their reduction are presented in Sec. \ref{sec:data}, along with the description of the sample selection. In Sec. \ref{sec:methods}, we introduce \geko\ and outline the techniques used to forward-model the NIRCam grism instrument and infer the spatially resolved kinematics of an emission line using a Bayesian inference framework and well-motivated choices of analytical models for the morphology and kinematics of the galaxy. We present our results on the redshift evolution of ionised gas and its relation to stellar populations in Sec. \ref{sec:res-kins}, analysing the observed trends and comparing them to studies at lower redshift. Finally, we interpret our results in the context of the fraction of rotationally supported systems and the build-up of disks with cosmic time (Sec. \ref{sec:discussion}), and we summarize our findings in Sec. \ref{sec:conclusions}. Throughout this work we assume $\Omega_0 =0.315$ and $H_0 = 67.4 \thinspace \text{km}\thinspace\text{s}^{-1}\thinspace\text{Mpc}^{-1}$ \citep{Planck-Collaboration:2020aa}.

\section{Observations and final sample}\label{sec:data}

We will begin by describing the data used in this work, along with its reduction, then proceed to outline the selection criteria used to obtain our final sample.

\subsection{NIRCam grism}
In this work, we use JWST/NIRCam wide field slitless spectroscopy (WFSS) observations in the F444W filter ($3.9-5.0 \thinspace\mu$m) in both GOODS fields obtained with FRESCO \citep{Oesch:2023aa} and in the F356W filter ($3-4 \thinspace\mu$m) with CONGRESS (PIs: Egami, Sun, PID: 3577) in GOODS-N. The observations for both surveys are all taken using the row-direction grisms (GRISMR mode) with modules A and B, and the spectral resolution is $R\sim 1400-1650$ across the filters used.

The data reduction for these observations is described in detail in \cite{Sun:2023aa} and \cite{Helton:2024ab}, so we will only provide a brief overview here. The raw data is first reduced to stage-1 products using the standard JWST pipeline. The stage-1 data is corrected by flat-fields and the sigma-clipped median sky background is subtracted from each individual exposure, and a world coordinate system (WCS) is assigned. We then use the corresponding long wavelength (LW) images taken after each grism exposure to assign every trace to an object and hence apply wavelength calibration to the 2D spectra of each galaxy. The LW images are astrometrically calibrated using the simultaneous SW exposures taken in parallel with the grism. The spectral tracing, grism dispersion, and sensitivity functions are obtained from commissioning and Cycle-1 calibration data \citep[PID: 1076, 1536,
1537, 1538;][]{Sun:2022tt, Sun:2023aa}. In order to obtain 2D continuum-subtracted emission-line maps from the grism data, we use row-by-row median filtering \citep{Kashino:2023wv} in order to remove residual continuum from the object itself and any neighbouring objects that may leave traces in the spectrum. The 2-step (iterative) median filtering used in this technique aims to ensure that the continuum is accurately removed but the emission line flux is not harmed.

Thanks to the $\sim 3-5\thinspace \mu$m range probed by the FRESCO and CONGRESS surveys, we are able to study star-forming galaxies, through their \Ha\ emission, at $z\approx 3.9-6.5$. Our sample is built following the technique from \citet{Helton:2024ab, Sun:2024aa} to accurately search for \Ha\ emitters at these redshifts. First, photometric redshifts are estimated using \textsc{EAZY} \citep{Brammer:2008aa} using the templates and parameters described in \citet{Helton:2024ab} and \citet{Hainline:2024aa}. We select sources based on their \textsc{EAZY} photometric redshifts $z_{\text{phot}}$ and confidence intervals $\Delta z_{\text{phot}}<1$, then extract their grism spectrum following the method outlined above. The extracted 2D spectra are then collapsed to 1D spectra using a boxcar aperture of $0.31''$ (five NIRCam LW pixels). For high signal-to-noise S/N>5 detections, we assign an emission line solution that minimizes the difference between the spectroscopic and photometric redshifts, then perform visual inspection to confirm the redshifts. For the final \Ha\ emitter catalogue, the 1D spectra are optimally extracted with Gaussian profiles to obtain accurate spectroscopic redshifts, emission line fluxes, and their corresponding integrated S/N. We obtain 393 galaxies in CONGRESS and 189 in FRESCO that have \Ha\ detections verifying $\rm S/N > 10$, for a total of 582 emitters \citep[][Lin et al. in prep]{Helton:2024aa}. 

\subsection{NIRCam imaging}\label{sec:imaging}
 
There is a wealth of NIRCam data in the GOODS fields. The FRESCO and CONGRESS grism programs obtained parallel imaging data in the F182M and F210M, and F090W and F115W filters, respectively. In addition, the FRESCO and CONGRESS footprints have a large overlap with JADES \citep{Eisenstein:2023aa,Rieke:2023aa}, a Guaranteed Time Observations (GTO) programme of the NIRCam and NIRSpec instrument teams that obtained imaging of an area of $\sim 175$ arcmin$^2$ in the GOODS-S and GOODS-N fields with an average exposure time of 20 hours with 8-10 filters. For 497 out of the 582 galaxies in our $\rm S/N > 10$ sample, we were able to complement the FRESCO and CONGRESS imaging with wide band $F090W, F115W, F150W, F200W, F277W, F356W$, and deeper $F444W$ observations, and medium band $F335M$ and $F410M$ observations from JADES. For certain regions in our sample, we also use additional medium bands $F182M, F210M, F430M, F460M$, and $F480M$ from the \textit{JWST }Extragalactic Medium Band Survey \citep[JEMS;][]{Williams:2023aa} in GOODS-S.
 Specifically, we use high-resolution data and photometric catalogues obtained from drizzled mosaics, and the full details on the data reduction, catalogue generation, and photometry can be found in \citet{Rieke:2023aa}, and are also detailed in \citet{Robertson:2024aa}. This wide range of filters allows us to obtain tight constraints on the stellar population from spectral energy distribution (SED) fitting (Sec. \ref{sec:modelling-SED}) and accurate morphology priors for our inference tool (Sec. \ref{sec:fitting}).  

\subsection{Sample selection}\label{sec:sample-selection}
\renewcommand{\arraystretch}{2}
\begin{table}
    \centering
    \begin{tabular}{c|c|c|p{1cm}|p{0.5cm}|p{0.5cm}|p{0.7cm}}
    Sample & $N$ & $\rm S/N$ & $|\rm PA_{morph}|$ $[^{\circ}]$ & $r_{\rm e}$ [''] & $\frac{v_{\rm re}}{\Delta v_{\rm re}}$ & $r_{\rm obs}$\\ \hline
    Gold & 41 & $>20$ & $<60$& $>0.12$& $> 1$ & - \\ \hline
    Silver & 132 & $>10$ & $<75$ &$>0.12$ & $> 1$ & - \\ \hline
    Unresolved & 99 &  $>10$&$<75$ &- & - & - \\ \hline \hline 
    Extended & 80 & $>10$ & $<75$& $>0.12$& $> 1$ & $> r_{\rm e}$\\
    \end{tabular}
    \caption{Summary of the sub-samples defined in this work and their selection criteria. The extended sample is a sub-samble of the gold and silver samples. For each category, we report the number of galaxies $N$ it contains, the S/N cut-off, the morphological position angle $\text{PA}_{\text{morph}}$ cut-off, the size $r_{\rm e}$ requirement. We add constraints on whether the velocity gradient at $r_{\rm e}$ has to be resolved, $\frac{v_{\rm re}}{\Delta v_{\rm re}} > 1$, where where $\Delta v(r=r_{\rm e})$ is the 1$\sigma$ lower limit, and constraints regarding the extent $r_{\rm obs}$ of the observed grism emission line map. }
    \label{tab:sample-selection}
\end{table}

\begin{figure}
    \centering
    \includegraphics[width=1\linewidth]{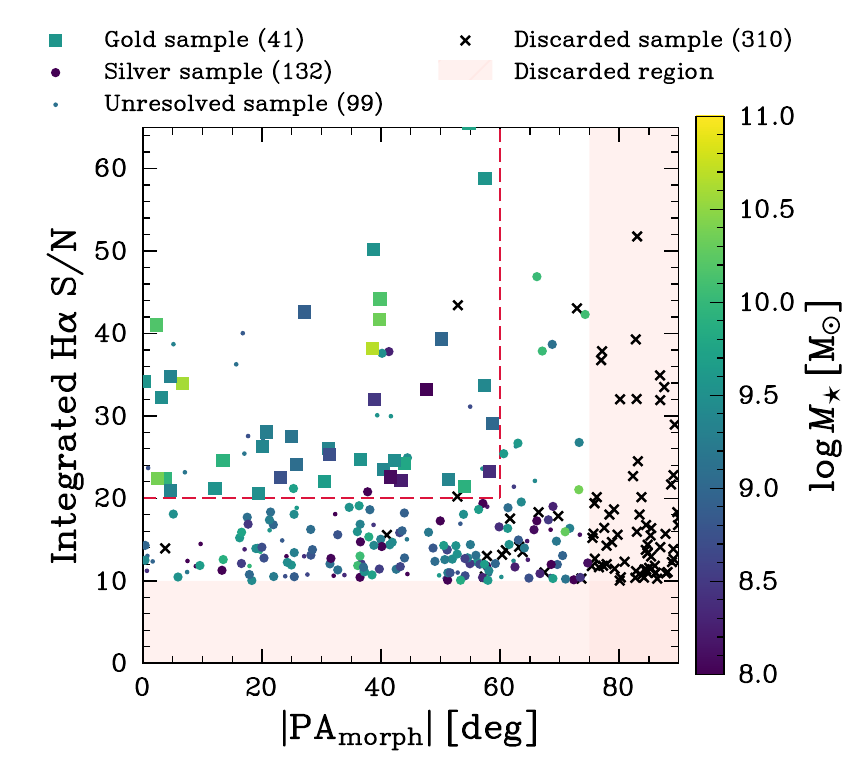}
    \caption{Selection of our sample in the plane of H$\alpha$ S/N and $|\text{PA}_{\text{morph}}|$, colour-coded by stellar mass. The discarded region highlights the parameter space in which galaxies are discarded due to low S/N and/or a PA that is too close to the dispersion direction ($\rm PA = 90^{\circ}$). The $\mathrm{S/N}=20$ and $|\text{PA}_{\text{morph}}| = 60^{\circ}$ cut-off for the gold sample is highlighted with the red dashed line. The small dots represent the unresolved sample of galaxies that have a spatial resolution below the FWHM of the F444W PSF and/or for which we do not measure a resolved velocity gradient. Based on the selection cuts described in Sec. \ref{sec:sample-selection}, we obtain 41 galaxies in the gold sample, 132 in the silver sample, and 99 in the unresolved sample (see Tab.~\ref{tab:sample-selection}).}
    \label{fig:sample-selection}
\end{figure}

\begin{figure}
    \centering
    \includegraphics[width=1\linewidth]{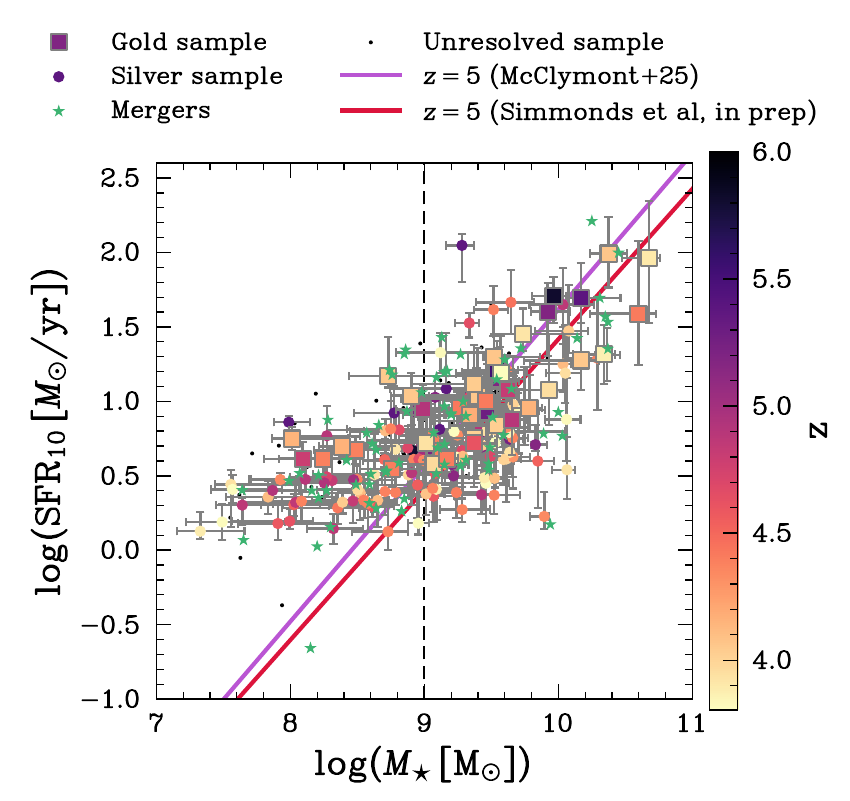}
    \caption{Star-formation rates (SFR$_{10}$) and stellar masses ($M_{\star}$) of our sample, derived with \textsc{Prospector}, colour-coded by the spectroscopic redshifts. The SFRs are averaged over 10 Myr. We compare our sample to SFMS prescriptions from \protect\citet{McClymont:2025aa} and Simmonds et al (in prep.). Our sample is representative of the star-forming galaxy population at $M_{\star}>10^{9}~M_{\odot}$ (indicated by the vertical dashed line). Below this stellar mass, because our sample selection is based on S/N in H$\alpha$ (Fig.~\ref{fig:sample-selection}), it is biased toward high SFRs relative to the SFMS. The discarded merger sample (green stars) is evenly spread across the parameter space and discarding it does not bias our results.}
    \label{fig:SFMS}
\end{figure}

From the emission line catalogue described above, there are $>1000$ \Ha\ detections in the FRESCO and CONGRESS grism data. We introduce a S/N cut for the integrated \Ha\ emission of $\rm S/N > 10$, following recovery tests done on mock data (Danhaive et al. in prep), narrowing the sample down to to a total of 582 galaxies. Below this S/N, we are not able to reliably constrain key kinematic properties $\disp$\ and $\rotsupp$, as we discuss in Appendix \ref{app:geko-tests} (see Fig. \ref{fig:geko-test-kin}), where we present results from the testing process of our tool \textsc{geko} (Sec. \ref{sec:modelling}). We then introduce additional cuts to separate our final sample into three categories, which are summarized in Tab. \ref{tab:sample-selection} and described below:

\begin{itemize}
    \item \textit{Gold sample:} This is the sample containing the galaxies for which we have the most reliable constraints. These galaxies have high integrated \Ha\ S/N ($\mathrm{S/N}>20$) and have a position angle (PA) distinct from the dispersion direction (which is at $\text{PA}_{\text{disp}} = 90^{\circ}$ \footnote{We adopt a convention where the position angles $|\text{PA}_{\text{morph}}|$ and $|\text{PA}_{\text{kin}}|$ are defined from the vertical spatial $y$-axis in a clockwise way.}): $|\text{PA}_{\text{morph}}| < 60^{\circ}$. $\text{PA}_{\text{morph}}$ is measured from the imaging data (see Sec. \ref{sec:morph-fit}). Throughout the paper, this sample of 41 galaxies is plotted with squares.
    \item \textit{Silver sample:} This sample contains the galaxies with $\text{S/N}\geq 10$ and has a looser PA cut $|\text{PA}_{\text{morph}}| < 75^{\circ}$. Throughout the paper, this sample of 132 galaxies is plotted with circles.
    \item \textit{Unresolved sample: }This sample contains the galaxies whose inferred \reff\ is smaller than half of the FWHM of the F444W PSF and/or whose velocity gradient is unresolved, i.e. its value is consistent with zero within $1\sigma$. We hence qualify galaxies as unresolved if $\rm|v(r=r_{\rm e})|/\Delta v(r=r_{\rm e}) < 1$, where $\Delta v(r=r_{\rm e})$ is the 1$\sigma$ lower limit. This final cut can of course only be done after the kinematic fitting is completed (Sec. \ref{sec:fitting}). This sample of 99 galaxies will be plotted with dots.

    \item \textit{Extended sample: } This sample contains 80 galaxies which are in the gold or silver sample and have grism data which extends beyond the best-fit effective radius of the \Ha\ emission. 
    \item \textit{Discarded sample: }This sample of 310 galaxies is not included in the paper. It contains galaxies whose fits provided large residuals (often due to clumpy/merging morphology) and hence need further analysis. It also contains galaxies that are almost parallel to the dispersion direction $|\text{PA}_{\text{morph}}| > 75^{\circ}$ and therefore cannot be kinematically constrained, and galaxies with a low \Ha\ integrated S/N ($\text{S/N}<10$). These cuts are all based on testing done on mock data (Danhaive et al. in prep). Finally, we also discard galaxies whose best-fit kinematic and morphological position angles are in disagreement $|\text{PA}_{\text{kin}} - \text{PA}_{\text{morph}}| > 30^{\circ}$.
\end{itemize}
The gold, silver, and unresolved categories make up a total sample of $272$ \Ha\ emitters. Fig. \ref{fig:sample-selection} shows the separation of the sample in these categories to highlight the cuts made.  In the rest of the paper, we will plot these different samples in similar ways as this plot to distinguish them.
In the 393 (CONGRESS) plus 189 (FRESCO) galaxies with S/N$>10$, we found (and discarded) 108 additional objects that showed a merging structure in the NIRCam imaging. These objects do not show clustering in their SFRs or stellar masses (Fig. \ref{fig:SFMS}), so discarding them should not impact our sample in this plane of parameter space. We discuss the impact of excluding these systems on our inferred disk fraction in Sec. \ref{sec:disk-fracs}. Additionally, 14 systems were classed as AGN due to their broad \Ha\ emission in the NIRCam grism spectra, indicative of an AGN broad line region (BLR). Out of these AGN candidates, seven are confirmed Little Red Dots (LRDs) \citep{Matthee:2024aa}. We exclude these objects from our sample as they require multiple component modelling to account for the BLR, which is an opportunity for future study but out of the scope of this work.

We obtain SFRs and stellar masses from \textsc{Prospector} modelling of both photometry from NIRCam imaging and emission line fluxes from NIRCam grism (details in Sec. \ref{sec:modelling-SED}), and find that our sample is representative of galaxies on the star-forming main sequence (SFMS) at $z\sim5$, as shown on Fig. \ref{fig:SFMS}, using the SFMS prescription from Simmonds et al. (in prep). This SFMS is derived from \textsc{Prospector} modelling of JADES galaxies from $z=3-9$, 90\% complete in stellar mass down to $\log (M_{\star}~\rm [M_{\odot}]) \approx 7.5$, presented in \citet{Simmonds:2024ab} for GOODS-S and extended to include galaxies in GOODS-N (Simmonds et al., in prep).

We also include the prescription from the \textsc{THESAN-ZOOM} simulations \citep{Kannan:2025aa}, presented in \citet{McClymont:2025aa}. Our sample lies on the relations above $M_{\star}\approx10^{9}~M_{\odot}$, but we note that, due to our S/N cut, the low-mass end ($M_{\star}\lesssim10^{9}~M_{\odot}$) of the SFMS is biased to systems that are brighter and hence more star-forming. 

\begin{figure}
    \centering
    \includegraphics[width=1\linewidth]{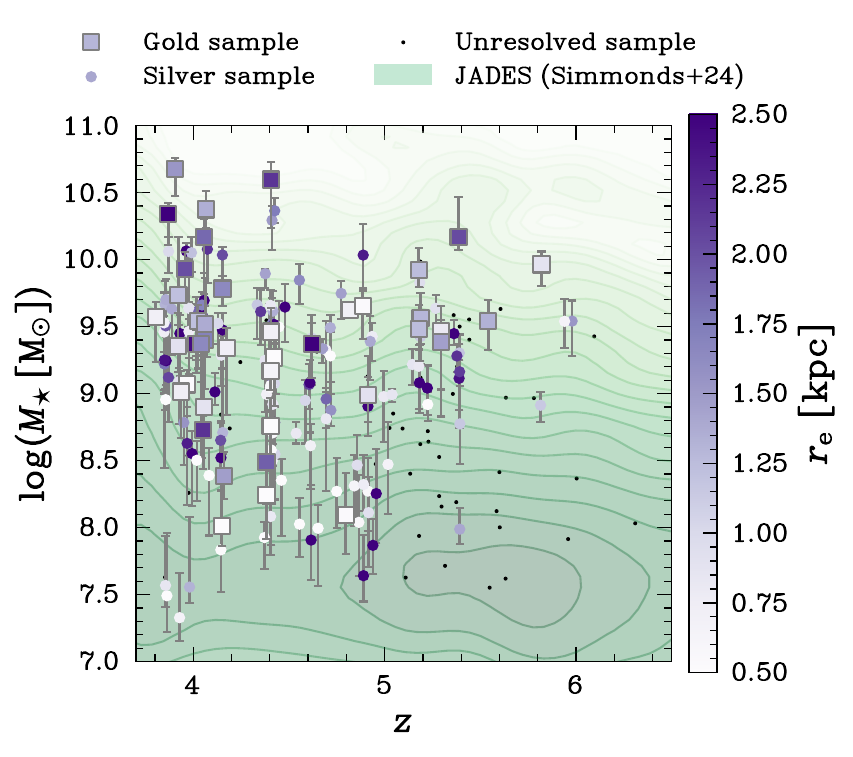}
    \caption{Distribution of our sample in the stellar mass ($M_{\star}$)-redshift plane. Our gold sample spans a wide range in stellar mass, but lies preferentially at $z<5$. There are overall fewer galaxies at $z>5$, with only few low mass systems ($M_{\star}\lesssim10^{9}~M_{\odot}$) falling in the gold and silver samples. Overall, our sample probes the high mass end of photometric candidates, as shown by the green contours representative of the JADES sample from \citet[][and in prep]{Simmonds:2024ab}, which is expected from our S/N cut. We colour-code galaxies in the gold and silver samples by their effective radius $r_{\rm e}$ in the rest-frame near-UV, highlighting that although more massive galaxies are typically larger, at fixed stellar mass, galaxies span a wide range of sizes.}
    \label{fig:Mstar-z}
\end{figure}

In terms of the redshift distribution of the galaxies in our sample, we show in Fig. \ref{fig:Mstar-z} that galaxies in the lower redshift region ($3.8<z<5$) have more galaxies and span a wider range in stellar mass. On the other hand, the higher redshift ($5<z<6.5$) galaxies are, in comparison, biased to higher masses, where most gold and silver sample systems lying above $M_{\star}\sim10^{9}~M_{\odot}$. The effective radii, measured from rest-frame near-UV imaging, indicate that our galaxies are representative of the population at $z\sim 4-6$, consistent with \citet{Allen:2024aa} (see also Danhaive et al. in prep). At fixed stellar mass, galaxies span a wide range of sizes, as can be seen from the colour-coding in Fig. \ref{fig:Mstar-z}. The distribution of the UV sizes in arcseconds for the gold and silver sample galaxies (see Appendix \ref{app:tables}, Fig. \ref{fig:r_eff_hist}) does not show a strong bias, although our gold sample galaxies do not probe the extended tail visible for the silver sample. Our $\rm S/N$ cut likely causes this bias, since at fixed stellar mass, more compact galaxies have a higher integrated $\rm S/N$. 

\section{Inferring stellar populations, morphology, and kinematics}\label{sec:methods}

In order to obtain a global view of the objects in our sample, we combine NIRCam imaging and spectroscopy, along with a wide range of techniques, to infer the stellar populations, morphologies, and kinematics of the galaxies in our sample. In Sec. \ref{sec:modelling-SED}, we describe our SED-fitting approach with \textsc{Prospector} to obtain constraints on the stellar mass and SFRs. Then, we briefly introduce our method for fitting one-component S\'{e}rsic models to our data with \textsc{pysersic} (Sec. \ref{sec:morph-fit}) in order to characterize the galaxy's morphology. Finally, Secs. \ref{sec:modelling} and\ref{sec:fitting} introduce a novel code \textsc{geko} and describes how we use it to infer key kinematic and morphological parameters for the emission lines probed by the grism data. 

\subsection{SED modelling}\label{sec:modelling-SED}

In order to characterize the stellar populations in the galaxies in our sample, we use the SED inference code \textsc{Prospector} \citep{Johnson:2021aa}. We fit all of the available photometry from the JADES survey, which contains all of the NIRCam wide and some medium band filters ($F070W, F090W, F115W$, $F150W, F162M, F200W$, $F250M, F277W, F300M, F335M, F356W, F410M$, and $F444W$), the MIRI filters $F770W$ and $F1280W$, and in certain regions also included medium bands from the JEMS and FRESCO surveys ($F182M, F210M, F430M, F460M$, and $F480M$). We also use photometry from the Hubble Space Telescope (HST) Advanced Camera for Surveys (ACS) bands $F435W, F606W, F775W, F814W$, and $F850LP$ \citep[the GOODS survey][]{Dickinson:2004aa,Giavalisco:2004aa}. Following \citet{Tacchella:2023aa}, we self-consistently model the photometry with \Ha\ line fluxes from NIRCam grism data from the FRESCO and CONGRESS surveys, and, for FRESCO galaxies in the GOODS-N field, we also include, when detected, $[\text{OIII}]5007$\AA, $[\text{OIII}]4959$\AA, H$\beta$, and [SIII]$9533$\AA\ fluxes from the CONGRESS survey. 

Within \textsc{Prospector}, there is the option to allow for an overall rescaling of the input emission line fluxes, in order to account for potential discrepancies in the flux calibrations of the different instruments. This rescaling factor then becomes a free parameter of the model, and a difference in calibrations would likely manifest as a similar ``constant'' best-fit factor found throughout our sample. However, we find no over-arching consistent rescaling factor, meaning the best-fit value changes significantly between objects. Because the best-fit values for stellar masses and SFRs remain consistent within the uncertainties in both cases, with and without rescaling, we adopt the model with no rescaling.  We also use the spectroscopic redshifts derived from the grism spectra to fix the redshift in the fits, which allows us to break the degeneracy between redshift and physical properties such as stellar mass.

In terms of priors, we assume a non-parametric continuity star formation history (SFH) prior \citep{Leja:2019aa} with six SFR bins. We assume a Student's t-distribution, with a width of $\sigma=0.5$, for the ratio of the SFR in each bin, which allows for more variable SFHs than the standard continuity prior \citep{Tacchella:2022tc}. We assume a two-component dust attenuation model \citep{Charlot:2000aa,Conroy:2009aa}, where the first component (a birth-cloud component) only affects the nebular and stellar emission from young stars formed in the last 10 Myr, and the second acts on the full emission from the galaxy (a diffuse component). We use the attenuation curve from \cite{Kriek:2013aa} for the diffuse dust component, where the dust bump at 2175 \AA\ is directly coupled to the dust index $n$, which is modelled as an offset from the \cite{Calzetti:2000uh} attenuation curve. We include self-consistent modelling of the nebular emission, coupling the number of Lyman-continuum photons to the SFH \citep{Byler:2017aa}. 

Using the wide range of photometry and the additional constraints from the grism data, we infer key properties such as stellar mass, SFR, and dust attenuation for all of the galaxies in our sample. \textsc{Prospector} also produces a best-fit SFH for each galaxy, which can help inform us on the burstiness of star formation and the overall mass assembly history. Finally, we note that the definition of stellar mass that we use in this paper is the mass of stars and remnants, and not the full integral of the SFH. This effectively means that we correct the SFH integral by removing mass that is put back in the ISM.

\subsection{Inferring morphology} \label{sec:morph-fit}

In order to obtain constraints on the morphological parameters of our galaxies, we use the fully Bayesian code \textsc{pysersic} \citep{Pasha:2023aa}. \textsc{pysersic} models imaging data from any filter to infer the best fit S\'{e}rsic profile(s) parameters. We fit photometry in relevant filters (see Sec. \ref{sec:fitting}) with a one-component S\'{e}rsic model for every galaxy in our sample to obtain estimates of the position angle $\theta$, ellipticity $e$, S\'{e}rsic index $n$, light centroid $(x_0,y_0)$, and half-light radius $r_{\rm e}$. We note that prior to fitting, the images from JADES (Sec. \ref{sec:imaging}) are PA-matched to the grism data in order to compare the morphological PA to the dispersion direction. Also, as described in Sec. \ref{sec:dyn-mod}, we assume an intrinsic axis ratio $q_0$ when deriving the inclination from the ellipticity. Through this modelling, we are able to highlight mergers through their strong residual in single component fits. We also fit the continuum-substracted grism emission line with \textsc{pysersic} in order to obtain an estimate of the total \Ha\ flux. Because of how we define the position angle from the vertical y-axis, we note the conversion $\text{PA}_{\text{morph}} = 90 - \theta$ degrees, where $\theta$ is the \textsc{pysersic} output angle computed from the positive x-axis.

\subsection{Inferring the kinematics} \label{sec:modelling}

In this section, we describe our methods for jointly inferring morphological and dynamical parameters from imaging and grism data using the forward-modelling and Bayesian inference tool \textsc{geko}. We follow a similar methodology to \citet{Price:2016uv}, which demonstrated such a forward modelling approach for kinematic inference with only one spatial dimension \citep[see also][]{Li:2023aa, de-Graaff:2024ab}. We will summarize the main aspects of the code here, but refer to an upcoming paper for a full description of the code and testing on mock and real data (Danhaive et al., in prep). We note that in this work, we focus on the \Ha\ emission, but this methodology can be extended to any other emission line.

\subsubsection{Dynamical modelling} \label{sec:dyn-mod}
The simplest assumption one can make about a galaxy's kinematics is to approximate the galaxy as a thin rotating disk, which is a good assumption for the disk galaxies we observe in the local Universe. Because these exponential disks are embedded in dark matter haloes, their kinematics are well modelled by the arctangent function \citep{Courteau:1997uw, Miller:2011aa}:
\begin{equation}
    V_{\text{rot}}(r_{\text{int}}, r_{\text{t}}, V_a) = \frac{2}{\pi}V_a\arctan{\frac{r_{\text{int}}}{r_{\text{t}}}},
\end{equation}
where $V_{\text{rot}}$ is the rotational velocity at a given radius $r_{\text{int}}$ in the intrinsic galaxy plane, $V_a$ is the asymptotic value that the arctangent rotation curve tends to at large radii $r_{\text{int}} \rightarrow \infty$ , and $r_{\text{t}}$ is the turn-around radius of the rotation curve.
To project this velocity on the observation plane, we need to account for the galaxy's inclination $i$:
\begin{equation}
        V_{\text{obs}}(x,y) = V_{\text{rot}}(r_{\text{int}}, r_{\text{t}}, V_a)\cdot\sin{i}\cdot\cos{\phi_{\text{int}}},
\end{equation}
where $\phi_{\text{int}}$ is the polar angle coordinate in the galaxy plane. 
Because galaxies are thought to be thicker at high-redshift \citep{van-der-Wel:2014aa}, we compute the inclination assuming an intrinsic disk axis ratio, which is the ratio of scale height to scale length, $(b/a)_0 = q_0 = 0.2$ \citep{Wuyts:2016aa, Genzel:2017aa, Price:2020wf,Ubler:2024aa}, where $q_0 = 0$ corresponds to the thin disk approximation. This gives for the inclination $i$
\begin{equation}
    \cos{i} = \left(\frac{q^2 - q_0^2}{1-q_0^2}\right)^{1/2},
\end{equation}
where $q=b/a$ is the measured axis ratio. We note that the choice of $q_0$ between common values of $q_0=0.0-0.2$ does not strongly affect the inferred kinematics properties \citep{Forster-Schreiber:2020aa,Price:2020wf}.
We model a constant isotropic velocity dispersion $\sigma_0$ across the disk based on deep adaptive optics imaging spectroscopy studies at lower redshift \citep{Genzel:2008ug, Forster-Schreiber:2018aa}. Although some high-resolution observations of molecular and atomic gas in the local Universe have shown radially declining velocity dispersions, the observed radial changes are well below the NIRCam grism spectral resolution \citep[for a more detailed discussion see][and references therein]{Ubler:2019vg}. We also introduce a velocity offset $v_0$, which acts as an effective redshift offset in the observed wavelength of H$\alpha$ by shifting the full 2D emission line map. For this disk model, the free parameters are therefore the inclination $i$, the position angle $\text{PA}_{\text{kin}}$, the asymptotic velocity $V_a$, the turn-around radius $r_t$, the intrinsic velocity dispersion $ \sigma_0$, and the net velocity offset $v_0$. We fix the centre of the velocity map to the centre of the light distribution $(x_0,y_0)$, which is fit simultaneously to the kinematics (Tab. \ref{tab:priors}). It is important to note that no assumption is made on the nature of the gravitational support (rotation
or pressure). The only assumption we do is that the gaseous disk is infinitely thin, without any constraint on the amplitude of the velocities or velocity dispersions.

\subsubsection{Morphological modelling}

In order to model the flux distribution of the chosen emission line, \textsc{geko} has a parametric and non-parametric option. For the scope of this study, the parametric flux modelling was chosen due to the small observed sizes of our galaxies ($r_{e,\text{H}\alpha}\lesssim 3$ kpc), and we will not expand on the non-parametric model here. We parametrize the flux distribution with a S\'{e}rsic profile \citep{Sersic:1968aa}:
\begin{equation}
    I(r) = I_e \exp\Biggl( -b_n\left[ \left(\frac{r}{r_{\rm e}}\right)^{1/n}-1\right]\Biggr),
\label{eq:S\'{e}rsic_profile}
\end{equation}
where $I_e$ is the intensity at the half-light radius $r_{\rm e}$ and $n$ is the S\'{e}rsic index. The radius $r$ is defined based on coordinates centred on $(x_0,y_0)$ on the sky, which are two additional free parameters of the model. For $b_n$, we use the polynomial expansion from \citet{Ciotti:1999aa}\footnote{$b(n) = 2n - 1/3 + 4/405n + 46/25515n^2 +131/1148175n^3 - 2194697/30690717750n^4$.}, which provides a good approximation \citep{Graham:2005aa}.
This morphological model also needs a position angle, \PAmorph, which we let differ from the kinematic position angle, and an inclination $i$ which we fix to the kinematic inclination. 

In order to correctly sample the inner radii of the S\'{e}rsic profile, we oversample our model grid by a factor of 125. In general, the model space has 5 times the spatial resolution of the observed images and 9 times the spectral resolution of the grism data. This is to ensure that we are accurately computing our models on a fine grid before sampling down to the instrument resolution. 

We obtain a 3D cube $I(x_0,y_0,\lambda)$ by convolving the arctangent velocity field with our flux profile, with two spatial dimensions and one spectral dimension. This cube can then be collapsed onto the 2D detector plane, going from the high-resolution 3D model space to the low-resolution 2D observation space thanks to a careful forward-modelling of the instrument.

\subsubsection{Instrument modelling}

The next step in the forward modelling of grism data is to project our data cube $I(x,y,\lambda)$ onto the NIRCam detector plane. The instrument is defined by a well-calibrated dispersion function $dx$ which dictates where a pixel at position $(x_0,y_0)$ on the image and emitting at a wavelength $\lambda_0$ will end up on the detector $x_d=x_0 + dx$ \citep[for full details, see][]{Sun:2023ab}. This function is calibrated for each filter/pupil/module combination. Due to the limited spectral extent of the region we are modelling (the emission line), we assume no spectral tracing, $dy = 0$. Before projecting the cube onto the detector grid D$(x,y)$, we convolve it with the filter-specific point spread function, PSF$(x,y)$ and the line-spread function LSF$(x,y)$ 
\begin{equation*}
    I(x,y,\lambda) \otimes \text{PSF}(x,y) \otimes \text{LSF}(\lambda) \rightarrow \text{D}(x,y) .
\end{equation*}
The convolution is done in the high dimensional 3D model space, before collapsing to the spatial x-axis and resampling the model down to the instrument resolution. We approximate the PSF in each filter using the model PSFs (mPSFs) from \citet{Ji:2024aa} constructed by mosaicing WebbPSF models repeatedly over the field identically to our exposure mosaics and then measuring the average PSF. The model PSFs are then resample down to the grism resolution. The wavelength-dependent one-dimensional LSF is modelled as the sum of two Gaussian distributions, to account for the broader wings.
It was calibrated using SMP-LMC-058, a compact planetary nebula in the LMC \citep[see also][]{Jones:2023aa} using extraction from JWST commissioning, Cycle-1 and 2 calibration observations (Sun et al., in prep.). The spectral resolution as function of wavelength is similar to the pre-launch measurements \citep{Greene:2017ui}. Based on the measurement for the four combination of modules (A or B) and dispersion directions (R or C), we estimate a typical error on the LSF of 10\%. We can now join our analytical models for galaxy morphology and kinematics with the forward modelling of the instrument in a Bayesian inference framework.

\subsection{Fitting}\label{sec:fitting}

\renewcommand{\arraystretch}{2}

\begin{table*}
    \centering
    \begin{tabular}{c|c|c|p{6cm}}
    Name & Parameter Description & Prior & Prior Description \\ \hline \hline 
$\text{PA}_{\text{morph}}$ & Position angle of \Ha\ morphology & Normal$(\mu_{\rm PA}, \sigma_{\rm PA})$ & Normal prior, in degrees, based on the image \textsc{pysersic} fit results. \\ \hline
$i$ & Inclination angle & TruncNormal$_{[0, 90]}(\mu_i, \sigma_i)$ & Truncated normal prior based on the image \textsc{pysersic} fit results for ellipticity. \\ \hline
$A$ & Amplitude of brightness & TruncNormal$_{[0, \infty]}(\mu_A, \sigma_A)$ & Normal prior, in degrees, based on the grism \textsc{pysersic} fit results. \\ \hline
$r_\text{e}$ & Effective radius & TruncNormal$_{[r_\text{min}, r_\text{max}]}(\mu_r, 2 \cdot \sigma_r)$ & Truncated normal prior in pixels based on the image \textsc{pysersic} fit results for effective radius with boosted uncertainty. \\ \hline
$n$ & Sérsic index & Normal$(\mu_n, 2 \cdot \sigma_n)$ & Normal prior, in degrees, based on the image \textsc{pysersic} fit results with boosted uncertainty. \\ \hline
$x_0$ & X--coordinate of center & Normal$(\mu_{x_0}, 2 \cdot \sigma_{x_0})$ & Normal prior, in degrees, based on the image \textsc{pysersic} fit results with boosted uncertainty. \\ \hline
$y_0$ & Y--coordinate of center & Normal$(\mu_{y_0}, 2 \cdot \sigma_{y_0})$ & Normal prior, in degrees, based on the image \textsc{pysersic} fit results with boosted uncertainty. \\ \hline \hline

$\text{PA}_{\text{kin}}$ & Position angle of kinematics & Normal$(\mu_{\rm PA}, \sigma_{\rm PA})$ & Same prior as $\text{PA}_{\text{morph}}$, but fit independently. \\ \hline
$V_a$ & Asymptotic rotational velocity & Uniform$(0, 1000)$ & Uniform prior from 0 to 1000 km/s, with modifiable upper limit. \\ \hline
$\sigma_0$ & Intrinsic velocity dispersion & Uniform$(0, 500)$ & Uniform prior from 0 to 500 km/s, with modifiable upper limit. \\ \hline
$r_t$ & Turnover radius & Uniform$(0, r_\text{e})$ & Uniform prior from 0 to $r_\text{e}$ pixels. \\ \hline
$v_0$ & Systemic velocity & Normal$(0, 50)$ & Normal prior with mean 0 and standard deviation 50 km/s. \\ \hline
    \end{tabular}
    \caption{Description of morphological (top section) and kinematic parameters (bottom section) being fit as free parameters in our \textsc{geko} modelling, along with their priors. For all of the parameters $p$ with Gaussian priors based on the Pysersic modelling, we refer to the best-fit value as $\mu_p$ and its uncertainty $\sigma_p$.}
    \label{tab:priors}
\end{table*}

\begin{figure*}
    \centering
    \includegraphics[width=1\linewidth]{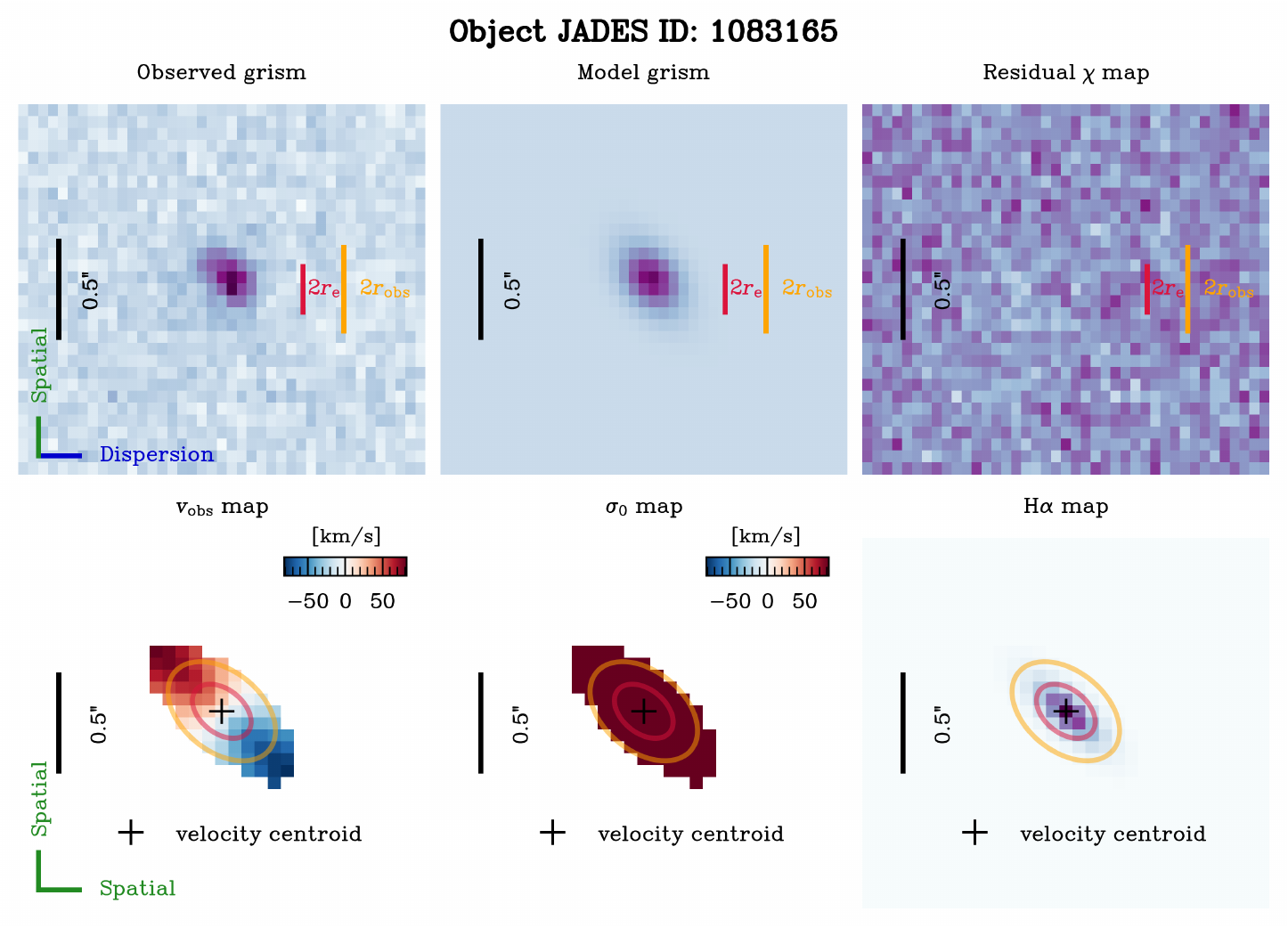}
    \caption{Example science products obtained with \textsc{geko} for a galaxy in our gold sample. In the top panels, we show the observed grism data, the best-fit model, and the corresponding residuals computed with the $\chi$ metric. The plotting range is $\chi = [-5, - 5]$. We also highlight the best fit effective diameter $D_e = 2r_{\rm e}$ for the \Ha\ emission, and the (projected) diameter of the $\text{S/N}>3$ observed map. In the bottom panel, we show the derived best-fit velocity field, velocity dispersion field (which has the same velocity scale as the velocity field plot), and the intrinsic H$\alpha$ emission map. In all of the panels on the left side, we show the scale in arcseconds (black) along with the velocity centroid (black cross). In each row, we show the physical meaning of the axes (spatial vs dispersion).}
    \label{fig:geko-summary}
\end{figure*}

\begin{figure*}
    \centering
    \includegraphics[width=1\linewidth]{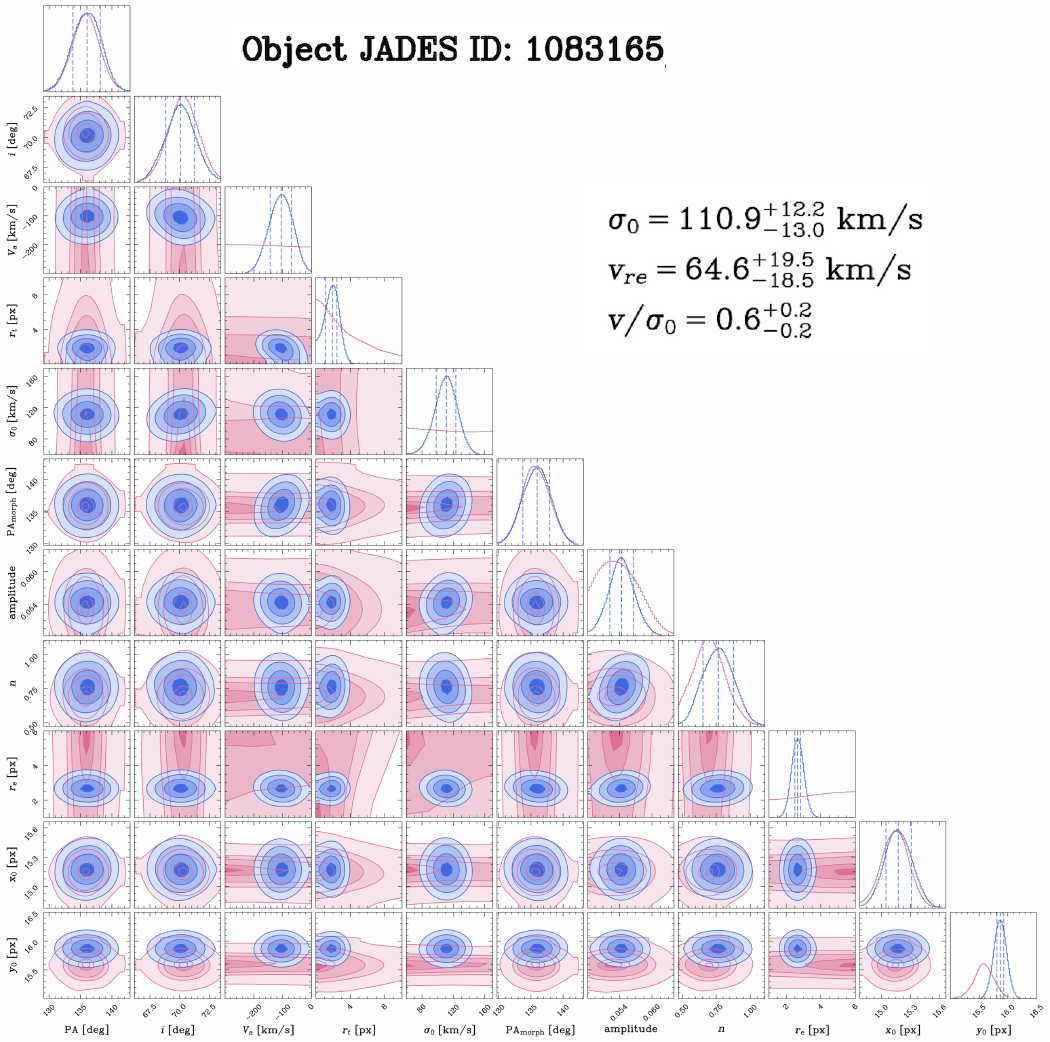}
    \caption{Prior (pink) and posterior (blue) distributions inferred with \textsc{geko} for the same gold sample galaxy as in Fig. \ref{fig:geko-summary}, for the free parameters of our model (Tab. \ref{tab:priors}). From these, we derive the posterior distributions for the rotational velocity at $v_{\rm re}$ and the rotational support $\rotsupp$, whose best fit values and uncertainties, computed from the $16^{\rm th}$ and $84^{\rm th}$ quantiles, are shown on the top right corner.}
    \label{fig:geko-corner}
\end{figure*}

To place the forward modelling in an inference framework, \textsc{geko} uses the Bayesian inference tool \textsc{Numpyro} \citep{Phan:2019wc}, which is based on \textsc{JAX}, a python library tailored for array-oriented numerical computation, with automatic differentiation and just-in-time (JIT) compilation, and optimized for running code on graphic processing units (GPUs). This state-of-the-art approach allows us to run \textsc{geko} on a large number of galaxies, with $\sim 20-60$ minutes per galaxy depending mainly on the S/N. 

\subsubsection{Physically motivated priors}

A key step when using Bayesian inference is choosing appropriate priors. In order to set informed priors for the morphological modelling, we run \textsc{pysersic} on the F115W (or F182M) imaging for all of our objects (Sec. \ref{sec:morph-fit}), obtaining PSF-deconvolved measurements of the effective radius \reff, the position angle $\text{PA}_{\text{morph}}$, the inclination $i$, the S\'{e}rsic index $n$, the total \Ha\ flux, and the light centroid $(x_0,y_0)$. In our redshift range $z\sim3.9-6.5$, F115W and F182M probe the near-UV continuum ($\lambda_{\rm rest} \sim 2000 \rm ~\mathring{A}$), so we use scaling relations to infer estimates of \reff\ for the \Ha\ emission based on Villanueva et al. (in prep). Specifically, for our prior, we boost the near-UV sizes by a factor of 1.58, which is also consistent with the measured increase of \Ha\ equivalent widths at large radii \citep{Matharu:2024aa,Nelson:2024aa}.

In Villanueva et al. (in prep), they carefully model and subtract the rest-optical continuum from the F356W photometry by using non-contaminated neighbouring medium band filters. They infer \Ha\ morphologies at $z\sim4-5$ and compare them to the near and far UV and optical morphologies. However, this method for inferring \Ha\ morphologies directly from the photometry only applies in the redshift ranges where \Ha\ falls in a wide band and outside neighbouring medium bands. Also, it would be less precise at $z>5$ where \Ha\ falls in the F444W filter, since most of the FRESCO and CONGRESS footprints have no F460M or F480M imaging, with the exception of the JEMS area where the overlap is minimal. For these reasons, we instead use the conversions they derive and use the near UV morphologies, with boosted sizes, so we can have a consistent prior for all the galaxies in our sample. Using a shorter wavelength filter also has the benefit of a smaller PSF and hence more accurate measurements.

We use Gaussian priors for all of these morphological parameters, doubling the uncertainty obtained from \textsc{pysersic} since we are not measuring the true emission-line morphology, and to account for scatter in the relations in Villanueva et al. (in prep).  For the kinematic parameters, we use uniform priors on $\disp$, $V_a$, and $r_t$, where this last parameter is bounded by \reff, motivated by lower redshift studies finding $r_t \sim 0.25 \thinspace r_{\rm e}$ \citep{Miller:2011aa}. The prior for the kinematic position angle \PAkin is the same as for \PAmorph. The free parameters of our model and their priors are summarized in Tab. \ref{tab:priors}.

For a given set of parameters, the likelihood is computed by comparing the grism model and observations and assuming a pixel-by-pixel normal distribution. In order to sample our posterior, we use the No U-Turn Sampler (NUTS; \citealt{Hoffman:2011wm}) which is an optimized Hamiltonian Monte Carlo (HMC) sampler. We fit the grism data, centred on the emission line of interest, to derive the posterior distributions of all of the free parameters of the model: $i$, \PAkin, $r_t$, $V_a$, $\disp$, $v_0$, \PAmorph, $r_{\rm e}$, $I_e$, $n$, $x_0$, and $y_0$. 

Figs. \ref{fig:geko-summary} and \ref{fig:geko-corner} show an example of the outputs of \textsc{geko}. On the top panel of Fig. \ref{fig:geko-summary}, we show the observed continuum subtracted grism data centred on \Ha, followed by our best fit model and the residual $\chi$ map which is defined by:
\begin{equation}
    \chi = \frac{\text{model} - \text{obs}}{\text{obs uncertainty}},
\end{equation}
and where the bounds for plotting are $\chi = [-5, 5]$. We also highlight the best-fit effective radius $r_{\rm e}$ of \Ha\ which is well probed by the grism data. On the bottom row we show the inferred velocity, velocity dispersion, and intrinsic flux maps for \Ha.  The centre of the velocity curve is marked by a black cross on all three plots. In Fig. \ref{fig:geko-corner} we show the posterior distributions of all of the free parameters of the model. We also highlight the best fit values and uncertainties, computed from the $16^{\rm th}$ and $84^{\rm th}$ quantiles, for the three key kinematic parameters studied in this work: the intrinsic velocity dispersion $\disp$, the rotational support $\rotsupp$, and the dynamical mass $M_{\text{dyn}}$. These last two are not free parameters of the model, and are instead derived from them. 

\subsubsection{Model-derived parameters}\label{sec:model-der-params}

We will now briefly outline our definitions for the model-derived parameters. In kinematic studies, a variety of different prescriptions are used to define rotation and dispersion, depending on how many spatial dimensions are probed by the data and on the focus of the work. In this work, and similar to \citet{Price:2016uv,Price:2020wf}, we define the intrinsic velocity dispersion $\disp$ as constant throughout the galaxy and measure it as a broadening that is equally applied across the system. In IFU studies, the velocity dispersion can be measured at (or averaged over) different radii, which can introduce differences in the derived relations \citep[e.g., see discussion in][]{Stott:2016aa}. 

For the rotational velocity $v_{\rm rot}$, and the derived rotational support $\rotsupp$, the prescriptions are even more varied. Some works measure $v_{\rm rot}$ at a certain radius, such as the effective radius $r_{\rm e}$ \citep[e.g.,][]{de-Graaff:2024ab}, $r=2.2r_{\rm s}$\footnote{$r_{\rm s}$ is the disk scale length $r_{\rm s}\approx 0.6 r_{\rm e}$ \citep{Burkert:2010aa, Miller:2011aa}.} \citep[e.g.,][]{Price:2016uv, Price:2020wf}, or $r=2.2r_{\rm e}$ \citep[e.g.,][]{Stott:2016aa}. Others instead use the maximal observed velocity gradient across the kinematic axis \citep[e.g.,][]{Wisnioski:2015vx}, or the maximum observed rotational velocity \citep[e.g.,][]{Rowland:2024aa}. Due to the grism sensitivity constraints, we do not probe \Ha\ emission to large enough radii and hence measurements at $r=2.2r_{\rm s}$ or $r=2.2r_{\rm e}$ would be extrapolations. In this work, we choose to measure the rotational velocity at the effective radius $r_{\rm e}$ because that is where we have constraints from our data. We also explore using the maximal observed rotational velocity $v_{\rm rot} = v_{\rm obs,max}$, but we do not find significant changes in our conclusions (see Appendix \ref{app:method-comparison} for full discussion), although the quantitative trends slightly change (by $\lesssim 10\%$).

In order to infer constraints on dynamical masses, we first need to compute circular velocities $v_{\text{circ}}$. The circular velocity is computed assuming an asymmetric drift correction to account for the pressure support \citep{Newman:2013aa, Wuyts:2016aa}, which is even more prominent in the turbulent high-redshift galaxies in our sample. We hence defined this term as in \cite{Price:2020wf} 
\begin{equation}
     v_{\text{circ}}(r) = \sqrt{v_{\text{rot}}^2(r) + 2(r/r_{\rm s})\sigma_0^2}.
    \label{eq:v_circ}
\end{equation}

At the effective radius \reff, where we compute the circular velocity $ v_{\text{circ}}(r_{\text{e}})$, we have $2(r_{\rm e}/r_{\rm s}) = 3.36$ under the assumption of an exponential disk. We compute the total dynamical mass following
\begin{equation}
    M_{\text{dyn}} = k_{\text{tot}}\frac{r_{\rm e}v^2_{\text{circ}(r_{\rm e})}}{G},
    \label{eq:dyn-mass}
\end{equation}
where $G$ is the gravitational constant and $k_{\text{tot}}$ is the virial coefficient \citep{Price:2020wf}. Because we have modelled our galaxies with $q_0=0.2$, we choose $k_{\text{tot}}=1.8$ as it is the coefficient for galaxies with $q_0=0.2$ and $n\sim 1-4$ \citep{Price:2022aa}. The virial coefficient $k_{\text{tot}}$ allows us to infer the total dynamical masses based on measurements out to $r_{\rm e}$.

\section{The interplay between kinematics and stellar populations from $z=0-6.5$}\label{sec:res-kins}
Studies out to $z<4$ describe the star-forming galaxy population as apparently evolving from turbulent thick disks at cosmic noon to thin disks with ordered rotation in the local Universe. In this section we extend relations between kinematics, redshift, and star formation out to $z\gtrsim4$ in order to understand how galaxies are growing in the early Universe. We present results from modelling with \textsc{geko} and \textsc{Prospector} on our final sample of $\sim 250$ \Ha\ emitters. We begin by discussing the redshift evolution of the velocity dispersion $\disp$ and rotational support $\rotsupp$ in our sample, and compare our findings to cosmological simulations and low-redshift measurements (Sec. \ref{sec:z-evol}). Then, in Sec. \ref{sec:sfr-kin}, we analyse the relation between these two key kinematic parameters and stellar population parameters, such as stellar mass and SFRs. The median values of $\disp$ and $\rotsupp$ presented in this section are summarized in Appendix \ref{app:tables} (Tabs. \ref{tab:medians-gold} and \ref{tab:medians-ext}), along with the kinematic and star-forming properties of every object in the gold sample (Tab. \ref{tab:results-gold}). The fits for the gold sample are shown in Appendix \ref{app:gold-fits} (Fig. \ref{fig:gold-summaries}).

\subsection{Redshift evolution of ionised gas kinematics from $z=0$ to $z\sim 6.5$ } \label{sec:z-evol}

\subsubsection{Kinematics at $z\sim 3.9-6.5$}\label{sec:res-1}

\begin{figure}
    \centering
    \includegraphics[width=1\linewidth]{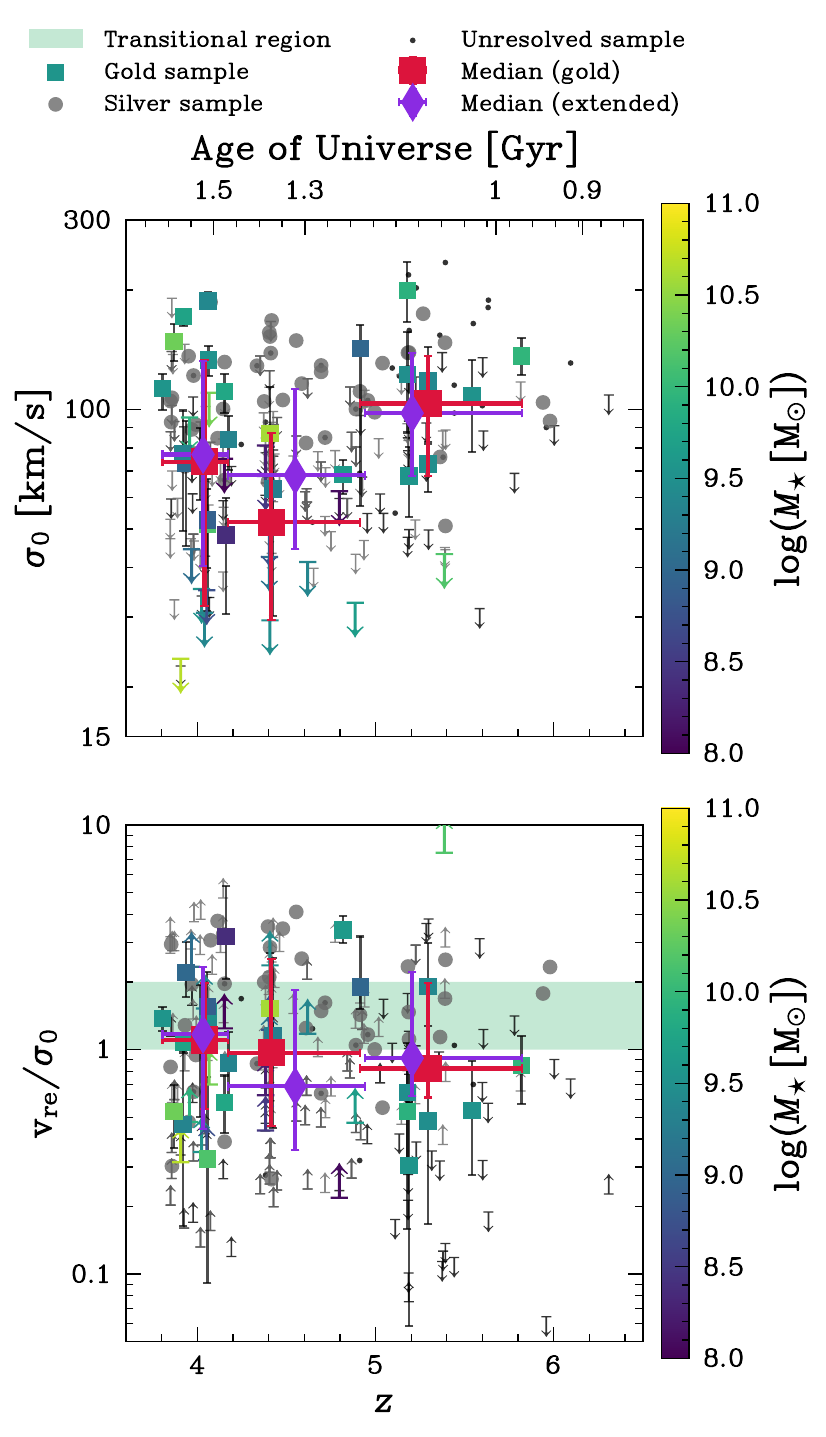}
    \caption{Redshift evolution of the intrinsic velocity dispersion $\sigma_0$ (top) and the rotational support $\rotsupp$ (bottom) for the H$\alpha$ emitters in the gold sample (squares), silver (big circles), and unresolved (small dots) samples. The medians for the gold (red squares) and extended sample (purple diamonds), show a mild redshift evolution, for both $\disp$ and $\rotsupp$, between $z\sim4$ and $z\sim6$. The errorbars on the medians represent the $16^{\rm th}$ and $16^{\rm th}$ quantiles along the y-axis, and the extent of the bin along the x-axis. Most of our sample lies in the dispersion supported regime ($\rotsupp<1$) and the transitional regions ($\rotsupp=1-2$; green shaded region), in part due to the high dispersions measured $\disp\approx 100$ km/s.}
    \label{fig:vsigma_sigma_zevol}
\end{figure}

Fig. \ref{fig:vsigma_sigma_zevol} shows the evolution of the intrinsic \Ha\ velocity dispersion $\disp$ and the rotational support $\rotsupp$ as a function of redshift for the galaxies in this work. Colour-coded by stellar mass, we plot objects in our gold, silver, and unresolved samples, including upper\footnote{For $\rotsupp$, we mark a measurement as a upper limit when the $16^{\text{th}}$ quantile of the posterior distribution of \vre\ includes zero. All of the unresolved objects fall into this category.  For $\disp$\ measurements, upper limits are defined in a similar way but we loosen the requirement to the $16^{\text{th}}$ quantile being smaller than 30 km/s, due to resolution limits.} and lower limit\footnote{If the $\disp$\ measurement is an upper limit for a given $\rotsupp$\ point, then this point is marked as a lower limit, since the ratio diverges when $\sigma_0 \rightarrow 0$. If a given $\rotsupp$\ verifies both upper and lower limits criteria, meaning its posterior distribution both coincides with zero and has a long tail, then we mark it as a lower limit since the velocity constraint is more secure than the velocity dispersion, which is rarely zero and likely only hits the limit due to resolution effects.} detections for which the value at $1\sigma$ is plotted instead. We compute (and report in Appendix \ref{app:tables}) the medians values of $\rotsupp$ and $\disp$, for three redshift bins probing $\approx 250$ Myr, for the gold sample only (Tab. \ref{tab:medians-gold}) and for the extended galaxies in the gold and silver samples (Tab. \ref{tab:medians-ext}). We define galaxies as extended when the grism data extends beyond the best-fit \Ha\ effective radius (see Sec. \ref{sec:sample-selection} and Tab. \ref{tab:sample-selection}). In both cases we report measured for $v_{\rm rot} = v_{\rm re}$ as well as $v_{\rm rot} = v_{\rm obs,max}$ in (Tabs. \ref{tab:medians-ext} and \ref{tab:medians-gold}).

We find a weak increase in the median values for the velocity dispersion $\disp$ from $z\sim 4$ to $z\sim 6$ for both the gold sample and the extended sample, with $\disp = 73 ^{+ 60}_{-41}$ km/s to $\disp = 103 ^{+ 35}_{-33}$ km/s, and with $\disp = 77 ^{+ 55}_{-36}$ km/s to $\disp = 98 ^{+ 40}_{-30}$ km/s, respectively. We do not find significant differences between the medians for these two subsamples. This is consistent given that they probe similar stellar mass distributions (Tab. \ref{fig:logmstar_hist}). Where the high and low redshift bins probe similar stellar masses, the median mass in the central bin ($4.2<z<5.0$) is $\sim0.3-0.4$ dex lower (Tabs. \ref{tab:medians-ext} and \ref{tab:medians-gold}), which could explain the lower median value of $\disp$ found (see Sec. \ref{sec:sfr-kin}). We investigate the overall mass dependence of the $\disp-z$ evolution but find no distinct trend, suggesting that the observed evolution is redshift-driven.

We find an overall mild decrease of the rotational support with redshift between $z\sim 4$ and $z\sim 6$ for both the gold sample and the extended sample, with $\rotsupp = 1.1 ^{+ 0.8}_{-0.5}$ to $\rotsupp = 0.8 ^{+ 1.2}_{-0.2}$ and $\rotsupp = 1.2 ^{+ 1.1}_{-0.7}$ to $\rotsupp = 0.9 ^{+ 1.3}_{-0.3}$, respectively. Similarly to the trends for $\disp$, we do not find significant differences in the median values for these two subsets. These low values of $\rotsupp$ highlight the turbulent nature of the galaxy population at high redshift, with medians falling in the dispersion-dominated regime by $z\sim6$. At $z\sim 4-5$, the medians fall in the transition region where galaxies have some rotational support but are unstable. 

The large observed scatter, spanning $\approx 0.6$ dex in $\disp$ and $\approx 2$ dex in $\rotsupp$, can in part be attributed to measurement uncertainties, which can reach $\approx 0.3$ dex in $\disp$ and $\approx 0.5$ dex in $\rotsupp$. However, it also highlights an intrinsic scatter in the galaxy population probed in this work, with resolved velocity dispersions spanning $\disp\sim 50-200$ km/s. This can also be seen in the measurements of $\rotsupp$, with many galaxies being consistent with a dispersion-dominated state, but others reaching values $\rotsupp\sim 2-4$ that are consistent with turbulent disks. 

Although many simulations predict that gas disks at high redshift are not highly rotationally dominated \citep{El-Badry:2018aa,Pillepich:2019aa}, consistent with what we see in this work, some works also highlight the key role of undisturbed co-planar and co-rotating gas accretion in building and sustaining rotation-dominated disks \citep{Dekel:2009aa,Sales:2012aa,Kretschmer:2020aa,Kretschmer:2022aa, Kohandel:2024aa}. These disks survive until they are destroyed by external effects such counter-rotating streams and mergers, which happen on the order of a few orbital times \citep{Dekel:2014aa,Zolotov:2015aa,Tacchella:2016aa,Tacchella:2016ab,Dekel:2020aa,Kretschmer:2022aa}. The effect of these processes on the measured rotational support of the gas could explain the range of $\rotsupp$ values we measure on Fig. \ref{fig:vsigma_sigma_zevol}, where we probe galaxies in different stages of these fluctuations between disk build-up and dismantling.

\subsubsection{The evolution of the ionised gas velocity dispersion}

\begin{figure*}
    \centering
    \includegraphics[width=1\linewidth]{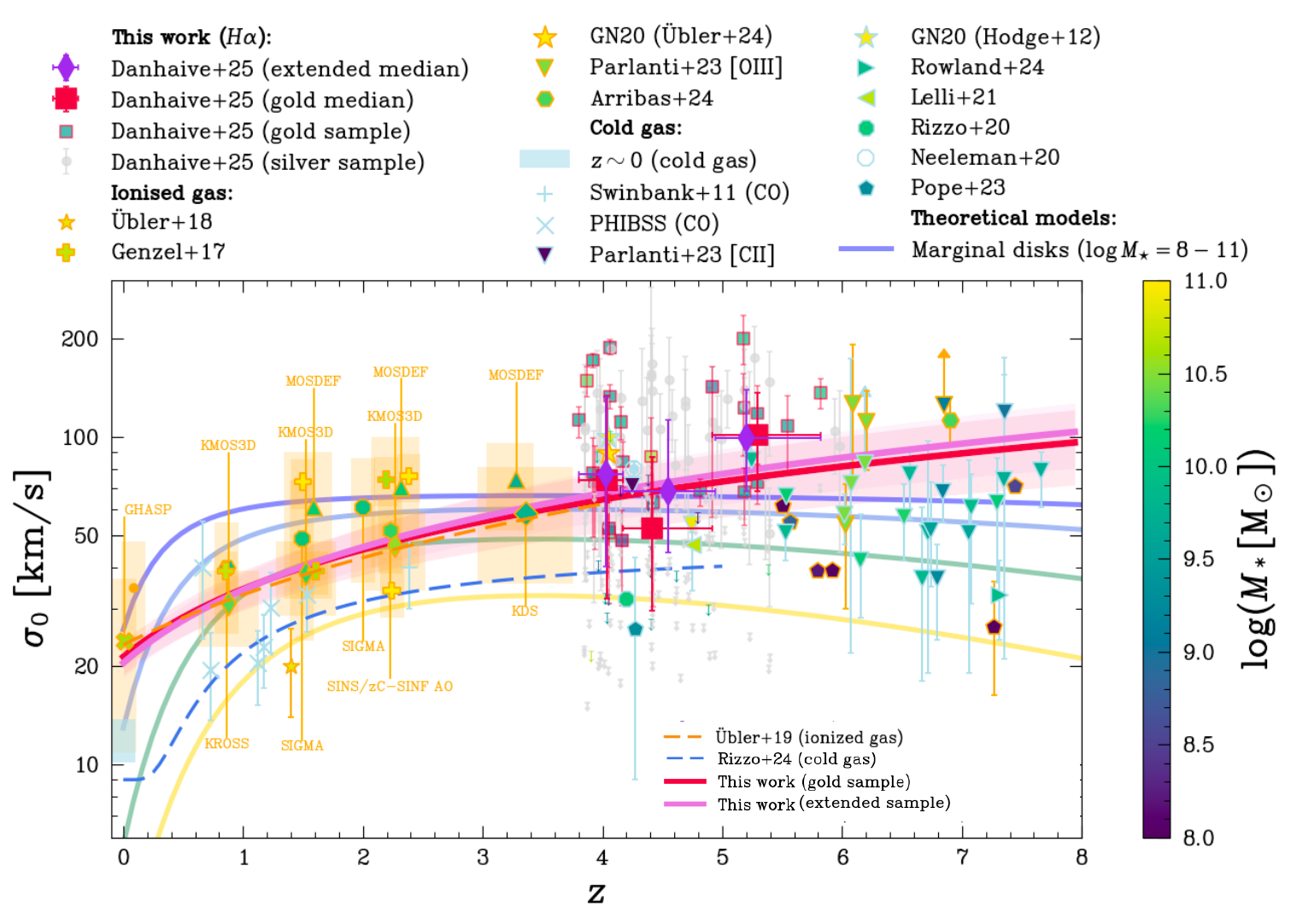}
    \caption{Intrinsic \Ha\ velocity dispersion $\disp$\ as a function of redshift for our gold (red outlined squares) and silver (light gray circles) in the context of studies of warm ionised gas (i.e. \Ha\ and [OIII]; orange outlines) and cold gas (i.e. HI, CO, and [CII]; blue outlines) kinematics across redshifts. The medians for the gold and extended samples are plotted in red squares and purple diamonds, respectively. The errorbars on the medians represent the $16^{\rm th}$ and $84^{\rm th}$ quantiles along the y-axis, and the extent of the bin along the x-axis. Our results are consistent with a decrease of $\disp$ with cosmic time when compared to  surveys from cosmic noon (KMOS3D \protect\citealt{Wisnioski:2015vx}, KROSS \protect\citealt{Johnson:2018aa}, MOSDEF \citealt{Price:2016uv, Price:2020wf}, KDS \protect\citealt{Turner:2017aa, Turner:2017ab}, SIGMA \protect\citealt{Simons:2016aa}, SINS/zC-SINF \protect\citealt{Forster-Schreiber:2006wk, Forster-Schreiber:2018aa}, PHIBSS \protect\citealt{Tacconi:2013aa}, and points from \protect\citealt{Swinbank:2011aa,Genzel:2017aa}) and the local Universe (GHASP \protect\citealt{Epinat:2010aa}, DYNAMO \protect\citealt{Green:2014aa}, EDGE-CALIFA \protect\citealt{Bolatto:2017aa}, HERACLES \protect\citealt{Leroy:2009aa}, THINGS \protect\citealt{Walter:2008aa}, and \protect\citealt{Dib:2006aa}; blue shaded region). Our best-fit relations (Eq.~\ref{eq:powerlaw}), with (purple) and without (red) including upper limits, agree with the fit from \protect\cite{Ubler:2019vg} (light red - long dashes), which is defined out to $z\sim4$. Similarly, we also include the exponential fit for cold gas measurements from \protect\citet{Rizzo:2024aa} (dark blue - long dashes) defined out to $z\sim5$, which highlight a similar evolution of $\disp$ tracing the cold gas, albeit with a factor 2-3 lower normalization. We compare our measurements with works at similar redshifts \protect\citep{Parlanti:2023ab,Arribas:2024aa,de-Graaff:2024ab}, as well as detection of dynamically cold systems at high-redshift \protect\citet{Neeleman:2020aa, Rizzo:2020aa, Lelli:2021aa, Pope:2023aa, Rowland:2024aa}.}
    \label{fig:sigma0-z-comp}
\end{figure*}

We now place our measurements in the context of the other studies targeting cold and warm ionized gas across cosmic time (see Appendix \ref{app:z-evol-comp} for details of each survey). For the rest of this paper, we use the term cold gas for tracers such as CO and [CII], in contrast with tracers of warmer (ionised) gas such as \Ha\ and [OIII]. We show the evolution of the velocity dispersion $\disp$, shown in Fig. \ref{fig:sigma0-z-comp}. The high values $\disp\gtrsim 70-100$ km/s we find lie above measurements from cosmic noon \citep{ Wisnioski:2015vx, Tiley:2016aa,Simons:2016aa, Turner:2017ab,Genzel:2017aa, Ubler:2019vg,Price:2020wf}. It is important to note that many of these works restrict their analyses to rotationally supported systems with $\rotsupp>1$, which could reduce the number of systems with high $\disp$ in their samples. 

In order to quantify the evolution of $\disp$ from $z\sim0$ to $z\sim 6$, we fit our medians at $z\sim4$, $z\sim5$, and $z\sim6$, for the gold and extended sample, in conjunction with medians from selected surveys \citep{Epinat:2010aa, Simons:2016aa, Tiley:2016aa,Simons:2017aa, Turner:2017aa} from $z=0$ to $z\sim 3.5$, following \citet{Ubler:2019vg}. We fit with a power law form
\begin{equation}
    \sigma_0 = \beta(1+z)^{\alpha},
    \label{eq:powerlaw}
\end{equation}
 and find the following best-fit parameters (summarized in Tab. \ref{tab:sigma_z_fit}): $\alpha_{\text{gold}} = 0.7 \pm 0.1, \thinspace \beta_{\text{gold}} = 21.4 \pm 2.5 \thinspace \text{km/s}$ and $\alpha_{\text{ext}} = 0.7\pm 0.1, \thinspace \beta_{\text{ext}} = 20.5\pm 2.0 \thinspace \text{km/s}$. We plot our best-fit relations on Fig. \ref{fig:sigma0-z-comp}. We find consistent results for the two sets of medians, indicating a soft, sub-linear, dependence of $\disp$ on $(1+z)$.  However, 75\% of galaxies from our gold sample (excluding upper limits) lie above the best-fit relation, which could suggest an even steeper increase. We note that our measurements span a large range of $\disp$, so we cannot draw clear conclusions. 

We compare our fit to the linear relation $\disp = az + b$ from \cite{Ubler:2019vg}, which is given by $a = 9.8 \pm 3.5 \thinspace \text{km/s} , \thinspace b = 23.3 \pm 4.9  \thinspace \text{km/s}$, and is computed using the same $z=0-3.5$ median measurements. We find excellent agreement with the extrapolation of this relation to the $z\sim 4-6$, which highlights the increase of turbulence with redshift. Our velocity dispersions are in the same ranges as probed by the ionised gas measurements from \citet{Parlanti:2023ab}, although they probe higher stellar masses in general, most of which also lie above our relation. \citet{de-Graaff:2024ab} find low values of $\disp$ that lie below the expected evolution of $\disp-z$, which could be driven by the low stellar masses probed $\log M_{\star} [\mathrm{M}_{\odot}] \approx 8$.

\begin{table}
    \centering
    \begin{tabular}{c|c|c}
         & Gold sample & Extended sample \\ \hline
       $\alpha$  & $0.7 \pm 0.1$ & $0.7 \pm 0.1$\\ \hline
        $\beta$ [km/s] & $21.4 \pm 2.5$ & $20.5\pm 2.0$
    \end{tabular}
    \caption{Best-fit parameters and uncertainties for the $\sigma_0 = \beta(1+z)^{\alpha}$ relation based on our sample medians and medians from surveys at lower redshifts \citep{Epinat:2010aa,  Simons:2016aa,Tiley:2016aa, Simons:2017aa, Turner:2017aa}.}
    \label{tab:sigma_z_fit}
\end{table} 

When looking at cold gas tracers (i.e. CO, [CI], [CII]), the best fit exponential relation from \citet{Rizzo:2024aa} also shows an increase in $\disp$ with redshift, with a factor of $2-3$ lower normalization. Although the curve shows a flatter tail at high redshift compared to our fit for ionised gas, [CII] measurements from \citet{Parlanti:2023ab} lie mostly above this relation, suggesting a steeper slope. The dynamically cold systems reported at high redshift \citep{ Swinbank:2011aa,Neeleman:2020aa, Rizzo:2020aa,Lelli:2021aa, Pope:2023aa,Rowland:2024aa} are in good agreement with the prediction from \citet{Rizzo:2024aa}, with low dispersions of $\disp\approx 20-50$ km/s, which are characteristic of cold gas tracers. 

In the framework of marginally stable disks, the expected evolution of $\rotsupp$\ and $\disp$\ with redshift can be computed directly from the gas fraction using the Toomre parameter $Q$ \citep{Toomre:1964aa}, which represents the instability of the disk through the turbulence in the ISM. In this empirical model, gas is accreted from the surrounding halo or cosmic web and is expelled through feedback \citep{Forster-Schreiber:2006wk, Genzel:2008ug, Dekel:2009aa, Lilly:2013ua}. The Toomre parameter $Q$ measures the equilibrium between the self-gravity of the gas and repelling forces due to turbulence and rotational support. Marginally stable disks are defined by $Q_{\text{crit}}\sim 1$, where  $Q_{\text{crit}}<1$ indicates an unstable disk. In this model, the redshift evolution of $\rotsupp$\ and $\disp$\ can be explained by the increase of gas fractions within these marginally stable disks \citep{Genzel:2011aa,Wisnioski:2015vx}:
\begin{equation}
    v/\sigma_0 = \frac{a}{f_{\text{gas}}(z)Q_{\text{crit}}},
    \label{eq:toomre-vs}
\end{equation}
where $a = \sqrt{2}$ for a disk with a constant rotational velocity. In order to qualitatively compare this model to our measurements and to measurements from the literature, we plot, on Fig. \ref{fig:sigma0-z-comp}, $\disp$ as predicted by Eq. \ref{eq:toomre-vs} for $Q=1$, $v = 100~\rm km/s$, and a range of stellar masses $\log M_{\star} [\mathrm{M}_{\odot}]= 8-11$. We derive the total gas fractions $f_{\text{gas}}=M_{\rm gas}/(M_{\rm gas} + M_{\star})$ from the empirical relation derived in \citet{Tacconi:2018aa,Tacconi:2020aa}, where $f_{\rm gas}(z)$ depends on total stellar mass, sSFR, and the distance from the star-forming main sequence. We plot this relation for galaxies on the main sequence.

Interestingly, the observed increase of $\disp$ with $z$ is not well reproduced by the marginal disk model, which shows an increase of $\disp$ until cosmic noon ($z\approx 2-3$) but then decreases slowly towards higher redshifts. When comparing our results to this model (Fig. \ref{fig:sigma0-z-comp}), it is important to note that the Toomre marginal disk curves are at fixed stellar masses. At fixed stellar masses, the gas fraction is predicted to decrease at redshifts beyond cosmic noon, which then drives the decrease in $\disp$ (Eq. \ref{eq:toomre-vs}). However, in the measurements from observations, we probe lower and lower stellar masses with increasing redshift, so the average gas fractions would stay approximately constant, or could even increase due to the low stellar masses probed. The observed increase could also point to sources of turbulence beyond the gravitational instabilities described in the marginal disk model, such as feedback and mergers (see Secs. \ref{sec:disk-fracs} and \ref{sec:gal-evol} for further discussion). 

\subsubsection{The evolution of the rotational support across cosmic time} \label{sec:form-disks}

\begin{figure*}
    \centering
    \includegraphics[width=1\linewidth]{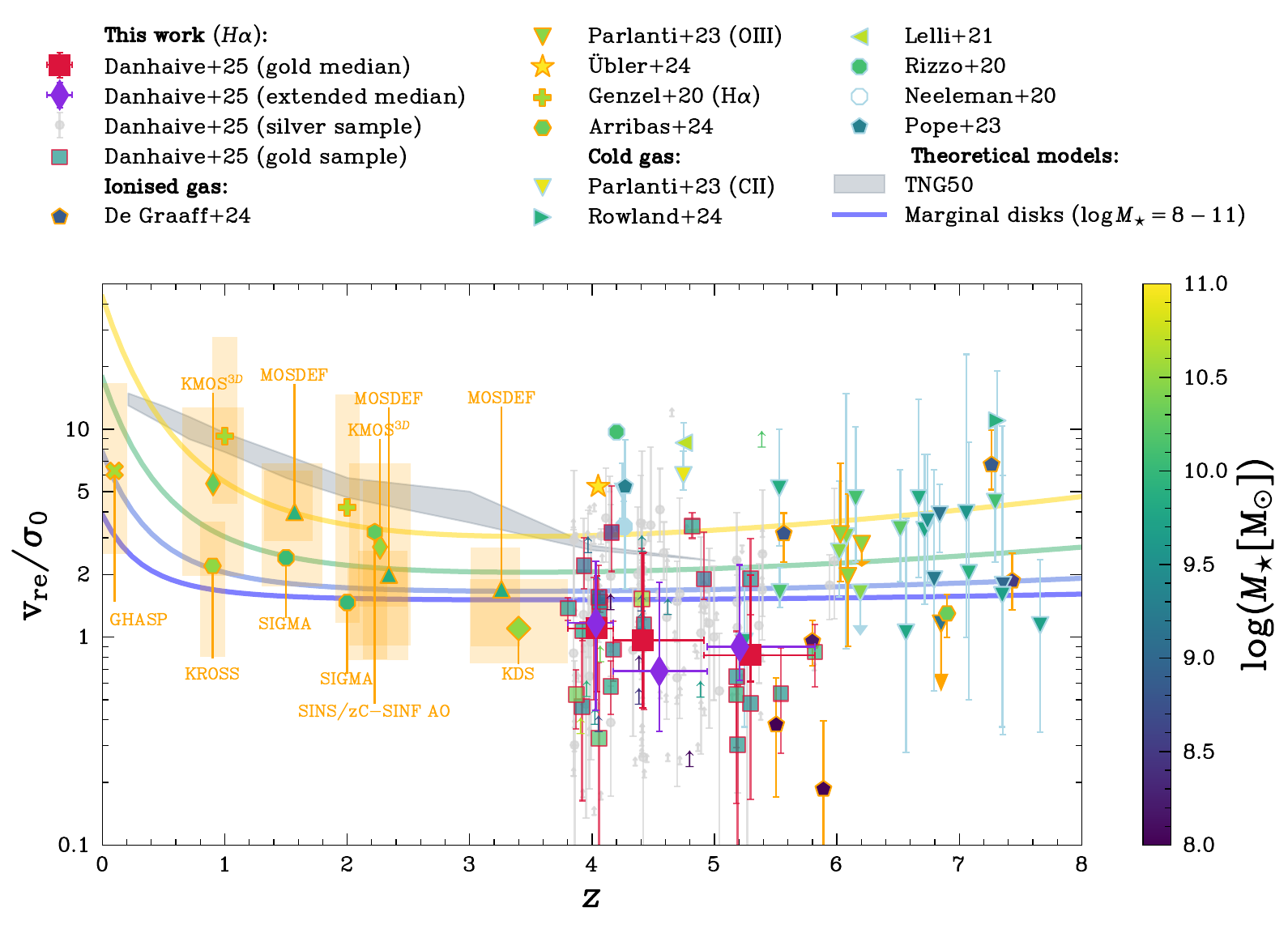}
    \caption{Evolution of the rotational support $\rotsupp$ with redshift for our gold (red outlined squares) and silver (light gray circles) in the context of studies of gas kinematics across redshifts. The medians for the gold and extended samples are plotted in red squares and purple diamonds, respectively. The errorbars on the medians represent the $16^{\rm th}$ and $16^{\rm th}$ quantiles along the y-axis, and the extent of the bin along the x-axis. All of our galaxies are $7<\log M_{\star} [\mathrm{M}_{\odot}] \lesssim 10$, and their kinematics are consistent with a decline of rotationally supported systems within our high-redshift sample, despite the presence of some rotation-supported systems (Fig. \ref{fig:vsigma_sigma_zevol}). They span similar $\rotsupp$ values as the ones of \citet{de-Graaff:2024ab}, despite probing slightly higher masses, and are overall consistent with cold gas ([CII]) measurements (blue outlines) from \citet{Parlanti:2023ab}, who find turbulent disks at $z\sim6-7$. Our results are placed in the context of surveys (shaded orange regions) of ionised gas (orange outline) from cosmic noon and the local Universe: MOSDEF \citep{Price:2016uv}, KMOS3D \citep{Wisnioski:2015vx}, SIGMA \citep{Simons:2016aa}, SINS/zC-SINF AO \citep{Forster-Schreiber:2018aa}, KROSS \citep{Stott:2016aa,Tiley:2016aa}, KDS \citep{Turner:2017aa, Turner:2017ab}, and GHASP \citep{Epinat:2010aa}. We also include medians from \citet{Genzel:2020aa}, and measurements from \citet{Ubler:2024aa}. We add single-object detections of dynamically cold systems in the early Universe \citep{Rizzo:2020aa, Neeleman:2020aa, Lelli:2021aa, Pope:2023aa, Rowland:2024aa}, which probe higher masses (i.e. more evolved systems) than our sample, and an unsettled disc in a protocluster \citep{Arribas:2024aa}. Our measurements are consistent with a decrease of $\rotsupp$ with redshift as predicted by the TNG50 cosmological simulations \citep{Pillepich:2019aa}, shown with the shaded gray region, and the Toomre (in)-stability model described in \citet{Wisnioski:2011us}, indicated with solid lines. Within the context of lower-z measurements, our results are consistent with a decline of rotational support with redshift.}
    \label{fig:vsigma_z_evol}
\end{figure*}

Our results for ionised gas at $z\sim 3.9-6.5$, shown on Fig. \ref{fig:vsigma_sigma_zevol}, are consistent with the smooth decline in the average rotational support of galaxies with redshift. As highlighted by the Toomre instability criterion (Eq. \ref{eq:toomre-vs}), the evolution of the rotational support is deeply linked to that of the velocity dispersion, which acts as a proxy for the turbulence of the ISM. Hence, this observed decrease in $\rotsupp$ with redshift is in part driven by the increase of $\disp$ with redshift.

As highlighted on Fig. \ref{fig:vsigma_sigma_zevol}, works from the local Universe \citep{Epinat:2010aa} to cosmic noon \citep{Wisnioski:2015vx, Simons:2016aa, Price:2020wf,Genzel:2020aa} show that $\rotsupp$\ declines from an average of $\rotsupp\sim10$ at $z\sim 0-1$ to $\rotsupp\sim2-4$ by $z\sim 2.5-3.5$ for galaxies with stellar masses $\approx 10^{9}-10^{11} ~\mathrm{M}_{\odot}$. In the local universe, star-forming galaxies have stellar masses up to $\approx 10^{11}~\mathrm{M}_{\odot}$, whereas studies at cosmic noon probe lower masses on average, partly driving the observed trend. These studies imply that star-forming discs are turbulent at cosmic noon and only settle into dynamically cold discs at later times. We find a further decrease to medians of $\rotsupp\approx1$ by $z\sim 3.9-6.5$, consistent with predictions of a smooth decline of $\rotsupp$ from the Illustris-TNG50 simulations, for the kinematics probed by \Ha\ \citep{Pillepich:2019aa}. 

Our sample probes a lower mass range than the studies in the local Universe and at cosmic noon, with the majority of galaxies in the $\log M_{\star} [\mathrm{M}_{\odot}] \sim 8-10$ range. The sample from \citet{de-Graaff:2024ab} probes even lower masses $\log M_{\star} [\mathrm{M}_{\odot}]\sim 7-9$ but still shows a range of kinematic states. We further compare our results to the sample of ionised ([OIII]) and cold ([CII]) gas observations from \citet{Parlanti:2023ab}, which reaches out to $z\sim 8$. Many studies have shown that cold gas is typically less turbulent than ionised gas, resulting in overall lower velocity in both observations \citep[e.g.,][]{Ubler:2019vg, Varidel:2020aa,Rizzo:2024aa} and simulations \citep[e.g.,][]{Kohandel:2024aa}. Given these intrinsically lower velocity dispersions, as well as their higher masses (which typically imply higher rotational velocities), the measurements from \citet{Parlanti:2023ab} are consistent with our observations of star-forming galaxies being turbulent at high redshift. 

Various high-redshift dynamically cold disks have been reported with ALMA and JWST observations \citep[][]{ Neeleman:2020aa, Rizzo:2020aa, Lelli:2021aa, Pope:2023aa, Rowland:2024aa, Ubler:2024aa}, and are plotted on Fig. \ref{fig:vsigma_z_evol}. The first thing to note is their stellar masses. In fact, all of the galaxies reported in these works, with the exception of \citet{Pope:2023aa}, have stellar masses $M_{\star} > 10^{10} \thinspace M_{\odot}$ comparable to the star-forming discs observed at $z<3$. Their high masses indicate that these systems are more evolved than their lower-mass counterparts and are not always representative of the bulk of the galaxy population at $z\sim 3.9-6.5$. For instance, the systems from \citet{Lelli:2021aa}, \citet{Rowland:2024aa}, and \citet{Ubler:2024aa} all lie close to or above the characteristic mass above which the stellar mass function at those redshifts drops exponentially \citep{Weibel:2024aa, Shuntov:2025aa, Harvey:2025aa}, indicative of their extreme nature. Furthermore, all of these measurements (with the exception of \citealt{Arribas:2024aa,Ubler:2024aa}) are based on cold gas tracers (CO and [CII]), which as mentioned previously typically have intrinsically lower velocity dispersions and hence higher rotational support. Finally, where we measure the rotational velocity at $r_e$, many of these works use different quantities such as the maximum observed velocity, which is typically higher \citep[e.g.]{Rowland:2024aa}.

In our sample, we find one system with comparable rotational support $\rotsupp\sim 10$ to these works, which is the massive galaxy previously reported in \citet{Nelson:2024aa} as a rapidly rotating system at $z=5.2$ with $v_{\rm rot}(r_{\rm e})= 240^{+50}_{-50}$ km/s. It is also discussed in \citet{Li:2023aa} where they find a similar value of $v_{\rm rot}(r_{\rm e})= 211^{+21}_{-18}$ km/s. Our best fit value of $v_{\rm rot}(r_{\rm e})= 255^{+21}_{-20}$ km/s is in good agreement with these works. However, both \citet{Nelson:2024aa} and \citet{Li:2023aa} find low values for the rotational support, $\rotsupp\sim2$, in contrast with our finding of $\rotsupp\sim 10$. This is most likely due to differences in the morphological and kinematic modelling. For example, the aforementioned works do not account for the instrument PSF and LSF when deriving the kinematics, which can heavily impact the derived velocity dispersion.

To compare our findings to the Toomre marginal disk model, we plot Eq. \ref{eq:toomre-vs} on Fig. \ref{fig:vsigma_z_evol} for $Q=1$, and a range of stellar masses $\log M_{\star} [\mathrm{M}_{\odot}]= 8-11$. After a sharp decline from $z=0$ to $z\sim 2$, this model predicts nearly constant values of $\rotsupp$ beyond $z>2$ at fixed stellar mass. This is driven by the strong dependence of gas fractions on stellar mass. In fact, most galaxies assemble the bulk of their stellar mass at cosmic noon where the cosmic SFR density peaks. At later cosmic times, galaxies are overall more massive and hence have lower gas fractions, which drives an increase in $\rotsupp$ as they settle into cold disks (Eq. \ref{eq:toomre-vs}). At $z>2$, the stellar masses probed are low on average $\log M_{\star} [\mathrm{M}_{\odot}] < 10$, which is reflected in the high gas fractions and low values of $\rotsupp$, and, importantly, they evolve slower.

Our medians are broadly consistent with predictions from this simple model, as is the evolution of $\rotsupp$\ at lower redshift. This suggests that gravitational instabilities due to the high gas fraction observed in high-redshift galaxies \citep{Tacconi:2018aa, Parlanti:2023ab} can drive the decrease in rotational support. Unstable disks are also qualified by fragmentation and clumpy star formation, which has been observed in this sample and other high-redshift observations \citep{Nakazato:2024aa}. Our dispersion-supported systems naturally lie below this relation and occupy a more unstable regime, with $Q<1$. The instability in these systems could be in part explained by stellar feedback and mergers, which are not accounted for in the Toomre model.

In summary, Fig. \ref{fig:vsigma_z_evol} shows that galaxies become increasingly turbulent at high redshift, with only the most evolved systems being able to settle into cold disks. Our findings underline high turbulence in high-redshift systems, with $\disp\approx 100$ km/s and $\rotsupp\approx 1-2$ at $z\sim 3.9-6.5$. To explore what could be driving this strong increase, we will study how $\rotsupp$\ and $\disp$\ are related to their stellar populations. 

\subsection{The effect of star formation on internal kinematics}\label{sec:sfr-kin}

At high redshift $z>3$, the bulk of the galaxy population is in its early growth stages, and is characterized by smaller masses and sizes. These systems are hence less stable to perturbations from mergers, gas accretion, and stellar outflows, than their more evolved, low redshift counterparts. One of the characteristics of the large sample of galaxies studied in this work is the large scatter in their kinematic properties. In order to investigate what is driving the observed scatter, and to further understand the main causes of the high turbulence observed at high redshift, we look at how the velocity dispersion and rotational support correlate with other physical properties of the galaxy, specifically stellar mass $M_{\star}$, SFR averaged over the past $10$ Myr ($\text{SFR}_{10}$), the corresponding specific SFR ($\text{sSFR}_{10}$) and SFR surface density ($\Sigma_{\text{SFR}_{10}}$). $\Sigma_{\text{SFR}_{10}}$ is computed assuming that half of the total SFR is arising from inside the half-light radius $r_{\rm e}$:
\begin{equation}
\Sigma_{\text{SFR}_{10}} = \frac{\text{SFR}_{10}/2}{\pi r_{\rm e}^2}.
    \label{eq:SFR_density}
\end{equation}
We use the best fit half-light radius of the \Ha\ emission, as it is the most representative of regions of recent star formation, as probed by $\text{SFR}_{10}$.
After presenting the results for our sample (Sec. \ref{sec:corr-z6}), we place them in the context of correlations found in other works (Sec. \ref{sec:disp-drivers}) and discuss the implications for the evolution of galaxies and the formation of disks at high redshift.

\subsubsection{Trends at $z\gtrsim4$} \label{sec:corr-z6}

\begin{figure*}
    \centering
    \includegraphics[width=1\linewidth]{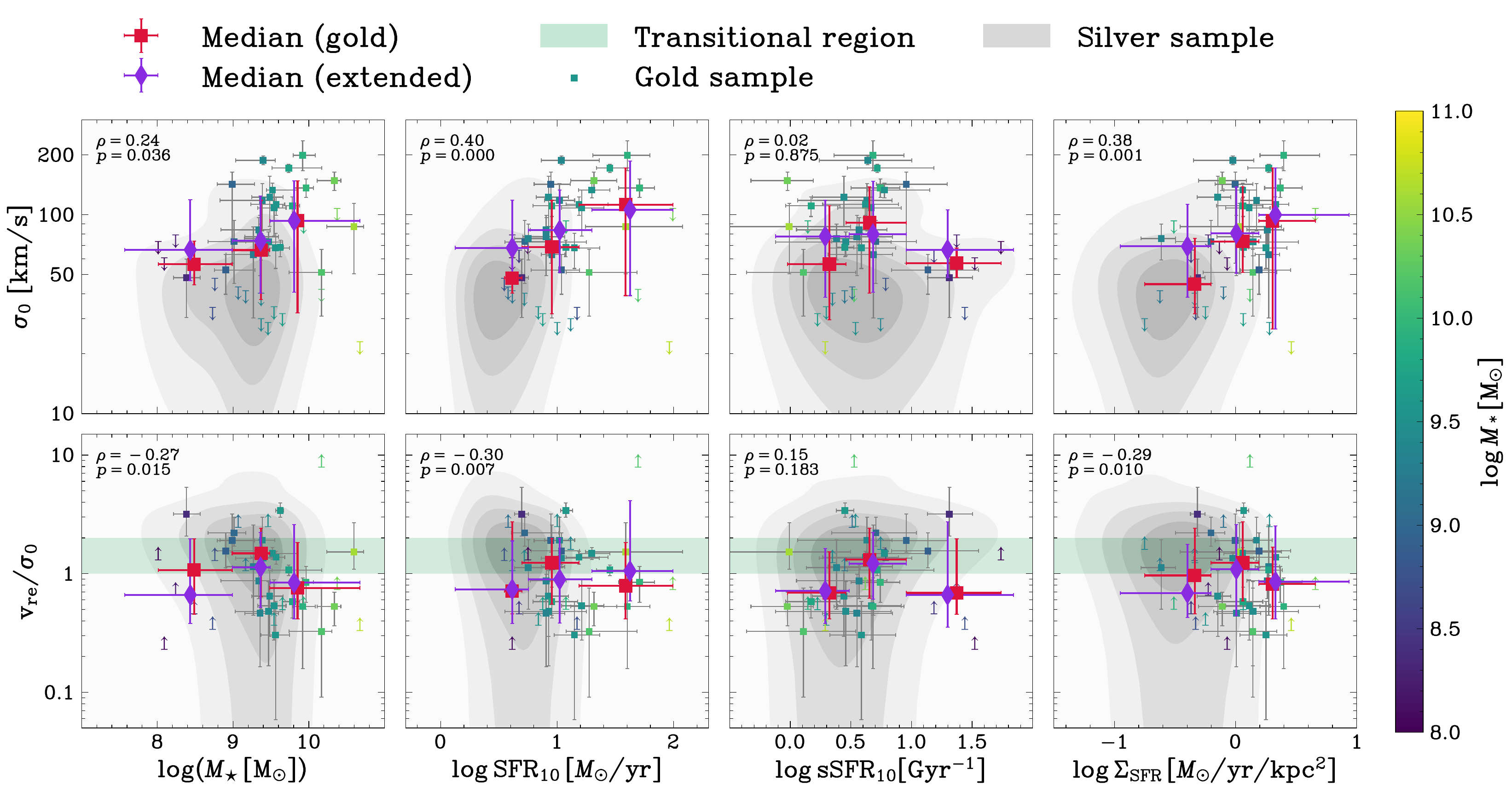}
    \caption{Dependence of the intrinsic velocity dispersion $\disp$ (top panels) and rotational support $\rotsupp$ (bottom panels) on $M_{\star}$, $\text{SFR}_{10}$, $\text{sSFR}_{10}$, and SFR surface density $\Sigma_{\text{SFR}_{10}}$ for our gold (squares) and silver sample (gray contours). We find a significant correlation between $\disp$ and $\text{SFR}_{10}$, which is highlighted by the running medians, for the gold sample (red squares) and for the extended sample (purple diamonds), and the Spearman rank coefficient $\rho$ and p-value $p$ for the resolved sample. The errorbars on the medians represent the $16^{\rm th}$ and $84^{\rm th}$ quantiles along the y-axis, and the extent of the bin along the x-axis. Both $\disp$ and $\rotsupp$ show a (weaker) correlation with $\Sigma_{\text{SFR}_{10}}$, highlighting the role of surface density in efficiently driving turbulence in the gas. The trends with $\text{SFR}_{10}$ and $\Sigma_{\text{SFR}_{10}}$ are stronger for $\disp$ than $\rotsupp$, which could be caused by the fact that we probing many dispersion-supported systems ($\rotsupp<1$) and systems in the transitional regions (green shaded region) between dispersion and rotation support, where stable rotation has not yet been established.}
    \label{fig:vsigma_sigma_mstar_sfr}
\end{figure*}

Fig. \ref{fig:vsigma_sigma_mstar_sfr} shows the evolution of $\disp$\ and $\rotsupp$\ with $M_{\star}$, $\text{SFR}_{10}$, $\text{sSFR}_{10}$, and $\Sigma_{\text{SFR}_{10}}$, with medians plotted for the gold and extended samples, and summarized in Tabs. \ref{tab:medians-ext} and \ref{tab:medians-gold}, respectively. The silver sample, highlighted in grey contours, lies on the lower portion of the $\disp$ distribution when compared to the gold sample. This causes the gold sample medians to lie above the bulk of the silver sample measurements. Galaxies in the silver sample typically have lower $\text{SFR}_{10}$ and $\Sigma_{\text{SFR}_{10}}$, but on average probe similar stellar masses and $\text{sSFR}_{10}$ to the gold sample. This is likely due to our $\rm S/N$ cut, which is more lenient for the silver sample. The gold sample has the brightest \Ha\ galaxies, which hence have higher SFRs and are tipycally more massive.

The strongest correlation\footnote{We only discuss the correlations and medians for the extended  sample, as presented on Fig. \ref{fig:vsigma_sigma_mstar_sfr}, but the medians of gold sample follow similar trends.} we observe is between $\disp$\ and $\text{SFR}_{10}$, with Spearman rank coefficient $\rho = 0.40$ and p-value $p<0.001$, with averages going from $\disp \approx 70-80$ at $\text{SFR}_{10} = 10 \thinspace M_{\odot}$/yr to $\disp \approx 100$ at $\text{SFR}_{10} = 20-100 \thinspace M_{\odot}$/yr. This correlation naturally arises from the main physical drivers of turbulence, namely feedback from star formation and the release of gravitational energy through the radial inflow of gas through the disk \citep{Krumholz:2016aa, Krumholz:2018aa}, as both are connected to star-formation activity. 

Our results at $z\gtrsim 4$ are broadly in agreement with predictions from gravitational-instability driven turbulence models (Figs. \ref{fig:sigma0-z-comp} and\ref{fig:vsigma_sigma_zevol}). This is consistent with studies finding that gravitational instabilities are the dominant source of turbulence at $1<z<3$ \citep{Krumholz:2016aa,Krumholz:2018aa,Ubler:2019vg}. However, stellar feedback could be a more important driver of turbulence at high-redshift than at cosmic noon and in the local universe (see Sec. \ref{sec:disp-drivers}), which could explain the high values of $\disp$ that lie above the Toomre marginally stable disks. 

We investigate this further, with the relevant results summarized in Appendix \ref{app:invest-sigma0-sfr}. We first study the relation between $\disp$ and the offset from the main sequence $\Delta \rm MS = SFR/SFR_{MS}$, but find no significant trend (Fig. \ref{fig:sigmao-vsigma-MS}). This is consistent with findings at cosmic noon \citep{Ubler:2019vg} and implies that at fixed stellar masses, galaxies with higher SFRs do not show enhanced $\disp$. This lack of enhancement does not support purely stellar feedback driven turbulence. These conclusions do not change if we include only galaxies in the conservative range of $\log (M_{\star} ~ \rm [M_{\odot}]) = 9-10$. We also find no trend between $\disp$ and $\text{sSFR}_{10}$. We explore the placement of galaxies around the MS, in Fig. \ref{fig:SFMS-comb}, in relation to their kinematic properties. Although we find no overarching trend, as indicated by the lack of correlations with $\Delta \rm MS$, it appears that many of the systems with high rotational support $\rotsupp>3$ lie close to the main sequence, particularly for the gold sample. An object-by-object analysis is needed to study the link between migrations around the MS and kinematics, but this is beyond the scope of this work.

The observed $\disp-\text{SFR}_{10}$ correlation ($\rho = 0.40$ and $p<0.001$) appears more significant than the one between $\disp$ and $M_{\star}$ ($\rho = 0.24$ and $p=0.036$), which suggests the latter is mainly driven by the SFMS. This is supported by predictions from theoretical models \citep{Pillepich:2019aa}. Nonetheless, the weak correlation between $\disp$ and $M_{\star}$, and the overall distribution of the gold sample points (as seen on Fig. \ref{fig:vsigma_sigma_mstar_sfr}) suggests that higher mass galaxies appear to have higher velocity dispersions in our sample. Although this is somewhat physically motivated by the SFMS, it could also be due, in part, to selection effects. The less massive galaxies are typically smaller, so we could be bias to low-$\disp$ systems at lower masses because the higher $\disp$ systems are too faint (see Sec. \ref{sec:geko-biases} for further discussion).

Finally, we find that $\disp$ correlates with  $\Sigma_{\text{SFR}_{10}}$ ($\rho = 0.38$ and $p=0.001$), although we probe a relatively small range of parameter space. This correlation supports the idea that more compact star-forming systems have increased turbulence, but it is most likely driven by the $\rm \disp-SFR$ relation as we find no strong correlation between size $r_{\rm e}$ and $\disp$. 

The trends for $\rotsupp$ with $M_{\star}$ and SFR are weaker, suggesting they are driven by the existing relations for $\disp$. This is supported by the fact that the rotational velocities do not strongly correlate with either of these properties (see Fig. \ref{fig:v_max_mstar_sfr10}). As discussed in Sec. \ref{sec:res-1}, at fixed stellar masses, galaxies probe a range of kinematic states depending on external influences such as gas accretion and mergers, which can either promote or completely disrupt the formation of the disk. These processes leave imprints on the rotational velocities and can create the observed scatter within galaxies sharing similar masses and star-forming properties. Hence, the measured rotational support $\rotsupp$ in our sample at $z\sim 4-6$ is mainly driven by the state of turbulence of the gas at the time of observation. As such, we expect that the kinematics of the ionised gas in these galaxies are, on average, not dominated by undisturbed circular motions, as can be seen by the low values of $\rotsupp$.

Although the correlation between $\rotsupp$ and $M_{\star}$ is not significant ($\rho = -0.27$ and $p=0.015$), measurements from the gold sample show a decrease of $\rotsupp$ with stellar mass, particularly at the high-mass end (Fig. \ref{fig:vsigma_sigma_mstar_sfr}). From studies out to cosmic noon, we would expect more massive galaxies to be more stable and hence more rotationally supported, which would result in a positive $\rotsupp-M_{\star}$ relation, as found in many observation and simulation-based works \citep[e.g.,][]{Wisnioski:2015vx,Pillepich:2019aa, Price:2020wf}. At high redshift, this is not necessarily intuitive given some key differences such as the increased occurrence of mergers and the smaller orbital timescales. As we discuss in Sec. \ref{sec:disp-drivers}, we would, independent of redshifts, expect a flattening in the $\rotsupp$-$M_{\star}$ relation in the lower mass end ($M_{\star}<10^{10} \rm ~M_{\odot}$) explored in this work. The medians for both the gold and extended samples are consistent with such a flattening. The observed decrease, when looking at the gold sample measurements, of $\rotsupp$ with mass $M_{\star}$ could be driven in part by observational biases, since, as previously discussed, we may be biased to low $\disp$ systems at low masses, which would decrease the number of low $\rotsupp$ objects detected at low masses. However, this trend is also driven by the higher mass ($\log M_{\star} [\mathrm{M}_{\odot}] > 9.3$) dispersion-dominated systems. These systems have very high velocity dispersions $\sigma_0\sim 100-200$ km/s, not comparable to the similar-mass galaxies seen in the local Universe.  It is possible that the highest-mass systems in our sample are the result of recent major mergers, as they probe the tail end of the stellar mass functions at $z\sim 4-6$. Such an event would heavily disrupt the disk and could explain the high $\disp$ values measured. We note that this trend decreases in strength and significance ($\rho = -0.01$ and $p=0.93$) when we only include rotationally supported systems $\rotsupp>1$ (see Appendix \ref{app:method-comparison}, including Tab. \ref{tab:corr-metho-comp}).

Overall, despite a large scatter in our measurements, we find that $\disp$ correlates with star formation already at $z\sim 4-6$. The high values of $\disp$ we find could be explained by gravitational instabilities in the disk caused by radial inflows of gas, as well as a potential increase in the role of stellar feedback in injecting turbulence in the ISM.

\subsubsection{Drivers of turbulence across cosmic time} \label{sec:disp-drivers}

\begin{figure*}
    \centering
    \includegraphics[width=1\linewidth]{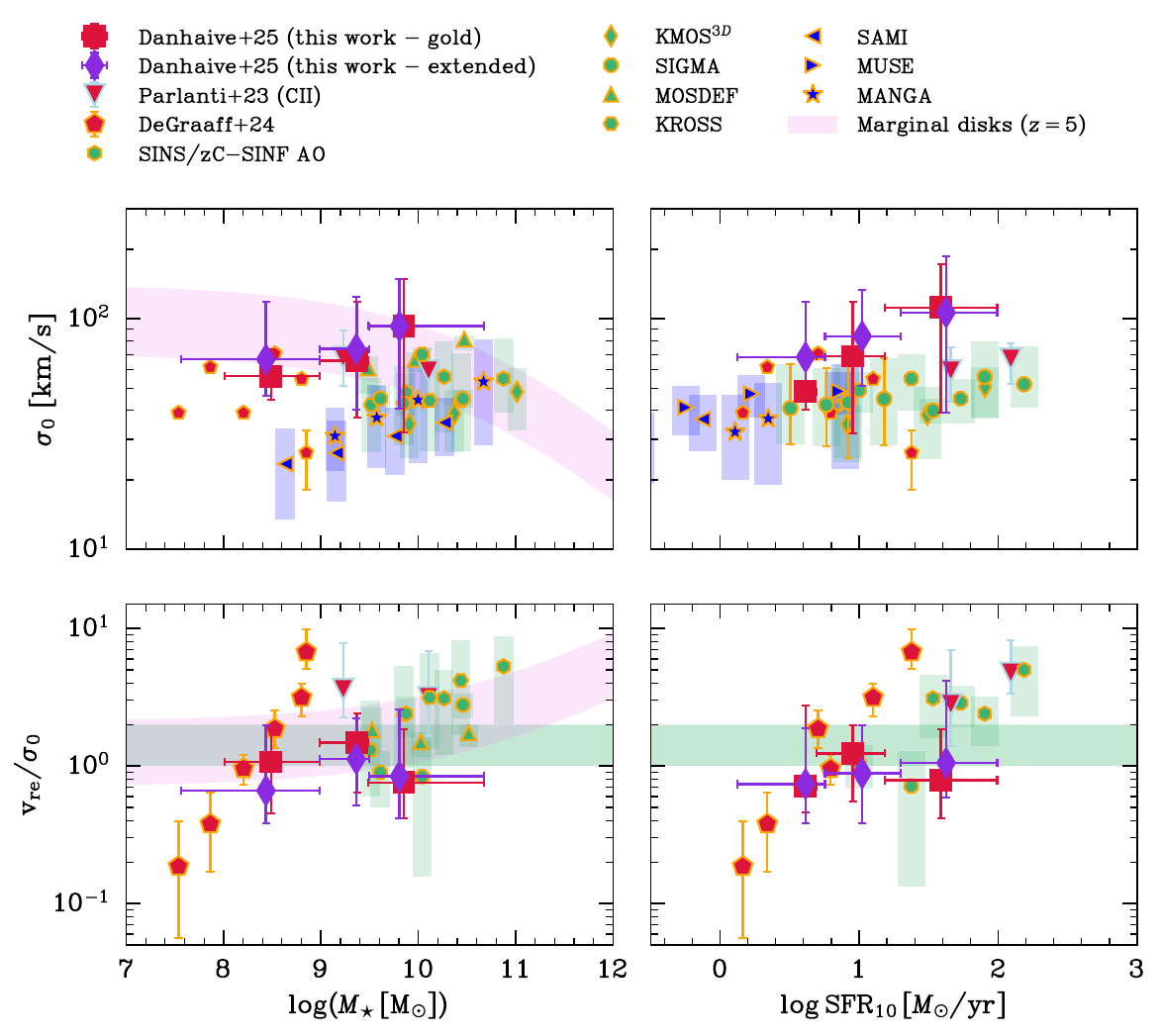}
    \caption{Dependence of $\disp$\ (top panels) and $\rotsupp$\ (bottom panels) on stellar mass and SFR for our sample (red squares and purple diamonds) in the context of other high-redshift works \citep[in red;][]{Parlanti:2023ab, de-Graaff:2024ab}, and works at cosmic noon (in green; see Fig. \ref{fig:sigma0-z-comp}) and the local Universe (in blue; MANGA \citealt{Yu:2019aa}, SAMI \citep{Allen:2015aa,Johnson:2018aa,Green:2018aa}, and MUSE \citep{Swinbank:2017aa}). The errorbars on our plotted medians represent the $16^{\rm th}$ and $16^{\rm th}$ quantiles along the y-axis, and the extent of the bin along the x-axis. The $\disp$-SFR relation holds across redshifts, for both ionised (\Ha; orange outline) and cold ([CII]; blue outline) gas, although we find higher values of $\disp$ and a steeper slope. Some works at lower redshift find a similar weak increasing trend for the $\disp-M_{\star}$ relation as our sample \citep{Wisnioski:2015vx,Forster-Schreiber:2018aa,Yu:2019aa,Price:2020wf}, while \citet{Simons:2017aa} finds no significant trend. The trends for $\rotsupp$ are less clear, but our results could indicate a flattening in the $\rotsupp-M_{\star}$ relation at low masses $\log M_{\star} [\mathrm{M}_{\odot}] \lesssim10$, which is consistent with other works in this mass range and which is predicted by the marginal disk model (purple shaded regions).}
    \label{fig:corr-comp}
\end{figure*}

We now put the correlations we find between kinematic parameters and stellar populations in the context of results reported by other works across redshifts, as shown in Fig. \ref{fig:corr-comp}. We focus on trends with $M_{\star}$ and SFR, and compare our running medians to medians from studies of the local Universe (MANGA, \citealt{Yu:2019aa}; SAMI, \citealt{Croom:2012aa, Johnson:2018aa}; MUSE, \citealt{Bacon:2010aa, Swinbank:2017aa, Johnson:2018aa}), cosmic noon (KMOS3D, \citealt{Wisnioski:2015vx}; MOSDEF, \citet{Price:2020wf}, SIGMA, \citealt{Simons:2016aa}; KROSS, \citealt{Johnson:2018aa}; SINS/zC-SINF AO,\citealt{Forster-Schreiber:2018aa}), and high redshift $z\sim 4-8$ \citep{Parlanti:2023ab,de-Graaff:2024ab}. We again note that many of these works restrict their analyses to rotationally supported systems $\rotsupp>1$, which could also reduce the number of systems with high $\disp$ in their sample. We compute correlations for the $\rotsupp>1$ galaxies in our sample and find no qualitative difference, although the significance of the correlations slightly varies (Tab. \ref{tab:corr-metho-comp}). 

As shown by the colour-coding of the points by approximate redshift, our values (red) of $\disp\approx 50-150 \thinspace \text{km/s}$ lie on average above values at cosmic noon (green) $\disp\approx 30-70 \thinspace \text{km/s}$, as well as measurements from the local Universe (blue) $\disp\approx 20-50 \thinspace \text{km/s}$ which have the lowest dispersions on average (see also Fig. \ref{fig:sigma0-z-comp}). 

We discussed in Sec. \ref{sec:corr-z6} that the $\disp \rm - SFR$ correlation is predicted by theoretical models in which gravitational instabilities and feedback drive turbulence in the disk. These processes act at all redshifts, and explain why we observe this trend in all the surveys shows in Fig. \ref{fig:corr-comp}. In the local Universe, several studies find a significant positive correlation of $\disp$\ with SFR and $\Sigma_{\text{SFR}}$  \citep{Arribas:2014aa,Yu:2019aa, Law:2022aa}. These trends seem to hold out to cosmic noon, albeit with higher observed velocity dispersions and weaker correlations, indicative of a turbulent ISM \citep{Forster-Schreiber:2009aa,Genzel:2011aa,Wisnioski:2015vx,Simons:2017aa,Johnson:2018aa, Ubler:2019vg,Price:2020wf}. In our work, we find that this trend is already in place at $z\sim 4-6$.

Also, the turbulence induced by these processes is expected to increase with redshift as galaxies are on average less massive, becoming less stable to gravitational instabilities and disruptions from stellar feedback \citep{Nelson:2019aa,Pelliccia:2020aa}. The overall decrease of gas fractions with cosmic time also promotes the decrease of turbulence in the ISM \citep{van-Donkelaar:2022aa}, as studies have shown that at high gas surface densities, and high star formation rate surface densities \citep{Rathjen:2023aa}, the turbulence injected into the ISM from supernovae explosions can rise from $\sigma\sim 5-20$ km/s \citep{Krumholz:2018aa} to $\sigma\sim 50-60$ km/s due to strong stellar feedback \citep{Hopkins:2011aa} and high gas fractions \citep{Gatto:2015aa}. All of these factors could explain why the $\disp \rm - SFR$ trend is steeper, and has a higher normalization, at high redshift. 

The plotted medians from \citet{Parlanti:2023ab} are from measurements of cold gas, which tend to have factors of $2-3$ lower velocity dispersions \citep{Ubler:2019vg,Rizzo:2024aa}. We also plot the six galaxies from \citet{de-Graaff:2024ab}, which probe the low-mass end. 

Our observed mild dependence of $\disp$\ with \Mstar\ is also in agreement with past works which find a weak positive \citep{Wisnioski:2015vx,Turner:2017aa,Forster-Schreiber:2018aa,Yu:2019aa,Ubler:2019vg,Price:2020wf} or non-existent \citep{Simons:2017aa} correlation.  The intrinsic correlation between stellar mass and SFR, i.e. the SFMS, implies that the $\disp$-SFR relation could explain most of the trend observed with stellar mass. At lower redshifts, many attribute the more fundamental correlation to SFR instead of \Mstar\ \citep{ Wisnioski:2015vx,Yu:2019aa}. We compare observational measurements to prediction from the marginal disk model at $z=5$, and find this empirical framework predicts a negative relation between $\disp$\ with \Mstar. This is in contrast with findings across cosmic time on Fig. \ref{fig:corr-comp},  which show a flat or increasing trend. The Toomre marginal disk model (Eq. \ref{eq:toomre-vs}) describes $\disp$ based on gas fractions, which are expected to decrease with mass. The observed increase of $\disp$ with $M_{\star}$ suggests either higher gas fractions, than predicted by the model, at high masses, or significant contributions from other sources of turbulence.

The results for $\rotsupp$ are more difficult to interpret. Starting with the $\rotsupp-M_{\star}$ relation, it is difficult to conclude on an overarching trend that holds at all masses. Our sample of galaxies at $z\sim 4-6$ probes lower masses than lower redshift counterparts, but lies slightly above the masses from \citet{de-Graaff:2024ab}. The works from the local Universe to cosmic noon have reported positive correlations, albeit with varying strengths. However, as can be seen on Fig. \ref{fig:corr-comp}, the trend weakens below $\log M_{\star} [\mathrm{M}_{\odot}]\sim 10$, and the medians appear to flatten. The medians from \citet{Parlanti:2023ab} also show no strong evolution between $\log M_{\star} [\mathrm{M}_{\odot}]\sim 9-10$. Overall, the different regimes in which the positive $\rotsupp-M_{\star}$ relations holds remain unclear, but this work provides evidence that the dependence of $\rotsupp$\ on $M_{\star}$ weakens at lower masses. This is in line with predictions from the marginal disk model (Eq. \ref{eq:toomre-vs}), as is also discussed in \citet{Simons:2016aa}.

We do not find a significant correlation between $\rotsupp$ and SFR, but measurements from the gold sample show a decreasing trend, which appears in contrast with measurements from cosmic noon on Fig. \ref{fig:corr-comp}. However, we probe lower SFRs and overall more turbulent systems than studies at than cosmic noon, and given the large scatter our measurement, we cannot draw a strong conclusion.

\subsection{Comparing stellar masses and dynamical masses}\label{sec:dyn-mass}

Using the results from our kinematic modelling, we are able to compute dynamical masses for the objects in our sample. Dynamical masses offer insights into the dark matter content of galaxies, but for the scope of this work we only discuss them in the context of a comparison with stellar mass. We defer a detailed study of the dark matter and baryonic mass contents of our sample to a follow-up paper (Danhaive et al, in prep). 

\begin{figure}
    \centering
    \includegraphics[width=1\linewidth]{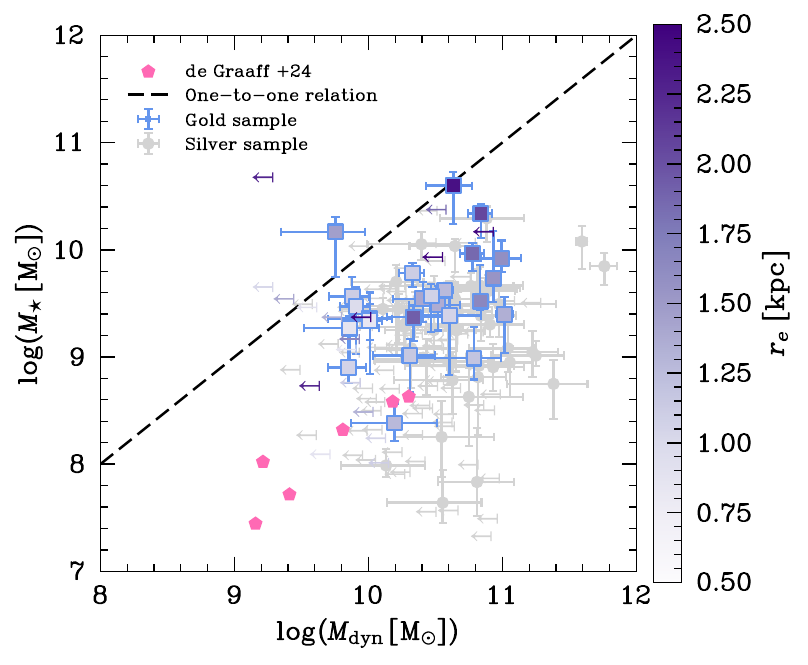}
    \caption{Comparison of the dynamical masses (\Mdyn) inferred from grism data and the stellar masses ($M_{\star}$) inferred from SED fitting for our gold (blue squares), colour-coded by \Ha\ size, and silver (gray circles) samples. The majority of our systems lie below the one-to-one relation (dashed line), consistent with a significant contribution to the dynamical mass from gas and/or dark matter. Six galaxies lie on the relation, potentially highlighting underestimated uncertainties on $\disp$ (see Sec. \ref{sec:geko-biases}). We compare our values to ionised gas measurements from \citet{de-Graaff:2024ab}.}
    \label{fig:mdyn-mstar}
\end{figure}

We use our posteriors inferred with \textsc{geko} to compute the posterior distribution of the dynamical mass for each system using Eqs. \ref{eq:v_circ} and \ref{eq:dyn-mass}. Eq. \ref{eq:dyn-mass} assumes that the system is in virial equilibrium and that the measured velocity gradient is not probing non-circular motions such as outflows. In Fig. \ref{fig:mdyn-mstar} we compare the dynamical masses \Mdyn\ inferred from grism data to the \textsc{prospector} stellar masses inferred from spectral energy distribution (SED) modelling. We expect the dynamical masses to lie above the stellar masses, as they should also incorporate the gas and the dark matter. However, both methods of inferring masses suffer from uncertainties (as can be seen from the large error bars) and caveats, so comparing the two offers a good benchmark to understand how reliable both estimates are.

Fig. \ref{fig:mdyn-mstar} shows a large scatter in dynamical masses, indicating varied kinematic states and gas (and dark matter) fractions. We plot points from \citet{de-Graaff:2024ab} for comparison, which lie well below the one-to-one relation, indicative of high gas and dark matter fractions.

We find consistent measurements $M_{\star}<M_{\text{dyn}}$ for the majority of the galaxies in our sample, with the exception of 5 systems which lie above or on the one-to-one relation. It is difficult to evaluate the cause for this discrepancy, as it could be due to the SED fitting, the kinematics fitting, or a mixture of both. We discuss possible sources of error in the stellar mass estimate in Sec. \ref{sec:caveat-mass}. For systems $\disp$\ is only an upper limits, the measurement of $M_{\text{dyn}}$ is more uncertain. For example, we could be under-estimating the uncertainties on our $\disp$\ measurements, which has a large impact on the derived upper limits (Sec. \ref{sec:geko-biases}). If the velocity dispersions are under-estimated, this would propagate to the dynamical masses and cause the observed discrepancy in these systems. 

We find an interesting system (JADES ID: 1082948) that is among the most massive in our sample, with $\log (M_{\star}~\rm [M_{\odot}]) = 10.6^{+0.1}_{-0.2}$, and yet shows no signs of rotation, resulting in a low dynamical mass of $\log (M_{\star}~\rm [M_{\odot}]) = 8.9^{+0.4}_{-0.6}$. We investigate possible reasons for this discrepancy. Regarding the kinematic modelling, although we find a good fit of the model to the data, there could be a faint, broad \Ha\ component that isn't being modelled well due to the overpowering brighter central region. Also, the best-fit results from \textsc{Prospector} show, despite the large uncertainties, that the galaxy has a declining SFH since the last $\sim 20$ Myr, after peaking at $\rm SFR\sim 270 ~M_{\odot}/yr$. This galaxy could be in a quenching epoch, as it also has a relatively small UV size, $r_{\rm e} = 1.5$ kpc, for its mass, lying below the predicted relations at $z\sim 4$ \citep{Allen:2024aa, McClymont:2025ab}. However, it lies close to the SFMS (Fig. \ref{fig:SFMS}), so its star formation is decreasing following a burst and it may now be entering a quenching episode. The imaging indicates that this galaxy could be in a post-merger phase. This system passed our visual inspection, where we discarded close pairs as mergers, because the clumps were too close to classify it as a merger. 

\section{Discussion} \label{sec:discussion}

Extensive studies of galaxy kinematics from the local Universe to cosmic noon have shown that the star-forming galaxy population evolves from gas-rich star-forming thick disks at $z\sim2-3$ to dynamically cold thin disks by $z\sim 0$. Despite their increased turbulence \citep{Glazebrook:2013aa,Wisnioski:2015vx, Ubler:2019vg, Forster-Schreiber:2020aa,Rizzo:2024aa}, the thick disks at cosmic noon can be described by many of the same key scaling relations, between kinematics and star formation related quantities, that are in place at $z\sim0$, albeit with some offsets and a larger scatter \citep[e.g.][]{Tiley:2016aa, Price:2020wf,Ubler:2017aa,Sharma:2024aa}. 

Our work at $z\sim 3.9-6.5$ shows that in the earlier stages of galaxy formation, the picture becomes complex. As we attempt to bridge the gap between the Epoch of Reionization (EoR) and cosmic noon, it is important to remember the importance of stellar mass. Studies from $z\sim0$ to $z\sim3$ predominantly probe galaxies with $M_{\star} \gtrsim 10^{10}\thinspace M_{\odot}$, with the typical sample distribution peaking around $M_{\star}\sim 10^{11}\thinspace M_{\odot}$ and $M_{\star}\sim 10^{10.5}\thinspace M_{\odot}$, respectively. At $z\sim 4-6$, we have a galaxy population dominated by systems in a lower mass slice of the parameter space $M_{\star} \sim 10^{8}-10^{10} \thinspace M_{\odot}$. Careful consideration must therefore be given to separate pure redshift evolution from evolution mainly driven by stellar mass. 

Studies have shown that even at high redshifts $z\sim 4-5$, some galaxies with stellar masses $\log M_{\star} [\mathrm{M}_{\odot}] \gtrsim 10$, comparable to those probed by the star-forming population at cosmic noon, are already able to settle into stable rotating disks \citep[e.g.,][]{Neeleman:2020aa, Lelli:2021aa}. However, it is not surprising that many of the low-mass systems at high redshift are turbulent and have not settled into a stable rotationally-supported state (Fig. \ref{fig:vsigma_z_evol}). Their ISM is characterized by elevated velocity dispersions (Fig. \ref{fig:sigma0-z-comp}), which are likely driven by increased gravitational instabilities promoted by their lower masses, high gas fractions, and the overall increase of mergers and starbursts in the early Universe. These instabilities, along with a potential contribution from stellar feedback, drive the observed correlations between SFR, and SFR surface density, and $\disp$ (Fig. \ref{fig:vsigma_sigma_mstar_sfr}).

Overall, the picture painted in this work is consistent with galaxies at high redshift being turbulent and changing on short timescales. In this case, we expect the prevalence of disks to be lower at $z\gtrsim4$ than in the local Universe and at cosmic noon. We investigate and discuss this in the next section.

\subsection{The prevalence of disks at $z>4$}\label{sec:disk-fracs}

\begin{figure*}
    \centering
     \includegraphics[width=1\linewidth]{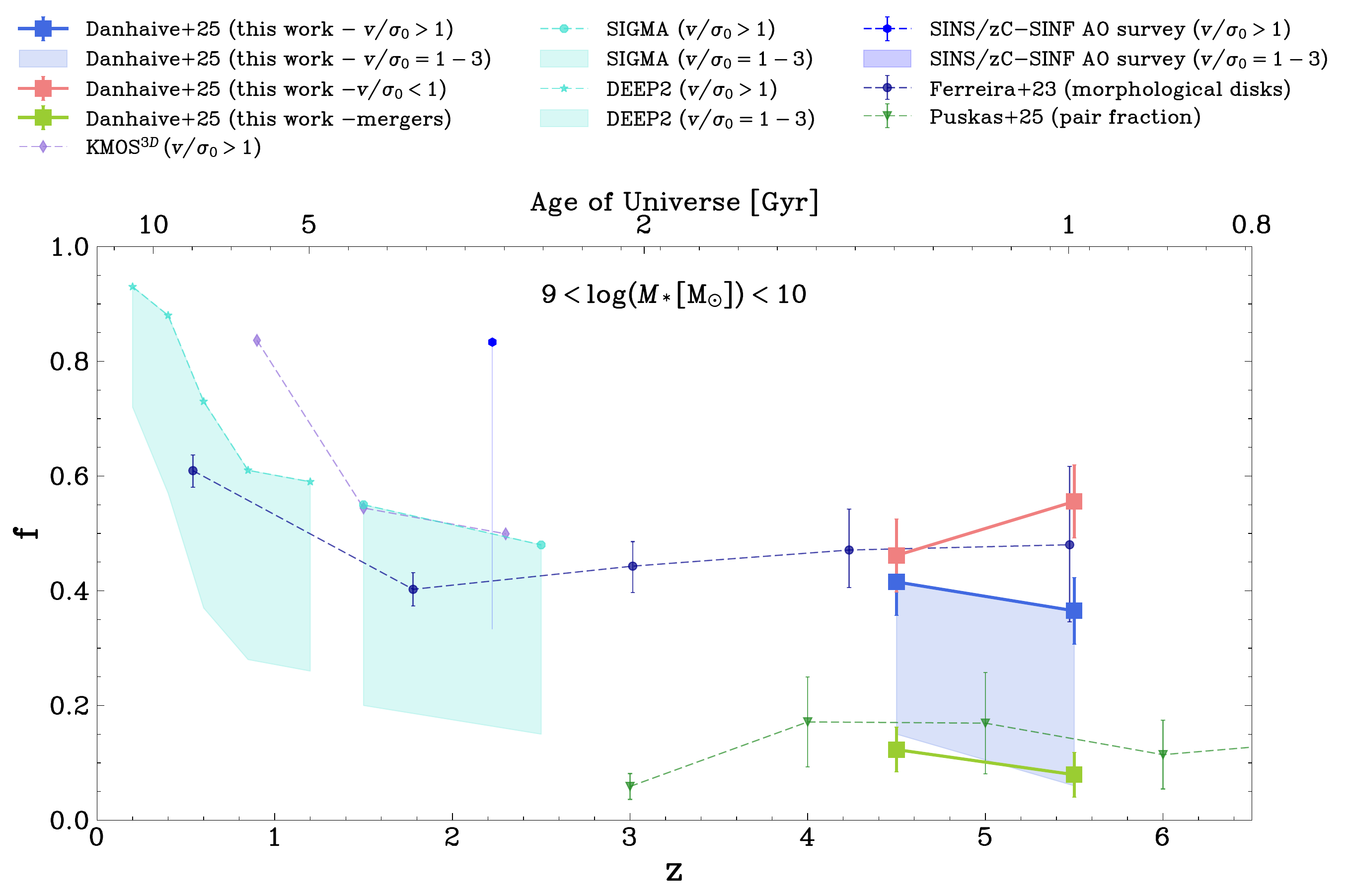}
    \caption{Evolution of the fraction of rotationally supported systems ($\rotsupp>1$) with redshift for our sample (blue lines) and for kinematic samples \citep{ Simons:2017aa, Forster-Schreiber:2018aa,Wisnioski:2019tg} and morphological sample \citep{Ferreira:2023aa} from the literature (blue dashed lines). We find a decrease in the fraction of systems with rotational support with redshift, consistent with predictions from lower redshifts, with fractions significantly lower if we consider a more conservative cut of $\rotsupp>1-3$ (blue shaded regions), with the lower edge at $\rotsupp>3$. This result contrasts the (on-average) constant fractions reported by \citet{Ferreira:2023aa}. We compare the fraction of pairs (green line), which we discard from our kinematic analysis, to the one reported in \citet{Puskas:2025aa} (green dashed line), and find broadly consistent results.}
    \label{fig:disk-frac}
\end{figure*}

With recent JWST and ALMA observations finding disk-like systems in the early Universe \citep[$z>4$; e.g.,][]{Lelli:2021aa,Pope:2023aa, Rowland:2024aa}, questions arise as to how these systems are able to form and stabilize in the turbulent, high-redshift environment. To answer the question of how representative these massive rotating disks are of the high redshift star-forming galaxy population, we investigate the fraction of rotationally supported systems $\rotsupp>1$ in our sample and place it in the context of lower redshift measurements (Fig. \ref{fig:disk-frac}). 
Without spatially resolved velocity maps (such as those obtained with IFUs), it is difficult to classify systems as disks and to compute a disk fraction. This is highlighted by recent IFU studies at high redshift, which find rotationally supported disks that are heavily perturbed by non-circular motions \citep{Arribas:2024aa,Ubler:2024aa}. This motivates our choice of not defining a disk fraction but instead focusing solely on the evolution of the fraction of rotationally supported systems $\rotsupp>1$ across cosmic time. However, we require systems to have kinematic and morphological position angles that are relatively aligned $|\text{PA}_{\text{kin}} - \text{PA}_{\text{morph}}|<30^{\circ}$. From our original full sample of \Ha\ emitters in the FRESCO and CONGRESS surveys, we remove all low S/N galaxies (S/N<10), all galaxies with a position angle too close to the grism dispersion direction, and all galaxies which are spatially and/or rotationally unresolved. This cut excludes all galaxies that are not in our gold or silver samples (Sec. \ref{sec:sample-selection}). We focus on the stellar mass range $9<\log M_{\star} [\mathrm{M}_{\odot}] < 10$ where our sample is most representative of the star-forming galaxy population (Fig. \ref{fig:SFMS}).

We also need to quantify a fraction of merging systems, which cannot be modelled with our single-component approach (see Sec. \ref{sec:modelling}). This includes both minor and major mergers, and could include clumpy systems that resemble mergers. We select systems with small separations $r_{\rm max}\sim10\rm ~kpc$ ($r\lesssim 1.5''$) as mergers, since the single component modelling would not be contaminated by pairs with larger separations. This differs from in-depth studies of merger fractions, which include much larger separations $r_{\rm max}\sim30-50 \rm ~kpc$ \citep{Duncan:2019aa,Duan:2024aa, Puskas:2025aa}. In order to consistently measure the pair fraction in our sample, we exclude galaxies which would have not made the cut for kinematic modelling, i.e., galaxies with a low S/N and galaxies with  $|\rm PA_{\text{morph}}|>75^{\circ}$ (Sec. \ref{sec:sample-selection}).

Our results are consistent with an increase of rotationally supported ($\rotsupp>1$) systems with cosmic time. In our sample, we find that this fraction increases slightly from $f = 0.36 \pm 0.06$ to $f = 0.41 \pm 0.06$ from $z\sim 5.5$ to $z\sim 4.5$. This measured increase is consistent with the slow build-up of disks over the $\approx 600 \rm ~ Myr$ probed by our sample. These fractions drop significantly, to  $f = 0.06 \pm 0.03$ and $f = 0.15\pm 0.04$ at $z\sim 5.5$ and $z\sim 4.5$, when considering systems with $\rotsupp>3$. This highlights that although a considerable fraction of the galaxies in our sample are rotationally supported, the majority of these systems are found in the transitional region of $\rotsupp=1-3$. These systems could be starting to build rotational support through co-planar accretion of gas, or conversely could have just undergone a merger or radial accretion event, which has disrupted the formation of a rotating disk. Only a small fraction $f\lesssim0.1$ of systems has significant rotational support, indicating that disk-like galaxies do not dominate the galaxy population at $z>4$ at stellar masses $9<\log M_{\star} [\mathrm{M}_{\odot}] < 10$, but that dispersion-supported systems are common. 

The increase of rotationally supported systems with cosmic time found in this work is consistent with works at lower redshift $z<4$ \citep{Simons:2017aa,Forster-Schreiber:2018aa,Wisnioski:2019tg}. In fact, both \citet{Wisnioski:2019tg} and \citet{Simons:2019aa} find a decrease from $f\approx 0.8-0.9$ at $z<1$ to $f\approx 0.5$ by cosmic noon ($z\sim 2-3$). This is also consistent with \citet{Pandya:2024aa} who classify systems from $z=0.5-8$ based on their 3D geometry and find an increase of oblate (disky) systems with cosmic time. In their highest redshift bin $z=3-8$, they find that the fraction of oblate systems is $f\sim0.2-0.4$ (depending on the modelling technique chosen) for galaxies with stellar masses $\log (M_{\star} ~ \rm [M_{\odot}]) \sim 9.25-9.75$.

When compared to the morphological analysis of disk fractions conducted in \citet{Ferreira:2022aa, Ferreira:2023aa}, we find lower fractions across both redshift bins. However, their work does not apply an upper mass cut-off, as we do here, so this could introduce a bias. More importantly, our fraction of rotationally supported systems is not equivalent to a disk fraction, as shown in part by the significant decrease when considering a higher threshold for rotational support. We also find that only a fraction of $\sim 0.33$ of our rotationally supported sample has an \Ha\ morphology with a S\'{e}rsic index $0.5<n<1.5$, consistent with the exponential profiles we see in local disks. Although systems with $0.5<n<1.5$ represent the peak of the $n$ distribution, we find a long tail with rotationally supported systems having S\'{e}rsic indices $n\sim 2-7$. The wide range of $n$ values offers insight into the perturbed morphologies of galaxies at $z\gtrsim 4$, making it more difficult to identify rotating disks akin to those seen in the local Universe. This is further supported by the predominantly low values of $\rotsupp$\ we find, with $\approx 65\%$ of the rotationally supported systems having $\rotsupp<3$. These galaxies could drop out of $\rotsupp>1$ disk criterion on short timescales if enough momentum is injected in their ISM, for instance through a merger, a starburst, or a disruptive accretion event.

\begin{figure*}
    \centering
    \includegraphics[width=1\linewidth]{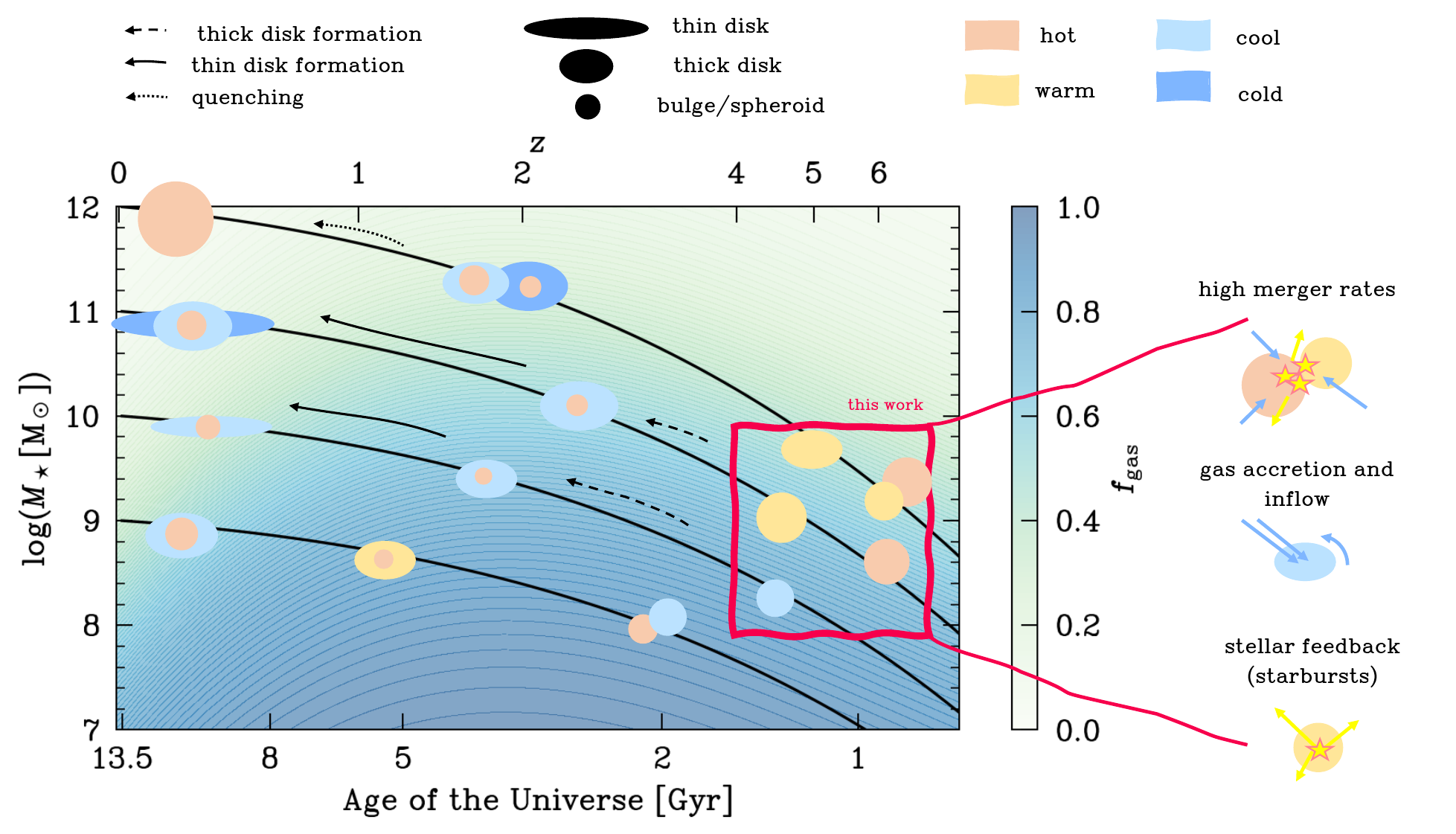}
    \caption{Schematic view of the dynamical evolution of galaxies from $z\sim 8$ to the local Universe, placing the results of our work (red square) in this broader picture. Galaxies are color-coded by dynamical state, from cold ($\rotsupp\gtrsim1$; blue) to hot ($\rotsupp\lesssim1$; orange), and shaped according to their thickness/diskiness. We show that as galaxies evolve along the main sequence (black curves), they settle into disk-like systems once they reach small enough gas fractions (inferred by the \protect\citealt{Tacconi:2020aa} relation), which depend on both mass and redshift. In the redshift and mass range probed by this work, galaxies are gas-rich and have a turbulent ISM, with high velocity dispersions driven by instabilities caused by mergers, gas inflows, and feedback from star formation.}
    \label{fig:cartoon}
\end{figure*}

Alongside the fraction of disks, we also plot in Fig. \ref{fig:disk-frac} our merger fraction obtained from the visual inspection of images of the galaxies in our sample, and compare it to the major merger pair fractions derived in \citet{Puskas:2025aa} \citep[see also][]{Duan:2024aa}. We find relatively consistent results, which offers a sanity check for our visual inspection. However, we note that our merger fraction selection differs from that of \citet{Puskas:2025aa}. As mentioned above, we probe significantly smaller separations, which can in part explain the lower fractions we find. In terms of merger timescales, which quantify how long galaxies appear as pairs, we find that our separation $r_{\rm max}\sim10\rm ~kpc$ implies timescales of $T_{\rm mg} \sim 0.3-0.5$ Gyr for $\log (M_{\star}\rm ~[M_{\odot}]) = 9-10$ at $z=5$, whereas the separation $r_{\rm max}\sim30\rm ~kpc$ from \citet{Puskas:2025aa} probes timescales of $T_{\rm mg} \sim 1.4-2.2$ Gyr for the same masses and redshift. These values are derived from a fitting formula based on cosmological dark matter simulations \citep{Husko:2022aa}. Also, \citet{Puskas:2025aa} only select major mergers (mass ratio > 1/4), whereas we include systems with smaller mass ratios as well. We not that our fraction of mergers is low $f\lesssim 0.15$, so discarding these systems does not strongly affect our inferred fractions or the main conclusions of this work.

Overall, the strong decrease of the fraction of disk-like systems from $z<1$ to $z = 6$ suggests that the majority of the galaxy population only settles into cold disks during and after cosmic noon, when the bulk of the stellar mass is formed. The fraction of rotationally supported systems increases smoothly from $z\sim 6$ until the start of cosmic noon $z\sim 2.5-3$, where this fraction becomes $f\gtrsim 0.5$, and disk-like galaxies begin to dominate the galaxy population.

\subsection{Implications for galaxy evolution across cosmic time}\label{sec:gal-evol}

We will now discuss how this work fits into the more global evolution of galaxies and propose explanations for the observed turbulence at high redshift. In Fig. \ref{fig:cartoon}, we show a schematic view of our understanding of galaxy evolution from a dynamical point of view, tracing the evolution of local Universe galaxies with $\log M_{\star} [\mathrm{M}_{\odot}] = 9,10,11$ and 12 back in time by integrating the main sequence from Simmonds at al. (in prep.), which is consistent with predictions from simulations \citep{Tacchella:2016ab, McClymont:2025aa}. We then colour the $M_{\star}-z$ plane by gas fraction $f_{\text{gas}}$ following the relation from \citet{Tacconi:2020aa}. 

In the local Universe, most star-forming galaxies have settled into disks, which are increasingly thin at higher masses, as highlighted by stellar kinematics \citep{van-der-Wel:2014aa,Cappellari:2016aa}. On the higher-mass end, some galaxies have quenched, and their rotational support decreases as they compress and their sizes decrease. At cosmic noon, the star-forming population is predominantly made up of gas-rich thick disks, with the fraction of rotationally supported systems, based on ionised gas measurements, dropping by $\sim 30\%$ from $z\sim 0$ for galaxies with stellar masses $9<\log M_{\star} [\mathrm{M}_{\odot}] < 10$ (Fig. \ref{fig:disk-frac}). As drawn in Fig. \ref{fig:cartoon}, these thick disks only settle in dynamically cold disks below $z\sim 1$, which also coincides with a decrease in gas fractions to below $f_{\text{gas}} \lesssim 0.5$. Because of the dependence of gas fractions on stellar mass, this happens sooner for higher-mass systems, which could promote the formation of thin disks with $M_{\star}\sim 10^{11} \thinspace M_{\odot}$ at cosmic noon \citep{Wisnioski:2015vx,Price:2020wf}. However, mass itself is a key factor in disk formation, as more massive galaxies are more stable to gravitational instabilities, especially when the mass profile becomes centrally concentrated \citep{Hopkins:2023aa}. Even at high gas fractions, massive galaxies can settle into disks if the gas accretion is somewhat smooth.

With the arrival of JWST, we are now able to further constrain the kinematics of galaxies beyond $z\sim4$ and complement the view of cold gas kinematics from ALMA. The picture that has been constructed in this work, and is supported by other works at high redshift \citep{Parlanti:2023ab,Arribas:2024aa}, is one of a turbulent epoch of early galaxy formation. Galaxies at masses below $M_{\star}\sim 10^{10} \thinspace M_{\odot}$ are characterized by high gas fractions $f_{\text{gas}} \gtrsim 0.7-0.8$, which increase at low masses (Danhaive et al., in prep.). Although the gas fractions are mildly lower at higher masses, it is likely that the highest-mass systems in our sample have recently undergone mergers. In fact, \citet{Puskas:2025aa} shows that merger fractions at $z\sim4-6$ are high, with $2-10$ major mergers per Gyr, and it is therefore reasonable to expect that the higher-mass systems were able to form quickly through major or minor mergers. This would justify why we find many dispersion-dominated systems on the massive end, as mergers can increase turbulence in the ISM and also boost star formation through gas inflows.

At lower masses, $\log M_{\star} ~\rm[M_{\odot}]\lesssim 9.5-10$, galaxies become more unstable to disruptions from gravitational instabilities and star formation. Specifically, as also discussed in \citet{de-Graaff:2024aa}, simulations predict that stellar feedback starts to play an important role in this mass range, with mass loading factors for stellar winds and supernovae reaching values $\sim 30$ times larger for low-mass ($\log M_{\star} ~\rm[M_{\odot}]\sim 8$) galaxies than for high-mass ($\log M_{\star} ~\rm[M_{\odot}]\sim 11$) galaxies \citep{Nelson:2019aa}. The shallow potential wells of low mass galaxies, compared to high mass galaxies, promote outflows as they need less energy to be driven out of the galaxy \citep{Bassini:2023aa}. Recent observational works are consistent with this view, detecting strong outflows in low mass galaxies out to $z\sim 9$ \citep{Zhang:2024aa, Carniani:2024aa}. The shallow potential wells of low mass galaxies also promote inflows of gas that reach the central regions of the galaxy \citep{Dekel:2014aa,Tacchella:2016aa,El-Badry:2016aa,Hopkins:2023aa,McClymont:2025aa}. As the gas compacts, it triggers bursts of star formation, whose strong feedback can expel a lot of the remaining gas. These bursts are hence regulated by inflows and outflows which disrupt disk formation and inject turbulence in the ISM, sustaining high velocity dispersions. Such processes perturb disk formation through misaligned accretion and spin flips \citep{Dekel:2020aa, Sales:2012aa}. 

Many galaxies in our sample have rotational support $\rotsupp \sim 2$ akin to massive quiescent galaxies at $z\sim 1-2$ \citep{Newman:2018aa,Ji:2024ab}, found from studies of stellar kinematics, which show slow rotation. This population is distinct from quiescent galaxies which appear to be still be sustaining rapid rotation \citep{DEugenio:2024ab}. Some of the galaxies we present here could be progenitors of these slow rotators, without having to undergo much dynamical transformation. However, it is also likely that many of them will evolve around the main sequence, build up their mass, and build rotational support through accretion of cold gas, becoming the star-forming disks we observe at cosmic noon. Because of the high merger rate at these redshifts, many of these galaxies will undergo numerous major mergers before being able to settle into disks.

Overall, the systems we study in this work are turbulent galaxies in their early stages of formation. Although disk settling is promoted through the smooth accretion of gas with conserved angular momentum, the low masses, high gas fractions, and high merger rates characteristic of the high redshift Universe ($z>4$) favour bursty and clumpy star formation. Simulations predict that bursty star formation in the early universe hinders disk formation, and the transition from bursty to smooth star formation coincides well with the transition from dispersion-dominated systems to disks \citep{Hopkins:2023aa, Semenov:2024aa}.

\subsection{Caveats}

\subsubsection{Modelling and observational biases} \label{sec:geko-biases}

In the development of \textsc{geko}, we have performed recovery tests to understand how robust the constraints are that can be placed on the kinematics in different scenarios, mainly focusing on the orientation of the galaxy with respect to the dispersion direction (position angle), the intrinsic kinematics of the galaxy, and the S/N. The key results from these tests are summarized in the Appendix \ref{app:geko-tests}, but the full description of the code and the tests will be done in Danhaive et al. (in prep). Specifically, Figs. \ref{fig:geko-test-kin} and \ref{fig:fig:geko-test-sn} highlight the uncertainties in the velocity dispersion measurement. It is important to note that grism observations inherently provides a weaker constraints on kinematics than IFU observations, which have been used for measurements in the local Universe \citep[e.g.,][]{Law:2022aa}, at cosmic noon \citep[e.g.,][]{Genzel:2011aa,Wisnioski:2015vx, Stott:2016aa,Forster-Schreiber:2018aa}, and out to $z\approx 11$ with the JWST/NIRSpec IFU \citep[e.g.,][]{Marconcini:2024aa,Scholtz:2024aa,Venturi:2024aa}.  This is firstly due to the line spread function, which in the F444W band is $R\sim 1600$, meaning a kinematic resolution of about $\Delta v \sim 80$ km/s. Even though we are able to constrain the velocity dispersion below $\Delta v$ in certain favoured scenarios (e.g., high S/N, see Fig. \ref{fig:geko-test-kin}), we still suffer from large uncertainties, especially at the low-$\disp$\ end, where we can often only place upper limits. Finally, the LSF is calibrated from JWST Cycle 0, 1, and 2 observations, and has an average uncertainty of 10\%. We do not marginalize over this uncertainty in the inference framework, which could boost errors by a factor of up to 1.5 (for the lowest values of $\disp$). The boost is minimal for dispersions larger than the LSF. This is particularly important when the velocity dispersion is only measured as an upper limit, since increasing the errors would automatically increase the value of the inferred upper limit. 

Our choice of an arctangent velocity curve model instead of other common functions such as the exponential disk model \citep{Freeman:1970aa, Wisnioski:2011us, Genzel:2017aa} should not strongly affect our measurements since the grism data does not probe the large radii where these curves differ \citep[see also discussion in ]{Price:2020wf}. Furthermore, the nature of the grism data is the morphology-kinematics degeneracy along the dispersion direction, which is the entire motivation behind \textsc{geko}. This means that our constraints are inferred from modelling, which comes with its own assumptions, and are not ``directly'' measured. Instruments like IFSs benefit from two spatial axes, where grism only has one \footnote{If both grism R and C are available \citep[e.g. SAPPHIRES;][and see an example therein]{Sun:2025aa} or if observations are taken with varying PAs \citep[e.g. ALT;][]{Naidu:2024aa}, more spatial information can be obtained.} , which allows them to have better constraints on the spatially resolved kinematics without strong priors on the emission-line maps or dynamical models.

While it is important to keep all of this in mind, the grism remains a crucial tool for studying the statistical properties of high-redshift galaxies. The power of Bayesian inference is in fully tracking uncertainties using the posterior distributions, which allows us to be realistic about the constraints we are making and the conclusions we are drawing for them. 

There are additional caveats which we discuss in terms of the biases they could introduce. The first is related to spatial resolution, as it is difficult to distinguish tightly merging galaxies from disks, and the former are hence often misclassified as disks \citep{Simons:2019aa}. This is one of the reasons why strict criteria need to be applied to define disks. It is also important to note that most of the galaxies in our sample are relatively low mass $M_{*}<10^{10} \thinspace M_{\odot}$, and we could be sensitive to non-circular motions. Their low masses and high gas fractions make them much more sensitive to stellar feedback, and combined with their turbulence and low rotational support, it is possible we are probing outflows alongside rotational velocities. This could bias our sample to high $\rotsupp$ at low masses, especially since the low masses systems are typically smaller and have less resolution elements. In general, if the thin-ish disk assumption underlying our models does not hold, it becomes difficult to interpret our measurements, especially the correlations. However, the main conclusion of turbulent ionized gas kinematics at $z\gtrsim4$ would still hold, since high enough rotational support would dominate any contribution from these effects. Overall, caution must be exercised when trying to categorise such systems as disks, even if rotationally-supported, and this is why we do not draw conclusions about the fraction of disks but only that of rotationally-supported systems. This fraction could however be overestimated if the measured velocities do not come from rotation. 

In our kinematic modelling, we assume that the velocity dispersion is isotropic and radially constant. However, some spatially resolved studies at lower redshift find a correlation between the local $\disp$\ and the local $\Sigma_{\text{SFR}}$ \citep{Yu:2019aa, Law:2022aa}. In cases of clumpy star formation, which are predicted to be more common at higher redshift \citep{Tan:2024aa, Nakazato:2024aa}, the constant $\disp$\ assumption may not hold as well. However, other works report no correlation in systems with highly star-forming clumps \citep[e.g.,][]{Ubler:2019vg}. Finally, although many studies show central peaks of velocity dispersion, an intrinsically flat $\disp$ profile is often recovered when beam smearing effects are accounted for in the kinematic modelling \citep[e.g.,][]{De-Breuck:2014aa}. A centrally peaked $\disp$ could bias our measurements to higher values when compare to studies that measure the intrinsic $\disp$ in the outskirts \citep[e.g.,][]{Wisnioski:2015vx}

Furthermore, our calculation of the rotational velocity is based on inclination corrections to the observed velocity, for which we need to assume an intrinsic thickness of the disk (we assume $q_0 = 0.2$). If the value of $q_0$ is over (under)-estimated, the inferred dynamical masses could be up to $\sim 0.02$ dex higher (lower) than the fiducial ones. This does not affect our conclusions, but could help reconcile some of the unphysical measurements in Fig. \ref{eq:dyn-mass}. We also note that the inclination of the galaxy is expected to affect the velocity and velocity dispersion measurements. For $\disp$, this is because the line-of-sight dispersion we measure is a mixture of its radial $\sigma_R$ and vertical $\sigma_z$ component, and as the galaxy becomes inclined we go from measuring predominantly $\sigma_z$ in face-on systems to measuring predominantly dispersion $\sigma_R$ in edge-on systems. Because we expect $\sigma_z<\sigma_R$ \citep{van-der-Kruit:2011aa}, we could be biased towards higher dispersions at higher inclinations \citep{Leroy:2008aa,Genzel:2011aa}. We have investigated this and see no trend of $\disp$ with inclination $i$. This is probably due to the limited resolution of our data, as was reported in various studies at cosmic noon \citep{Wisnioski:2015vx, Ubler:2019vg}. Mock observations of simulations can help test these various assumptions \citep{Kohandel:2020aa,Ubler:2021aa}.

\subsubsection{Uncertainties in the stellar and dynamical masses}\label{sec:caveat-mass}

In order to estimate the mass content of galaxies, we often need to rely on scaling relations linking photometry and kinematics to the underlying baryons and dark matter, and make assumptions on the galaxy's shape and size. We will now discuss the choices made in this work and how they affect our results.
We infer stellar masses through SED modelling with \textsc{Prospector}, a method with many advantages but also many degeneracies. We attempt to break some of the degeneracies by fixing the redshift to the grism spectroscopic redshift, and by simultaneously fitting the photometry with the line fluxes of the available emission lines (see \ref{sec:modelling-SED}).  
Although we are not fitting spectra and only have access to some of the brighter emission lines, \textsc{Prospector} has been shown to successfully recover emission line fluxes, when compared to NIRCam grism data, even when fitting with photometry only (Simmonds et al., in prep.). Furthermore, our S/N cut implies selecting higher equivalent width galaxies, meaning galaxies whose young stellar population and nebular emission is dominating the SEDs \citep{Boyett:2024ab,Endsley:2024aa}. This can make it difficult to detect the underlying population of older stars, and hence to constrain the full SFH. This effect is called ``outshining'' and can cause an underestimate of the stellar masses, effectively only attributing the stellar mass to the more visible young stellar population \citep{Tacchella:2022tc,Papovich:2023aa,Tacchella:2023aa, Whitler:2023aa}. 

Moving to the dynamical masses, the circular velocities needed to compute them are inferred assuming a virialised rotating disk with an exponential light profile. For galaxies with larger S\'{e}rsic indices, this assumption can lead to biased results. Furthermore, for pressure-dominated systems, the choice of the pressure support term multiplying the $\disp$\ in Eq. \ref{eq:v_circ} can lead to over- or under-estimations of the circular velocity \citep[see][for detailed analysis]{Price:2022aa}. Also, any uncertainties (and caveats) from the kinematic modelling (Sec. \ref{sec:geko-biases}) are propagated to the dynamical masses through the circular velocity. Finally, due to the increased merger rates at high redshift, it is possible we are probing velocity offsets between two systems or velocity gradients induced by the gravitational interaction in a pre- or post-merger phase \citep[see][for an in-depth discussion]{de-Graaff:2024ab}.

 \subsubsection{The impact of dust on kinematics}\label{sec:dust}

An important caveat in this work is that we do not directly measure dust attenuation. Although the early Universe was initially believed to be essentially dust free, recent works have revealed a rapid build-up of dust with detections out to $z\sim 7$ with both JWST and ALMA \citep{Pallottini:2023aa,Witstok:2023aa,Sawant:2025aa}, although these galaxies are typically more massive than the galaxies probed in this work. Furthermore, numerous works have indicated that a significant fraction of the star formation activity in the early Universe is actually occurring in dusty galaxies \citep{Laporte:2017aa,Sugahara:2021aa,Inami:2022aa}, with \citet{Algera:2023aa} finding a $\sim 30 \%$ contribution at $z\sim 7$ and $\sim66\%$ contribution at $z\sim5$ \citep{Sun:2025ab}. Still, other works find results consistent with a low dust content at high redshift \citep{Sandles:2024aa}.

In our sample, we infer dust attenuation using SED modelling with \textsc{Prospector}. Because of the inherent degeneracies in SED fitting, this could bias the SFRs we infer. Our fits show that the amount of dust attenuation increases with stellar mass, and that the relation between S/N and stellar mass has a large scatter, meaning that there are numerous massive galaxies, which also have higher dust content, with low S/N \Ha\ detections. This can however also be caused by low SFRs. An example of such a galaxy is the well-known HST dark galaxy HDF850.1 \citep{Sun:2024aa,Herard-Demanche:2025aa}, where the H$\alpha$ from the main component is heavily (>99\%) dust-obscured.. Because we are studying the kinematics of ionised gas through \Ha, we may be missing galaxies on the massive end due to their high dust content, and this could bias our correlations. However, studies at lower redshift suffer from the same biases, so our comparisons should be self-consistent.

\section{Summary \& Conclusions }\label{sec:conclusions}


In this work, we focus on a sample of $272$ \Ha\ emitters in GOODS-S and GOODS-N at redshift $z\approx 3.9-6.5$ detected using JWST NIRCam slitless spectroscopy from the FRESCO and CONGRESS surveys. We select galaxies with high S/N (integrated S/N $>10$) and a large enough PA offset from the dispersion direction ($|\text{PA}_{\text{morph}}| < 75^{\circ}$) in order to study their ionised gas kinematics and the properties of their stellar populations. Our final sample probes two orders of magnitude in stellar mass ($M_{\star}\approx 10^{8-10} \thinspace M_{\odot}$) and star formation rate ($\text{SFR}_{10} \approx 0.3-100 \thinspace M_{\odot}/$yr), and representative of the galaxy population down to $M_{\star}\approx 10^{9} \thinspace M_{\odot}$.

We use the SED-inference tool \textsc{Prospector}, with the redshift and \Ha\ emission line fluxes obtained from the grism observations and photometry from JADES, to infer SFRs and stellar masses for our galaxies. We use a novel grism forward-modelling tool \textsc{geko} to infer the \Ha\ morphology and kinematics and measure rotational velocities, velocity dispersions, and dynamical masses. We use morphological priors from NIRCam imaging from the JADES survey to alleviate the morphology-kinematics degeneracy of the grism data. This method fully forward models the instrument, accounting for LSF, PSF, and pixelisation. 

We summarize our key results below:
\begin{itemize}
    \item The star-forming galaxies in our sample at $z\approx 3.9-6.5$ are characterized by a highly turbulent ISM, with median ionised gas velocity dispersions of $\disp\approx 100$ km/s, but also a large scatter with velocity dispersions ranging from $\disp=30-200$ km/s. These systems also have an average rotational support $\rotsupp \approx 2$, with $>50\%$ of the galaxies being dispersion dominated ($\rotsupp <1$). Comparing to measurements at lower redshifts ($z<4$), we find that the $\disp$ ($\rotsupp$) increases (decreases) with redshift, broadly consistent with predictions from the Toomre model for marginal disks and from cosmological simulations.
    \item We analyse the correlations between $\rotsupp$\ and $\disp$\ with $M_{\star}$, SFR, sSFR, and SFR surface density, and find that the strongest correlations are between $\disp$\ and SFR ($\rho = 0.40, p<0.001$) and $\disp$\ and SFR surface density ($\rho = 0.38, p=0.001$). This is consistent with gravitational instabilities and feedback from star formation which perturb the build-up of rotational support and inject energy into the ISM. Although it is difficult to distinguish between these two main sources of turbulence, we find no strong evidence for a dominant role of feedback-driven turbulence. The $\disp$-SFR correlation is in agreement with results from the literature at $z<4$, albeit with an offset towards higher $\disp$. In comparison to works at lower redshifts, we find that the relations show more scatter and weaker correlations on average, which we attribute to the large diversity of the galaxy population. 
    \item We find a weak, decreasing trend of $\rotsupp$\ with $\log M_{\star}$, which is likely driven by observational biases at the low-mass end. However, a flattening of the $\rotsupp$-$\log M_{\star}$ relation below $M_{\star} \sim 10^{10} \thinspace M_{\odot}$ is not in tension with results from studies at $z<4$ and is predicted by the marginal disk framework. 
    \item The fraction of rotationally-supported ($\rotsupp>1$) systems in our sample increases with cosmic time, from $f = 0.36\pm 0.06$ to $f = 0.41 \pm 0.06$ from $z\sim 5.5$ to $z\sim 4.5$ for galaxies with stellar masses $9<\log M_{\star} [\mathrm{M}_{\odot}] < 10$. This increase is consistent with trends from $z<4$, which suggest that stable cold disks only form below $z<1$, with a smooth increase of rotational support with cosmic time. 
\end{itemize}

Overall, our work describes a turbulent phase in the evolution of galaxies in the early Universe. This epoch constitutes the dawn of disks, where galaxies start to be sustained by rotation but will go through several episodes of instability, caused by internal (feedback, radial inflows, bursty star formation) and external effects (non-smooth accretion, mergers), before settling into the cold disks we observe in the local Universe.

Our work has demonstrated that grism data is a powerful tool for population analyses of galaxy kinematics at high redshifts. Using this dataset in harmony with a novel Bayesian inference forward-modelling tool, we were able to paint the first statistically significant picture of ionised gas kinematics at $z>4$ on a large sample of galaxies. It however remains important to complement our work and future grism kinematics works with high-redshift multi-wavelength interferometric and IFU observations from ALMA and NIRSpec IFU to study the complete 2D kinematics structure in more detail and place stronger constraints on the formation of stable disks in the early Universe.

\section*{Acknowledgements}

We thank Emily Wisnioski and Trevor Mendel for the insightful discussions. ALD thanks the University of Cambridge Harding Distinguished Postgraduate Scholars Programme and Technology Facilities Council (STFC) Center for Doctoral Training (CDT) in Data intensive science at the University of Cambridge (STFC grant number 2742605) for a PhD studentship. ALD and ST acknowledge support by the Royal Society Research Grant G125142. H\"U acknowledges funding by the European Union (ERC APEX, 101164796). Views and opinions expressed are however those of the authors only and do not necessarily reflect those of the European Union or the European Research Council Executive Agency. Neither the European Union nor the granting authority can be held responsible for them. EE, FS, ZJ, MR, and CNAW acknowledge support from the JWST/NIRCam contract to the University of Arizona NAS5-02015.  SA acknowledges grant PID2021-127718NB-I00 funded by the Spanish Ministry of Science and Innovation/State Agency of Research (MICIN/AEI/ 10.13039/501100011033). AJB acknowledges funding from the "FirstGalaxies" Advanced Grant from the European Research Council (ERC) under the European Union’s Horizon 2020 research and innovation programme (Grant agreement No. 789056). SC acknowledges support by European Union’s HE ERC Starting Grant No. 101040227 - WINGS. GCJ acknowledges support by the Science and Technology Facilities Council (STFC), ERC Advanced Grant 695671 "QUENCH". RM, CS, and JS acknowledge support by the Science and Technology Facilities Council (STFC),and by the ERC through Advanced Grant 695671 “QUENCH”. RM and CS are also supported by the UKRI Frontier Research grant RISEandFALL. RM also acknowledges funding from a research professorship from the Royal Society. WM thanks the Science and Technology Facilities Council (STFC) Center for Doctoral Training (CDT) in Data Intensive Science at the University of Cambridge (STFC grant number 2742968) for a PhD studentship. NCV acknowledges support from the Charles and Julia Henry Fund through the Henry Fellowship. WMB gratefully acknowledges support from DARK via the DARK fellowship. This work was supported by a research grant (VIL54489) from VILLUM FONDEN. DJE is supported as a Simons Investigator and by JWST/NIRCam contract to the University of Arizona, NAS5-02015. DP acknowledges support by the Huo Family Foundation through a P.C. Ho PhD Studentship. BER acknowledges support from the NIRCam Science Team contract to the University of Arizona, NAS5-02015, and JWST Program 3215. The research of CCW is supported by NOIRLab, which is managed by the Association of Universities for Research in Astronomy (AURA) under a cooperative agreement with the National Science Foundation. 

This work is based on observations made with the NASA/ESA Hubble Space Telescope and NASA/ESA/CSA James Webb Space Telescope. The data were obtained from the Mikulski Archive for Space Telescopes at the Space Telescope Science Institute, which is operated by the Association of Universities for Research in Astronomy, Inc., under NASA contract NAS 5-03127 for JWST. These observations are associated with program \#1180, 1181, 1210 (JADES), \#1895 (FRESCO), \# 1963 (JEMS) and \#3577 (CONGRESS).
Support for program \#3577 was provided by NASA through a grant from the Space Telescope Science Institute, which is operated by the Association of Universities for Research in Astronomy, Inc., under NASA contract NAS 5-03127. The authors acknowledge the FRESCO team for developing their observing program with a zero-exclusive-access period. The authors acknowledge use of the lux supercomputer at UC Santa Cruz, funded by NSF MRI grant AST 1828315.






\bibliographystyle{mnras}
\bibliography{main} 




\appendix

\section*{Affiliations}
\noindent
{\it $^{1}$Kavli Institute for Cosmology, University of Cambridge, Madingley Road, Cambridge, CB3 0HA, UK \\
$^{2}$Cavendish Laboratory, University of Cambridge, 19 JJ Thomson Avenue, Cambridge, CB3 0HE, UK \\
$^{3}$Max-Planck-Institut f\"ur extraterrestrische Physik (MPE), Gie{\ss}enbachstra{\ss}e 1, 85748 Garching, Germany \\
$^{4}$Max-Planck-Institut f\"ur Astronomie, K\"onigstuhl 17, D-69117, Heidelberg, Germany \\
$^{5}$Steward Observatory, University of Arizona, 933 N. Cherry Avenue, Tucson, AZ 85721, USA \\
$^{6}$Center for Astrophysics $|$ Harvard \& Smithsonian, 60 Garden St., Cambridge MA 02138 USA \\
$^{7}$Centro de Astrobiolog\'ia (CAB), CSIC–INTA, Cra. de Ajalvir Km.~4, 28850- Torrej\'on de Ardoz, Madrid, Spain \\
$^{8}$Department of Physics, University of Oxford, Denys Wilkinson Building, Keble Road, Oxford OX1 3RH, UK \\
$^{9}$Scuola Normale Superiore, Piazza dei Cavalieri 7, I-56126 Pisa, Italy \\
$^{10}$The University of Texas at Austin, Department of Astronomy, 2515 Speedway, Stop C1400, Austin, TX 78712-1205, USA \\
$^{11}$DARK, Niels Bohr Institute, University of Copenhagen, Jagtvej 128, DK-2200 Copenhagen, Denmark \\
$^{12}$Department of Astronomy, Tsinghua University, Beijing 100084, China
$^{13}$Department of Astronomy and Astrophysics, University of California, Santa Cruz, 1156 High Street, Santa Cruz, CA 95064, USA \\
$^{14}$NSF National Optical-Infrared Astronomy Research Laboratory, 950 North Cherry Avenue, Tucson, AZ 85719, USA} 

\section{Sample statistics}\label{app:tables}
In this section, we present some summary tables for the key measurements of this work, as well as provide further insight into the physical distribution of our various sub-samples. 

In Tabs. \ref{tab:medians-ext} and \ref{tab:medians-gold}, we summarize the median values, for our extended and gold samples, of $\disp$ and $\rotsupp$ in bins of redshift (Figs. \ref{fig:vsigma_sigma_zevol}, \ref{fig:sigma0-z-comp}, and \ref{fig:vsigma_z_evol}), and stellar mass, SFR, sSFR, and SFR surface density (Figs. \ref{fig:vsigma_sigma_mstar_sfr} and \ref{fig:corr-comp}). For each bin, we specify the range probed by the bin, the median value of the binned quantity, the number of galaxies $N$ contained in the bin, and the average stellar mass $<\log (M_{\star} \rm [M_{\odot}])>$. For completeness, we also include the average value of the rotational support when computed using the maximum observed velocity $v_{\rm obs,max}$ instead of the velocity at the effective radius $v_{\rm re}$. We find consistent values, within the uncertainties, for the two measures of rotational support/ However we note that  $v_{\rm obs,max}/\disp$ is on average higher than $v_{\rm re}/\disp$.

\begin{table*}
    \centering
    \begin{tabular}{c|c|c|c|c|c|c}
      Bin & Bin median & N & $<\log (M_{\star} \rm [M_{\odot}])>$ & $<v/\sigma_0>$ & $< v_{\rm obs,max}/\sigma_0>$ & $<\sigma_0 \rm ~ [km/s]>$ \\ \hline
        $3.8<z<4.2$ & 4.03 & 32 & 9.43 & $1.2^{+1.1}_{-0.7}$ & $1.2^{+1.2}_{-0.7}$ & $77^{+55}_{-36}$\\
        $4.2<z<5.0$ & 4.55  & 32& 9.07 & $1.0^{+1.5}_{-0.5}$ & $0.7^{+1.2}_{-0.3}$ & $68^{+44}_{23}$ \\
        $5.0<z<6.1$ & 5.21 & 16 & 9.46 &  $0.8^{+1.2}_{-0.2}$ & $0.9^{+1.3}_{-0.3}$ & $98^{+40}_{-30}$\\ \hline

        $7.5<\log (M_{\star} ~[\rm M_{\odot}])<9.0$  & 8.43 & 22 & 8.41 & $0.6^{+1.3}_{-0.2}$&$0.7^{+1.3}_{-0.3}$ & $67^{+52}_{-20}$\\ 
        $9.0<\log (M_{\star} ~[\rm M_{\odot}])<9.5$ & 9.36 & 28 & 9.32 &$1.1^{+1.1}_{-0.6}$ & $1.1^{+1.1}_{-0.6}$& $74^{+50}_{-33}$\\ 
        $9.5<\log (M_{\star} ~[\rm M_{\odot}])<10.5$ &9.8  & 30 & 9.9 &$0.8^{+1.7}_{-0.4}$ & $0.8^{+1.7}_{-0.4}$ & $93^{+55}_{-52}$\\ \hline

        $0<\log \rm (SFR_{10} ~[M_{\odot}/yr])<0.8$ & 0.61& 35 & 8.82& $0.7^{+1.1}_{-0.3}$& $0.7^{+1.1}_{-0.3}$& $68^{+50}_{-21}$\\ 
        $0.8<\log \rm (SFR_{10} ~[M_{\odot}/yr])<1.3$& 1.02 & 33 & 9.47 &$0.9^{+1.1}_{-0.5}$ & $0.9^{+1.0}_{-0.6}$ & $84^{+48}_{-33}$\\ 
        $1.3<\log \rm (SFR_{10} ~[M_{\odot}/yr])<2.5$ & 1.62 & 12 & 10.16&$1.0^{+3.0}_{-0.5}$ &  $1.0^{+3.0}_{-0.5}$ &$107^{+79}_{-67}$\\ \hline

        $0<\log \rm (sSFR_{10} ~[M_{\odot}/yr])<0.5$ & 0.29 & 27 & 9.61& $0.7^{+0.9}_{-0.3}$& $0.7^{+0.9}_{-0.3}$& $77^{+40}_{-39}$\\ 
        $0.5<\log \rm (sSFR_{10} ~[M_{\odot}/yr])<1.0$ & 0.68 & 31 & 9.36 &$1.2^{+1.2}_{-0.6}$ & $1.2^{+1.0}_{-0.6}$ & $79^{+59}_{-39}$\\ 
        $1.0<\log \rm (sSFR_{10} ~[M_{\odot}/yr])<2.0$ & 1.30& 16 & 8.34&$0.7^{+2.0}_{-0.3}$ & $0.7^{+2.1}_{-0.3}$ & $67^{+39}_{-18}$\\ \hline

        $-1<\log \rm (\Sigma_{SFR_{10}} ~[M_{\odot}/yr/kpc^2])<-0.2$ & -0.40 & 32 & 8.90 & $0.7^{+1.1}_{-0.3}$& $0.7^{+1.1}_{-0.3}$ &  $69^{+43}_{-31}$\\ 
        $-0.2<\log \rm (\Sigma_{SFR_{10}} ~[M_{\odot}/yr/kpc^2])<0.2$ & 0.01 & 37 & 9.48 &$1.1^{+1.5}_{-0.7}$ & $1.1^{+1.5}_{-0.7}$ & $80^{+58}_{-30}$\\ 
        $0.2<\log \rm (\Sigma_{SFR_{10}} ~[M_{\odot}/yr/kpc^2])<1.0$ & 0.33 & 11 & 9.77 &$0.9^{+1.6}_{-0.4}$ & $0.9^{+1.7}_{-0.4}$ & $100^{+71}_{-73}$\\
        
    \end{tabular}
    \caption{Summary table of the medians for the extended galaxies in the gold and silver sample, for bins of redshift (Figs. \ref{fig:vsigma_sigma_zevol}, \ref{fig:sigma0-z-comp}, and \ref{fig:vsigma_z_evol}), and stellar mass, SFR, sSFR, and SFR surface density (Figs. \ref{fig:vsigma_sigma_mstar_sfr} and \ref{fig:corr-comp}). For each bin, we specify the range probed by the bin, the median value of the binned quantity, the number of galaxies $N$ contained in the bin, and the average stellar mass $<\log (M_{\star} \rm [M_{\odot}])>$. We also include the average value of the rotational support when computed using the maximum observed velocity $v_{\rm obs,max}$ instead of the velocity at the effective radius $v_{\rm re}$.}
    \label{tab:medians-ext}
\end{table*}

\begin{table*}
    \centering
    \begin{tabular}{c|c|c|c|c|c|c}
      Bin & Mean value &  N & $<\log (M_{\star} \rm [M_{\odot}])>$ & $<v_{\rm re}/\sigma_0>$ & $<v_{\rm obs,max}/\sigma_0>$ & $<\sigma_0 \rm ~ [km/s]>$  \\ \hline
        $3.8<z<4.2$ & 4.04 & 21 & 9.46 & $1.1^{+0.8}_{-0.6}$ & $1.1^{+0.8}_{-0.6}$& $74^{+60}_{-41}$\\
        $4.2<z<5.0$ & 4.41 & 12& 9.14 & $1.0^{+1.5}_{-0.5}$ & $1.0^{+1.5}_{-0.5}$ & $52^{+35}_{-22}$\\
        $5.0<z<6.1$ & 5.29 & 8 & 9.69 & $0.8^{+1.1}_{-0.2}$ & $0.8^{+1.2}_{-0.2}$& $103^{+33}_{-35}$\\ \hline
        
        $7.5<\log (M_{\star} ~[\rm M_{\odot}])<9.0$ & 8.49 & 22 & 8.51& $1.2^{+0.8}_{-0.7}$& $0.7^{+1.3}_{-0.3}$ & $56^{+17}_{-11}$\\ 
        $9.0<\log (M_{\star} ~[\rm M_{\odot}])<9.5$ & 9.37  & 28 & 9.32&$1.5^{+0.9}_{-0.8}$ & $1.1^{+1.1}_{-0.6}$& $66^{+52}_{-28}$\\ 
        $9.5<\log (M_{\star} ~[\rm M_{\odot}])<10.5$ & 9.85 & 30 & 9.93&$0.8^{+1.1}_{-0.8}$ & $0.8^{+1.7}_{-0.4}$ & $93^{+55}_{-61}$\\ \hline

        $0<\log \rm (SFR_{10} ~[M_{\odot}/yr])<0.8$ & 0.61 & 35 & 8.60& $0.7^{+2.0}_{-0.2}$&$1.1^{+0.8}_{-0.6}$  & $49^{+12}_{-8}$\\ 
        $0.8<\log \rm (SFR_{10} ~[M_{\odot}/yr])<1.3$ & 0.95 & 33 & 9.33&$1.2^{+0.7}_{-0.6}$ & $1.2^{+0.7}_{-0.7}$ &$68^{+49}_{-36}$\\ 
        $1.3<\log \rm (SFR_{10} ~[M_{\odot}/yr])<2.5$ & 1.59 & 12 & 10.01&$0.8^{+1.0}_{-0.3}$ & $0.8^{+1.0}_{-0.4}$& $112^{+59}_{-72}$\\ \hline

        $0<\log \rm (sSFR_{10} ~[M_{\odot}/yr])<0.5$ & 0.32 & 27 & 9.69& $0.7^{+0.8}_{-0.1}$& $0.7^{+0.8}_{-0.3}$ & $56^{+54}_{-26}$\\ 
        $0.5<\log \rm (sSFR_{10} ~[M_{\odot}/yr])<1.0$ & 0.65 & 31 & 9.49&$1.3^{+1.1}_{-0.7}$ & $1.3^{+1.1}_{-0.7}$ &$92^{+46}_{-51}$\\ 
        $1.0<\log \rm (sSFR_{10} ~[M_{\odot}/yr])<2.0$ & 1.37 & 16 & 8.41&$0.7^{+1.2}_{-0.2}$ & $0.7^{+1.3}_{-0.2}$ & $57^{+11}_{-8}$\\ \hline

        $-1<\log \rm (\Sigma_{SFR_{10}} ~[M_{\odot}/yr/kpc^2])<-0.2$ & -0.34 & 32 &9.03 & $1.0^{+1.5}_{-0.5}$& $1.0^{+1.4}_{-0.5}$ & $44^{+31}_{-12}$\\ 
        $-0.2<\log \rm (\Sigma_{SFR_{10}} ~[M_{\odot}/yr/kpc^2])<0.2$ & 0.06 & 37 & 9.45&$1.2^{+0.7}_{-0.6}$ & $1.2^{+0.7}_{-0.6}$ & $73^{+60}_{-22}$\\ 
        $0.2<\log \rm (\Sigma_{SFR_{10}} ~[M_{\odot}/yr/kpc^2])<1.0$ & 0.30 & 11 & 9.79&$0.8^{+0.8}_{-0.4}$ & $0.8^{+0.8}_{-0.4}$ & $93^{+78}_{-66}$\\
        
    \end{tabular}
    \caption{Summary table of the medians for the gold sample, following the same format as Tab. \ref{tab:medians-ext}.}
    \label{tab:medians-gold}
\end{table*}

In Tab. \ref{tab:results-gold}, we summarize the physical properties of the galaxies in the gold sample, along with their best-fit kinematic measurements.

\begin{table*}
    \centering
    \begin{tabular}{c|c|c|c|c|c|c|c}
         JADES ID & $z$ & $\log (M_{\star} \rm [M_{\odot}])$ & $\rm \log (SFR ~ [M_{\odot}/yr])$  & $r_{\rm e}$ [kpc] &  $v/\sigma_0$ & $\sigma_0$ [km/s] & $\log (M_{\rm dyn} \rm [M_{\odot}])$\\ \hline
1028887 &     5.19 & $9.56^{+0.19}_{-0.19}$  & $1.15^{+0.22}_{-0.17}$ & $1.12^{+0.11}_{-0.10}$ & $0.3^{+0.3}_{-0.2}$ & $68^{+12}_{-14}$  & $9.88^{+0.13}_{-0.17}$ \\
 1002030 &     5.18 & $9.48^{+0.18}_{-0.24}$  & $0.93^{+0.17}_{-0.11}$ & $1.37^{+0.19}_{-0.18}$ & $0.6^{+0.4}_{-0.3}$ & $122^{+34}_{-35}$ & $10.5^{+0.17}_{-0.21}$ \\
  191250 &     5.39 & $10.17^{+0.30}_{-0.09}$ & $1.69^{+0.24}_{-0.17}$ & $2.45^{+0.06}_{-0.06}$ & $13.9^{+5.0}_{-5.0}$ & $<39$  & $10.8^{+0.07}_{-0.08}$ \\
 1002222 &     5.29 & $9.47^{+0.18}_{-0.44}$  & $0.93^{+0.32}_{-0.38}$ & $0.98^{+0.17}_{-0.13}$ & $0.5^{+0.4}_{-0.3}$ & $73^{+9}_{-11}$   & $9.91^{+0.10}_{-0.12}$ \\
  214966 &     5.54 & $9.54^{+0.16}_{-0.21}$  & $1.21^{+0.19}_{-0.17}$ & $1.41^{+0.10}_{-0.10}$ & $0.5^{+0.3}_{-0.3}$  & $108^{+25}_{-30}$ & $10.4^{+0.15}_{-0.22}$ \\
 1025527 &     5.3  & $9.39^{+0.16}_{-0.56}$  & $1.02^{+0.13}_{-0.13}$ & $1.06^{+0.18}_{-0.14}$ & $1.9^{+1.1}_{-0.6}$ & $118^{+25}_{-35}$ & $10.6^{+0.24}_{-0.23}$ \\
 1001674 &     5.18 & $9.92^{+0.17}_{-0.13}$  & $1.60^{+0.20}_{-0.30}$ & $1.60^{+0.18}_{-0.17}$ & $0.5^{+0.4}_{-0.4}$ & $199^{+36}_{-33}$ & $11.0^{+0.15}_{-0.15}$ \\
  201125 &     5.82 & $9.96^{+0.10}_{-0.16}$  & $1.71^{+0.13}_{-0.18}$ & $1.86^{+0.16}_{-0.15}$ & $0.8^{+0.3}_{-0.3}$ & $136^{+15}_{-14}$ & $10.8^{+0.09}_{-0.09}$ \\
 1094616 &     4.06 & $9.40^{+0.16}_{-0.36}$  & $1.04^{+0.32}_{-0.17}$ & $1.35^{+0.12}_{-0.10}$ & $1.3^{+0.2}_{-0.2}$  & $187^{+9}_{-8}$   & $11.0^{+0.06}_{-0.06}$ \\
 1000989 &     3.97 & $9.07^{+0.11}_{-0.24}$  & $0.58^{+0.23}_{-0.12}$ & $0.87^{+0.07}_{-0.06}$ & $5.4^{+5.0}_{-2.6}$  & $<39$  & $9.76^{+0.14}_{-0.16}$ \\
 1088814 &   4.41 & $10.60^{+0.13}_{-0.36}$ & $1.59^{+0.49}_{-0.35}$ & $2.39^{+0.08}_{-0.08}$ & $1.5^{+1.2}_{-0.4}$  & $87^{+27}_{-36}$  & $10.6^{+0.14}_{-0.21}$ \\
 1087148 &     4.38 & $8.24^{+0.28}_{-0.33}$  & $0.61^{+0.14}_{-0.11}$ & $1.05^{+0.24}_{-0.19}$ & $2.4^{+3.3}_{-1.6}$  & $<63$  & $9.82^{+0.72}_{-0.62}$ \\
 1015956 &     4.02 & $9.54^{+0.12}_{-0.35}$  & $0.84^{+0.20}_{-0.20}$ & $1.40^{+0.15}_{-0.14}$ & $1.7^{+4.6}_{-1.3}$  & $<32$  & $9.08^{+0.43}_{-0.64}$ \\
 1085659 &     4.06 & $10.17^{+0.13}_{-0.42}$ & $1.27^{+0.41}_{-0.20}$ & $1.47^{+0.11}_{-0.10}$ & $0.3^{+0.4}_{-0.2}$  & $51^{+16}_{-21}$  & $9.76^{+0.22}_{-0.41}$ \\
 1083165 &     4.15 & $9.78^{+0.09}_{-0.13}$  & $0.96^{+0.22}_{-0.16}$ & $1.12^{+0.07}_{-0.07}$ & $0.6^{+0.2}_{-0.2}$  & $111^{+12}_{-13}$ & $10.3^{+0.09}_{-0.01}$ \\
 1065488 &     4.15 & $8.01^{+0.42}_{-0.15}$  & $0.75^{+0.08}_{-0.08}$ & $1.10^{+0.14}_{-0.14}$ & $2.1^{+1.5}_{-0.6}$  & $<68$  & $9.94^{+0.20}_{-0.26}$ \\
 1025101 &     4.91 & $8.99^{+0.29}_{-0.20}$  & $0.95^{+0.17}_{-0.14}$ & $1.19^{+0.16}_{-0.13}$ & $1.9^{+1.3}_{-0.4}$  & $142^{+21}_{-85}$ & $10.8^{+0.20}_{-0.45}$ \\
 1090526 &     3.96 & $9.93^{+0.12}_{-0.24}$  & $1.08^{+0.38}_{-0.23}$ & $2.47^{+0.13}_{-0.13}$ & $1.4^{+3.1}_{-0.9}$  & $<86$  & $10.1^{+0.37}_{-0.39}$ \\
 1077545 &     4.42 & $9.27^{+0.11}_{-0.36}$  & $0.96^{+0.11}_{-0.09}$ & $0.87^{+0.09}_{-0.09}$ & $1.1^{+1.4}_{-0.7}$  & $63^{+24}_{-33}$  & $9.86^{+0.22}_{-0.34}$ \\
 1089583 &     3.93 & $9.01^{+0.16}_{-0.35}$  & $0.72^{+0.29}_{-0.12}$ & $1.16^{+0.15}_{-0.13}$ & $2.2^{+0.8}_{-0.5}$  & $73^{+20}_{-28}$  & $10.3^{+0.19}_{-0.28}$ \\
 1077261 &     3.92 & $9.36^{+0.24}_{-0.20}$  & $0.91^{+0.16}_{-0.20}$ & $1.12^{+0.19}_{-0.16}$ & $0.5^{+0.5}_{-0.3}$  & $77^{+22}_{-28}$  & $10.0^{+0.17}_{-0.31}$ \\
 1095186 &     4.02 & $9.37^{+0.10}_{-0.22}$  & $0.75^{+0.17}_{-0.12}$ & $1.92^{+0.30}_{-0.27}$ & $1.1^{+0.8}_{-0.6}$  & $76^{+14}_{-17}$  & $10.3^{+0.22}_{-0.20}$ \\
 1091580 &     4.8  & $8.09^{+0.31}_{-0.29}$  & $0.61^{+0.11}_{-0.09}$ & $0.88^{+0.18}_{-0.16}$ & $1.0^{+3.4}_{-0.8}$  & $<56$  & $9.45^{+0.32}_{-0.53}$ \\
 1086992 &     4.04 & $9.37^{+0.23}_{-0.46}$  & $1.12^{+0.26}_{-0.27}$ & $3.42^{+0.20}_{-0.20}$ & $4.7^{+5.0}_{-2.9}$  & $<28$   & $9.84^{+0.34}_{-0.36}$ \\
 1090054 &     4.82 & $9.62^{+0.04}_{-0.06}$  & $1.07^{+0.06}_{-0.04}$ & $1.27^{+0.06}_{-0.06}$ & $3.4^{+0.5}_{-0.4}$  & $68^{+6}_{-7}$    & $10.6^{+0.06}_{-0.06}$ \\
 1014130 &     4.38 & $8.49^{+0.22}_{-0.31}$  & $0.67^{+0.08}_{-0.09}$ & $1.32^{+0.22}_{-0.20}$ & $1.9^{+3.9}_{-1.4}$  & $<61$  & $9.78^{+0.47}_{-0.05}$ \\
 1082948 &     3.91 & $10.68^{+0.08}_{-0.20}$ & $1.96^{+0.38}_{-0.44}$ & $2.25^{+0.13}_{-0.12}$ & $1.4^{+4.4}_{-1.0}$  & $<22$   & $8.87^{+0.43}_{-0.60}$ \\
 1000110 &     4.06 & $9.52^{+0.34}_{-0.08}$  & $1.30^{+0.15}_{-0.11}$ & $1.66^{+0.09}_{-0.08}$ & $1.5^{+0.2}_{-0.2}$  & $133^{+11}_{-11}$ & $10.8^{0.05}_{-0.05}$ \\
 1091153 &     4.07 & $10.37^{+0.14}_{-0.14}$ & $1.99^{+0.24}_{-0.23}$ & $1.85^{+0.09}_{-0.10}$ & $1.4^{+2.0}_{-0.5}$  & $<98$  & $10.3^{+0.25}_{0.34}$ \\
 1094903 &     3.87 & $10.34^{+0.08}_{-0.23}$ & $1.32^{+0.20}_{-0.20}$ & $2.06^{+0.11}_{-0.10}$ & $0.5^{+0.2}_{-0.2}$  & $148^{+15}_{-16}$ & $10.8^{+0.08}_{-0.09}$ \\
 1079264 &     4.05 & $8.73^{+0.56}_{-0.30}$  & $1.17^{+0.26}_{-0.08}$ & $2.24^{+0.17}_{-0.15}$ & $1.6^{+4.2}_{-1.2}$  & $<32$  & $9.24^{+0.44}_{-0.65}$ \\

    \end{tabular}
    \caption{Kinematic measurements and star formation properties for the galaxies in the gold sample.}
    \label{tab:results-gold}
\end{table*}

\begin{table*}
\ContinuedFloat
    \centering
    \begin{tabular}{c|c|c|c|c|c|c|c}
         JADES ID & $z$ & $\log (M_{\star} \rm [M_{\odot}])$ & $\rm \log (SFR ~ [M_{\odot}/yr])$ & $r_{\rm e}$ [kpc] &  $v/\sigma_0$ & $\sigma_0$ [km/s] & $\log (M_{\rm dyn} \rm [M_{\odot}])$\\
\hline
 1086406 &     4.41 & $9.46^{+0.18}_{-0.32}$  & $1.00^{+0.27}_{-0.25}$ & $0.91^{+0.14}_{-0.12}$ & $6.3^{+5.0}_{-3.5}$  & $<27$   & $9.48^{+0.28}_{-0.34}$ \\
 1013488 &     3.8  & $9.57^{+0.11}_{-0.33}$  & $1.19^{+0.25}_{-0.11}$ & $1.07^{+0.07}_{-0.07}$ & $1.4^{+0.2}_{-0.2}$  & $113^{+10}_{-13}$ & $10.5^{+0.09}_{-0.10}$ \\
 1028072 &     4.4  & $9.17^{+0.08}_{-0.14}$  & $0.62^{+0.12}_{-0.09}$ & $1.66^{+0.23}_{-0.20}$ & $3.7^{+5.0}_{-2.5}$  & $<39$  & $9.76^{+0.37}_{-0.50}$ \\
 1090742 &     4.62 & $9.37^{+0.16}_{-0.26}$  & $0.72^{+0.27}_{-0.30}$ & $1.34^{+0.16}_{-0.14}$ & $3.7^{+5.0}_{-2.3}$  & $<37$  & $9.61^{+0.33}_{-0.46}$ \\
 1091236 &     4.17 & $9.34^{+0.15}_{-0.50}$  & $0.91^{+0.17}_{-0.14}$ & $0.84^{+0.09}_{-0.08}$ & $0.9^{+0.3}_{-0.3}$  & $84^{+12}_{-14}$  & $10.0^{+0.09}_{-0.11}$ \\
 1009935 &     4.89 & $9.65^{+0.11}_{-0.25}$  & $0.88^{+0.20}_{-0.24}$ & $1.02^{+0.12}_{-0.11}$ & $2.1^{+5.0}_{-1.5}$  & $<29$   & $8.99^{+0.42}_{-0.59}$ \\
 1085494 &     4.06 & $8.90^{+0.39}_{-0.15}$  & $1.04^{+0.15}_{-0.26}$ & $1.06^{+0.10}_{-0.10}$ & $1.6^{+0.6}_{-0.4}$  & $53^{+10}_{-13}$  & $9.85^{+0.13}_{-0.15}$ \\
 1090891 &     3.92 & $9.74^{+0.44}_{-0.22}$  & $1.45^{+0.17}_{-0.11}$ & $1.55^{+0.07}_{-0.07}$ & $1.1^{+0.1}_{-0.1}$  & $172^{+8}_{-9}$   & $10.9^{+0.05}_{-0.05}$ \\
 1008197 &     4.4  & $8.76^{+0.21}_{-0.24}$  & $0.55^{+0.15}_{-0.12}$ & $1.11^{+0.19}_{-0.16}$ & $3.9^{+5.0}_{-2.4}$  & $<44$  & $9.77^{+0.35}_{-0.47}$ \\
 1029814 &     4.16 & $8.38^{+0.18}_{-0.17}$  & $0.70^{+0.06}_{-0.06}$ & $1.28^{+0.26}_{-0.21}$ & $3.2^{+2.1}_{-1.1}$  & $48^{+12}_{-18}$  & $10.2^{+0.32}_{-0.33}$ \\
    \end{tabular}
    \caption{Continued.}
\end{table*}

\begin{figure}
    \centering
    \includegraphics[width=1\linewidth]{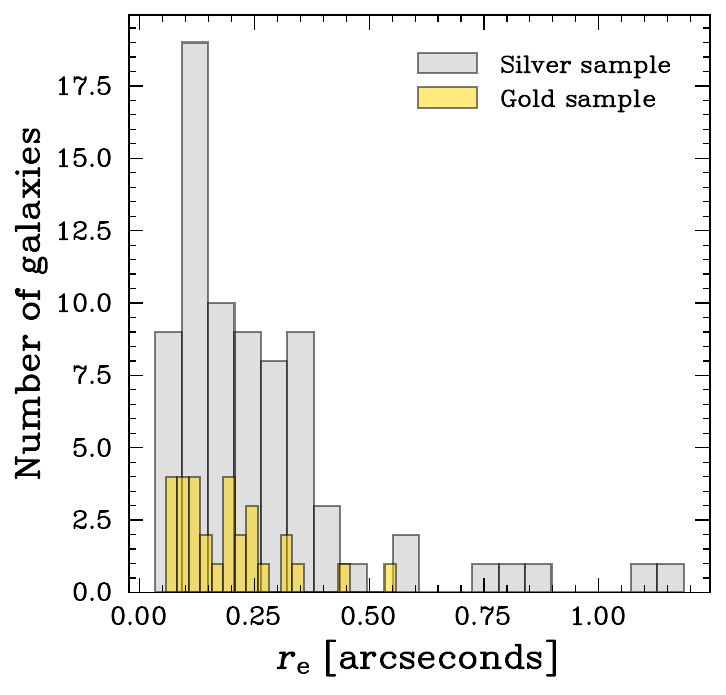}
    \caption{Effective radii $r_{\rm e}$, in arcseconds, of the galaxies in our gold and silver sample.}
    \label{fig:r_eff_hist}
\end{figure}

In order to study possible biases in the sizes probed by galaxies in our sub-samples, we plot the distribution of the best-fit \Ha\ effective radius $r \rm _{e}$ in arcseconds, for the gold and samples, on Fig. \ref{fig:r_eff_hist}. We find that the two sample follow similar distributions, with the silver sample probing some larger systems at the tail end of the distribution. The cuts made to obtain the gold sample to not bias it in size compared to the silver one. 

\begin{figure}
    \centering
    \includegraphics[width=1\linewidth]{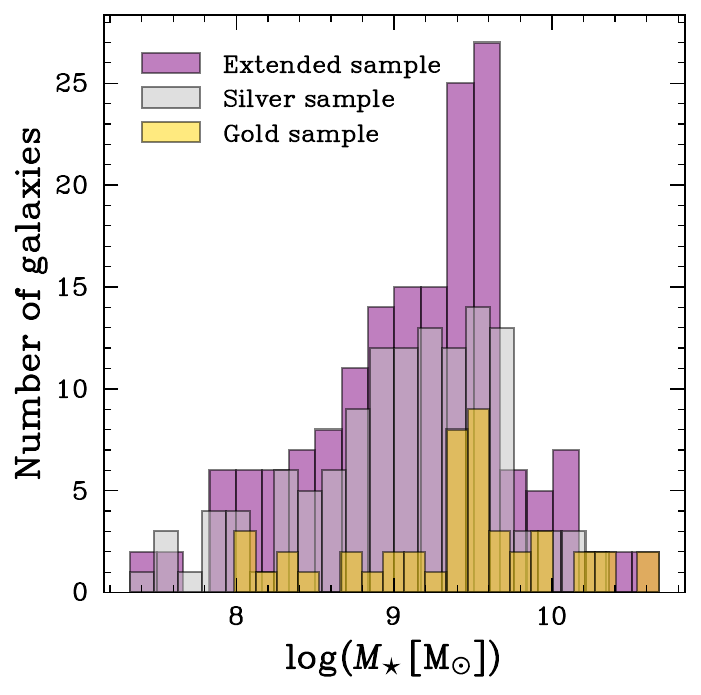}
    \caption{Stellar mass distribution of the galaxies in our gold, silver sample, and extended samples.}
    \label{fig:logmstar_hist}
\end{figure}

We further investigate the stellar mass distribution of galaxies in the gold, silver, and extended samples on Fig. \ref{fig:logmstar_hist}. The gold and extended samples, which we use to compute our medians and correlations, have similar distributions, with the resolved sample having a slightly longer tail on the low mass end. 

\section{Impact of selection cuts on inferred correlations} \label{app:method-comparison}

We will now discuss how the correlations presented in Secs. \ref{sec:corr-z6} and \ref{sec:disp-drivers} change if we only select rotation-supported galaxies and if we compute the rotational support $\rotsupp$ with $v_{\rm obs,max}$ instead of the velocity at the effective radius $v_{\rm re}$. The results are summarized on Tab. \ref{tab:corr-metho-comp}, for correlations between $\disp$ and $\rotsupp$ and $\log (M_{\star} ~[\rm M_{\odot}])$, SFR, sSFR, and SFR surface density. We find that the quantitative results do not change between scenarios, but the strengh ($\rho$) and significance ($p$) of the correlations change. For the $\rotsupp>1$ case, the correlations typically get weaker, but this is mostly driven by the significant decrease in sample size. For the $v_{\rm rot} = v_{\rm obs,max}$ case, the coefficients $\rho$ stay consistent within $10\%$. 

\renewcommand{\arraystretch}{1.5} 
\setlength{\tabcolsep}{3pt} 

\begin{table*}
    \centering
    \begin{tabular}{c|c|c|c|c|c}
        \hline
\textbf{Sample} & $\rotsupp$\ (full) & $\rotsupp$\ ($\rotsupp$$>1$) & $\rotsupp$\ ($v_{\rm rot} = v_{\rm obs,max}$) &  $\disp$\  (full)  & $\disp$\ ($\rotsupp$$>1$)\\
        \hline
        $\log (M_{\star} ~[\rm M_{\odot}])$ & $\begin{array}{c} \rho = -0.27 \\ p = 0.015 \end{array}$ & $\begin{array}{c} \rho = -0.38 \\ p = 0.019 \end{array}$ & $\begin{array}{c} \rho = -0.23 \\ p = 0.044 \end{array}$ &  $\begin{array}{c} \rho = 0.24 \\ p = 0.036 \end{array}$ & $\begin{array}{c} \rho = 0.28 \\ p = 0.094 \end{array}$ \\
        \hline
       $\text{SFR}_{10}$& $\begin{array}{c} \rho = -0.30 \\ p = 0.007 \end{array}$ & $\begin{array}{c} \rho = -0.08 \\ p = 0.653 \end{array}$ & $\begin{array}{c} \rho = -0.27 \\ p = 0.016 \end{array}$ & $\begin{array}{c} \rho = 0.40 \\ p = 0.000 \end{array}$ & $\begin{array}{c} \rho = 0.39 \\ p = 0.017 \end{array}$ \\
        \hline
        $\text{sSFR}_{10}$ & $\begin{array}{c} \rho = 0.15 \\ p = 0.183 \end{array}$ & $\begin{array}{c} \rho = 0.48 \\ p = 0.003 \end{array}$ & $\begin{array}{c} \rho = 0.11 \\ p = 0.326 \end{array}$ & $\begin{array}{c} \rho = -0.02\\ p = 0.875 \end{array}$ & $\begin{array}{c} \rho = 0.07 \\ p = 0.692 \end{array}$ \\
        \hline
         $\Sigma_{\text{SFR}}$ & $\begin{array}{c} \rho = -0.29 \\ p = 0.010 \end{array}$ & 
         $\begin{array}{c} \rho = -0.10 \\ p = 0.558 \end{array}$ & $\begin{array}{c} \rho = -0.26 \\ p = 0.018 \end{array}$ &  $\begin{array}{c} \rho = 0.38 \\ p = 0.001 \end{array}$ & $\begin{array}{c} \rho = -0.06 \\ p = 0.712 \end{array}$ \\

    \end{tabular}
    \caption{Spearman correlation coefficients ($\rho$) and p-values ($p$) for the four parameters cited in the paper, for $\rotsupp$\ and $\disp$\ in the case where the full (extended) sample is used, when only rotationally supported galaxies $\rotsupp>1$ are used, and when we use $v_{\rm rot} = v_{\rm obs,max}$ instead of $v_{\rm rot} = v_{\rm re}$.}
    \label{tab:corr-metho-comp}
\end{table*}

\section{Further investigation of the effect of SF on kinematics}\label{app:invest-sigma0-sfr}

We present in this section additional tests done to study the origin of the $\disp-\rm SFR$ correlation shown on Figs. \ref{fig:vsigma_sigma_mstar_sfr} and \ref{fig:corr-comp}. As discussed in Sec. \ref{sec:corr-z6}, this correlation naturally arises from models where gravitational instabilities, driven by radial transport of gas through the disk, and stellar feedback inject turbulence into the ISM. Both of these processes are linked to star-formation, hence producing the observed trend. However, it is unclear which of these processes is the dominant driver at high redshift. The low masses probed $\log M_{\star} ~ \rm [M_{\odot}]\lesssim 10$ make galaxies more unstable to gravitational instabilities, but also increase the impact of stellar feedback on the gas. 

\begin{figure*}
    \centering
    \includegraphics[width=1\linewidth]{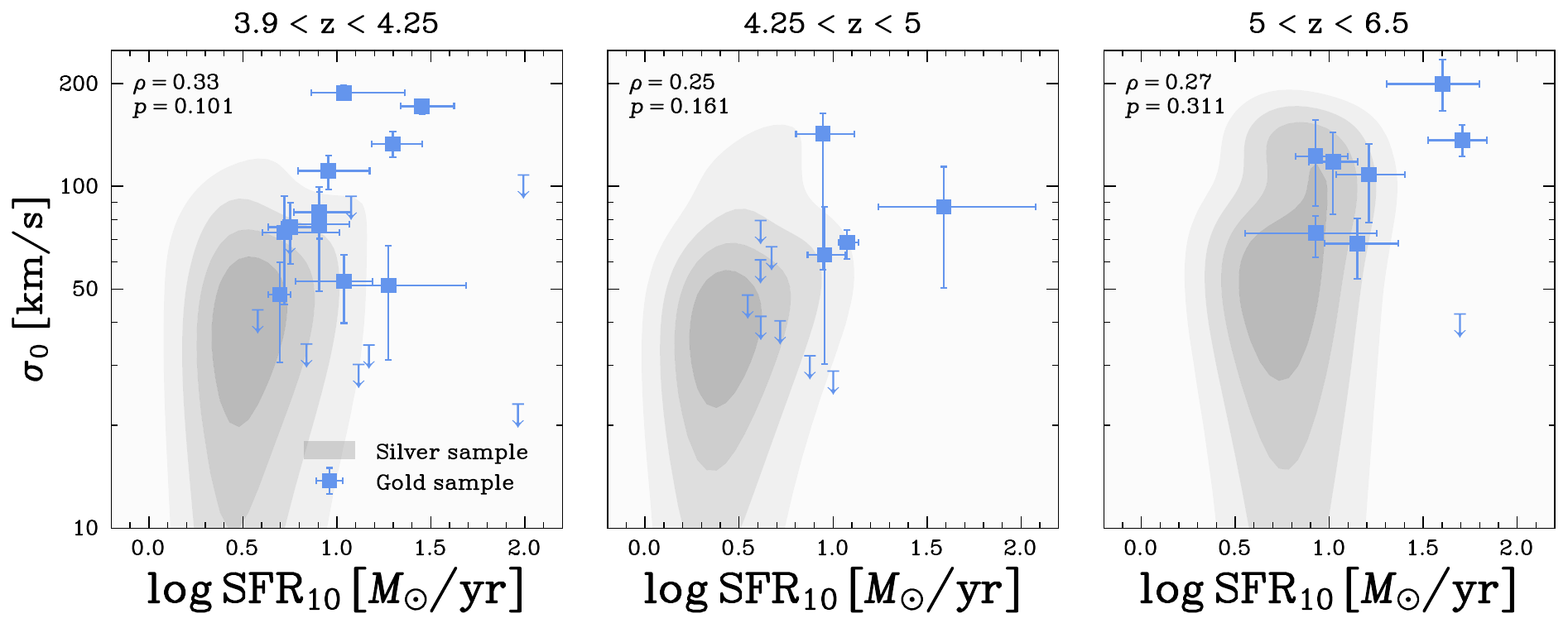}
    \caption{$\disp-\rm SFR$ relation for our gold (blue squares) and silver (gray contours) samples in three redshift bins spanning $\approx 0.2$ Gyr each. We show the Spearman rank correlation coefficients ($\rho$) and p-values ($p$) computed from the extended galaxies only, and find no significant correlations in the individual bins. }
    \label{fig:sigma0-sfr-zbins}
\end{figure*}

On Fig. \ref{fig:sigma0-sfr-zbins} we show the $\disp-\rm SFR$ relation for the gold and silver sample in three redshift bins, along with the Spearman rank coefficient $\rho$ and p-value $p$, computed for the resolved galaxies only. We see that in each bin, which all probe $\approx 0.2$ Gyr, we do not find significant correlations. This implies that the $\disp-\rm SFR$ relation shown in Fig. \ref{fig:vsigma_sigma_mstar_sfr}, which is calculated for the extended sample across the full redshift range $z\sim 4-6$ probed by the work, could be mainly explained by the redshift evolution of the SFR and the stellar masses. However, we note that these bins suffer from reduced sample sizes compared to the full sample, so this naturally decreases the significance of any correlations. 

\begin{figure}
    \centering
    \includegraphics[width=1\linewidth]{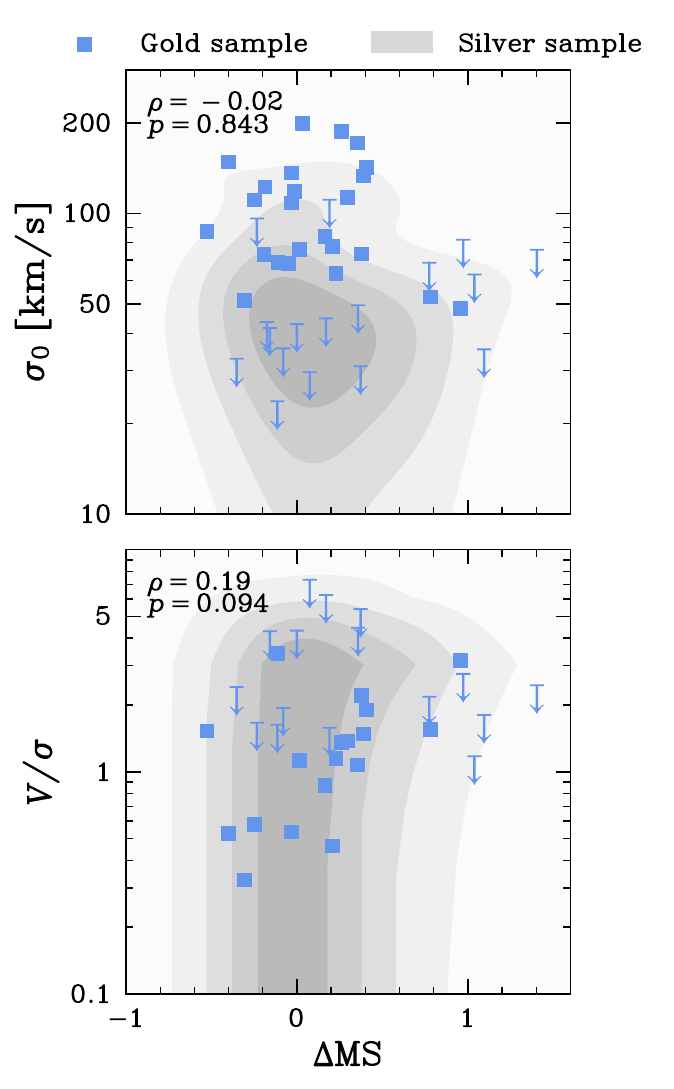}
    \caption{Evolution of $\disp$ (top panel) and $\rotsupp$ (bottom panel) with the offset from the SFMS (Simmonds et al, in prep) $\Delta \rm MS = SFR/SFR_{MS}$. We plot our gold (blue squares) and silver (gray contours) samples, and find no significant trends, as highlighted by the Spearman correlation coefficients $\rho$ and p-values $p$, which are computed for the extended sample. }
    \label{fig:sigmao-vsigma-MS}
\end{figure}

We further investigate the role of stellar feedback and gravitational instabilities by exploring trends with the offset from the main sequence $\Delta \rm MS = SFR/SFR_{MS}$, as shown on Fig. \ref{fig:sigmao-vsigma-MS} for our gold and silver samples. We use the SFMS prescription from Simmonds et al., in prep (Fig. \ref{fig:SFMS}). Although we find no correlation between $\disp$ and $\Delta \rm MS$, we find a slight positive correlation between $\disp$ and $\Delta \rm MS$ ($\rho = 0.19$), although the significance $p = 0.094$ is low. However, the trend is not significant enough to draw conclusions. This is overall consistent with findings from \citet{Ubler:2019vg}.

\begin{figure}
    \centering
    \includegraphics[width=1\linewidth]{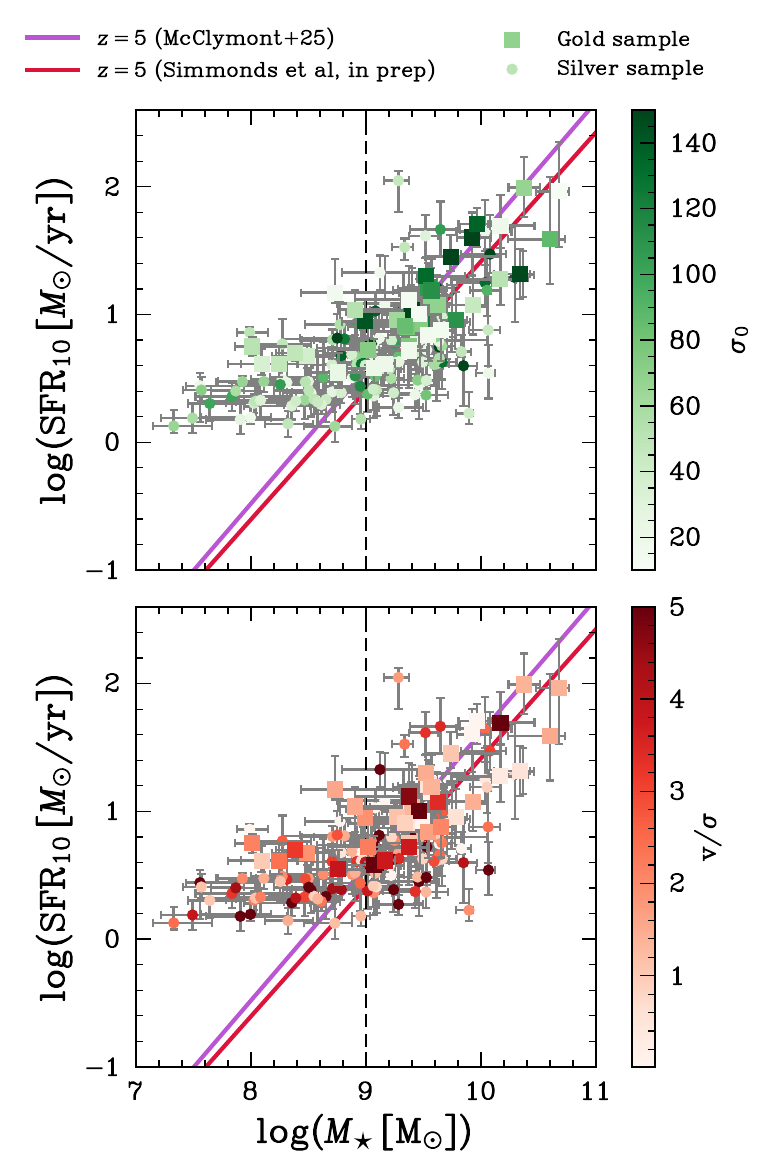}
    \caption{Distribution of our gold (squares) and silver (circles) samples on the $M_{\star}-\rm SFR$ plane, colorcoded by intrinsic velocity dispersion $\disp$ (top panel) and rotational support $\rotsupp$ (bottom panel). We plot the SFMS prescriptions from \citet{McClymont:2025aa} and Simmonds et al (in prep). }
    \label{fig:SFMS-comb}
\end{figure}

Finally, we explore the distribution of $\disp$ and $\rotsupp$ on the $M_{\star}-\rm SFR$ plane and plot our results on Fig. \ref{fig:SFMS-comb}. There is no clear trend relating the kinematics to this plane, it appears that on average, galaxies with high rotational support $\rotsupp>3$ lie close to the SFMS. An in-depth study of these results would require the study of individual galaxies on this plane in order to try to uncover the drivers of rotational support in each case, but this is beyond the scope of this work. 

\begin{figure*}
    \centering
    \includegraphics[width=1\linewidth]{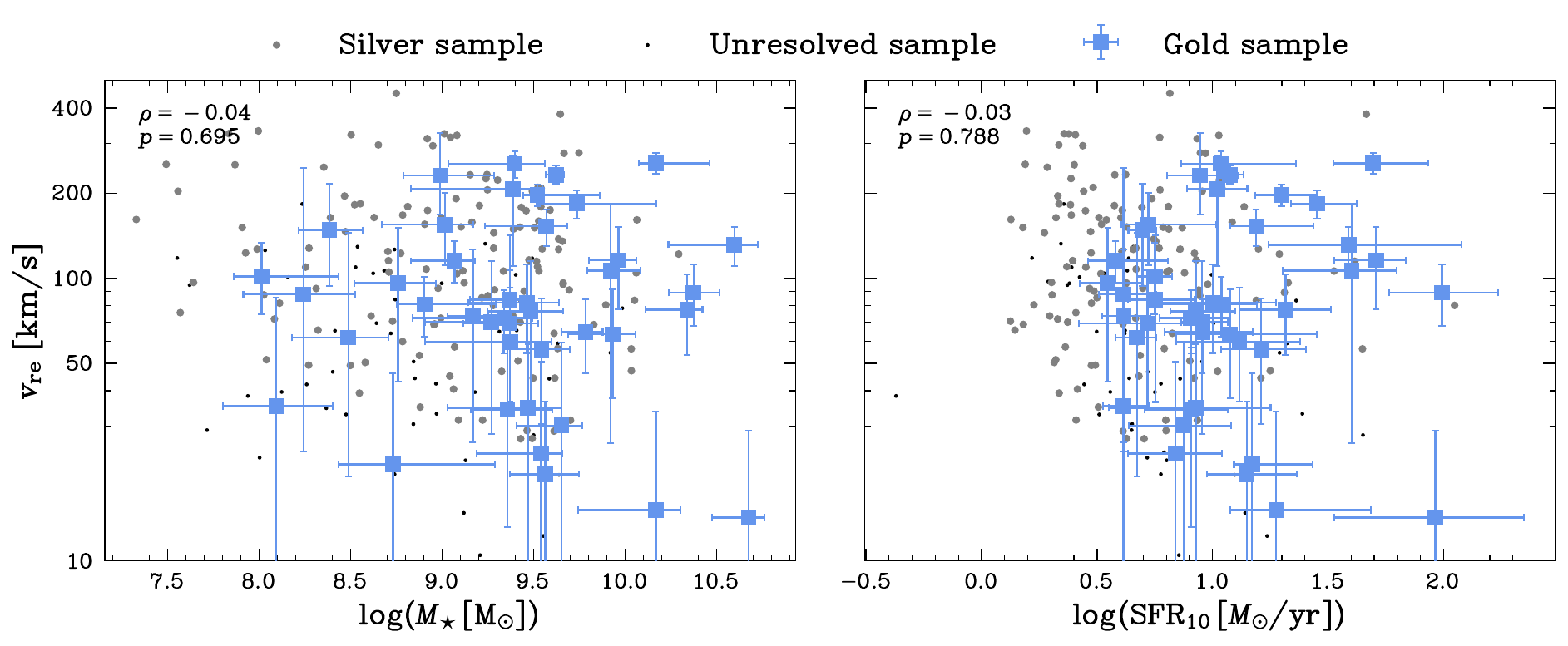}
    \caption{Evolution of the rotational velocity $v_{\rm rot} = v_{\rm re} $ with stellar mass $\log M_{\star} ~ \rm [M_{\odot}]$ (left panel) and SFR (right panel), for the gold (blue squares), silver (gray circles), and unresolved (black dots) samples. }
    \label{fig:v_max_mstar_sfr10}
\end{figure*}

When studying correlations between $\rotsupp$ and physical properties such as SFR and stellar masses, it is interesting the study the role of both $v=v_{\rm rot}$ and $\disp$ in driving the observed trends. On Fig. \ref{fig:v_max_mstar_sfr10} we show the relations between the rotational velocity and $\log M_{\star} ~ \rm [M_{\odot}]$ and SFRs. We find no correlation in both cases, meaning that at these redshifts $z\gtrsim4$ and stellar masses $\log M_{\star} ~ \rm [M_{\odot}] \lesssim 10$, rotational velocity does not strongly depend on stellar mass. This implies that the correlations measured on Fig. \ref{fig:vsigma_sigma_mstar_sfr} are mainly driven by the evolution of $\disp$.

\section{\textsc{geko} recovery tests}
\label{app:geko-tests}

\begin{figure}
    \centering
    \includegraphics[width=1\linewidth]{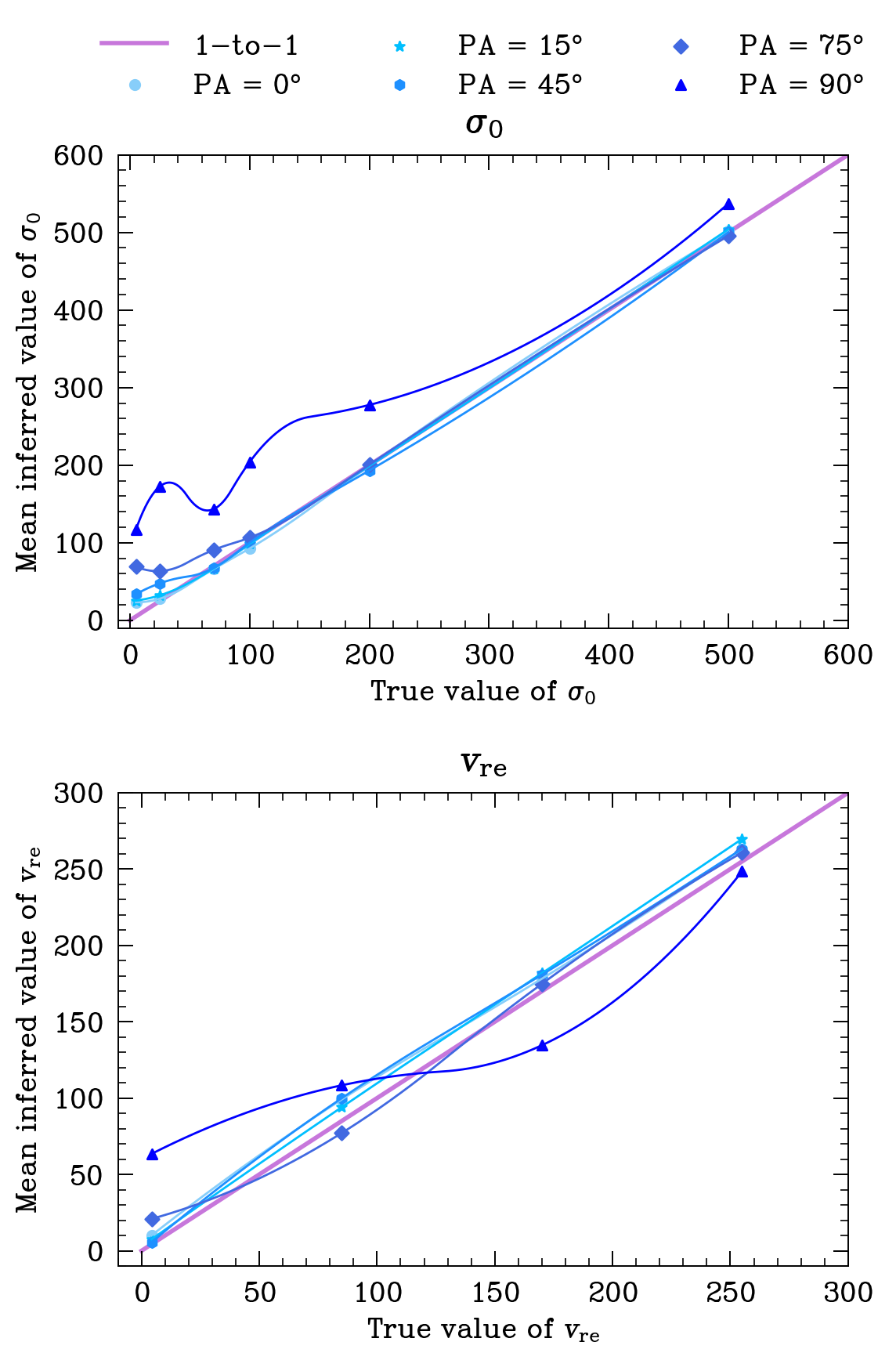}
    \caption{Results of recovery tests for $\disp$ (top panel) and $v_{\rm re}$ (bottom panel), for mock galaxies with a range of position angles $\rm PA=0-90^{\circ}$. The mean values are computed by averaging over tests done with a range of rotational velocities $v_{\rm re} = 5 - 250$ km/s, for $\disp$, and velocity dispersions $\disp = 5-500$ km/s, for $v_{\rm re}$.}
    \label{fig:geko-test-kin}
\end{figure}

In this section, we present the key recovery tests done for our tool \textsc{geko}. The full presentation of the code and the tests will be done in an upcoming dedicated paper (Danhaive et al., in prep). On Fig. \ref{fig:geko-test-kin} we show the results for the recovery of the key model-derived parameters $\disp$ and $v_{\rm re}$ (which are used to compute $\rotsupp$), for different angles of mock galaxies with respect to the dispersion direction $\rm PA = 90^{\circ}$. These tests are done on mock data, with an integrated $\rm S/N = 40$, and a range of velocity dispersions $\disp = 5-500$ km/s and rotational velocities $v_{\rm re} = 5 - 250$ km/s. The values plotted on Fig. \ref{fig:geko-test-kin} are averaged over all values of $v_{\rm re}$ for the $\disp$ recovery test and vice-versa. When the galaxy is parallel to the dispersion direction, we are not able to accurately recover $\disp$, which is artificially boosted. This unphysical value of $\disp$ is driven by the degeneracy between morphology,velocity, and velocity dispersion when the galaxy is parallel to the slit. For this reason, we discard galaxies with $\rm PA > 75^{\circ}$. As shown on Fig. \ref{fig:geko-test-kin}, our results converge to the true values, within uncertainties, when $\rm PA < 75^{\circ}$, which is why we use this cut to define our sample. We obtain the best recovery when galaxies as perpendicular to the dispersion direction, $\rm PA = 0^{\circ}$. We can further see that, for cases $\rm PA < 15^{\circ}$, we are able to recover values of $\disp\approx 20$ km/s, which is well below the instrument LSF $\disp\approx 80$ km/s.

\begin{figure}
    \centering
    \includegraphics[width=1\linewidth]{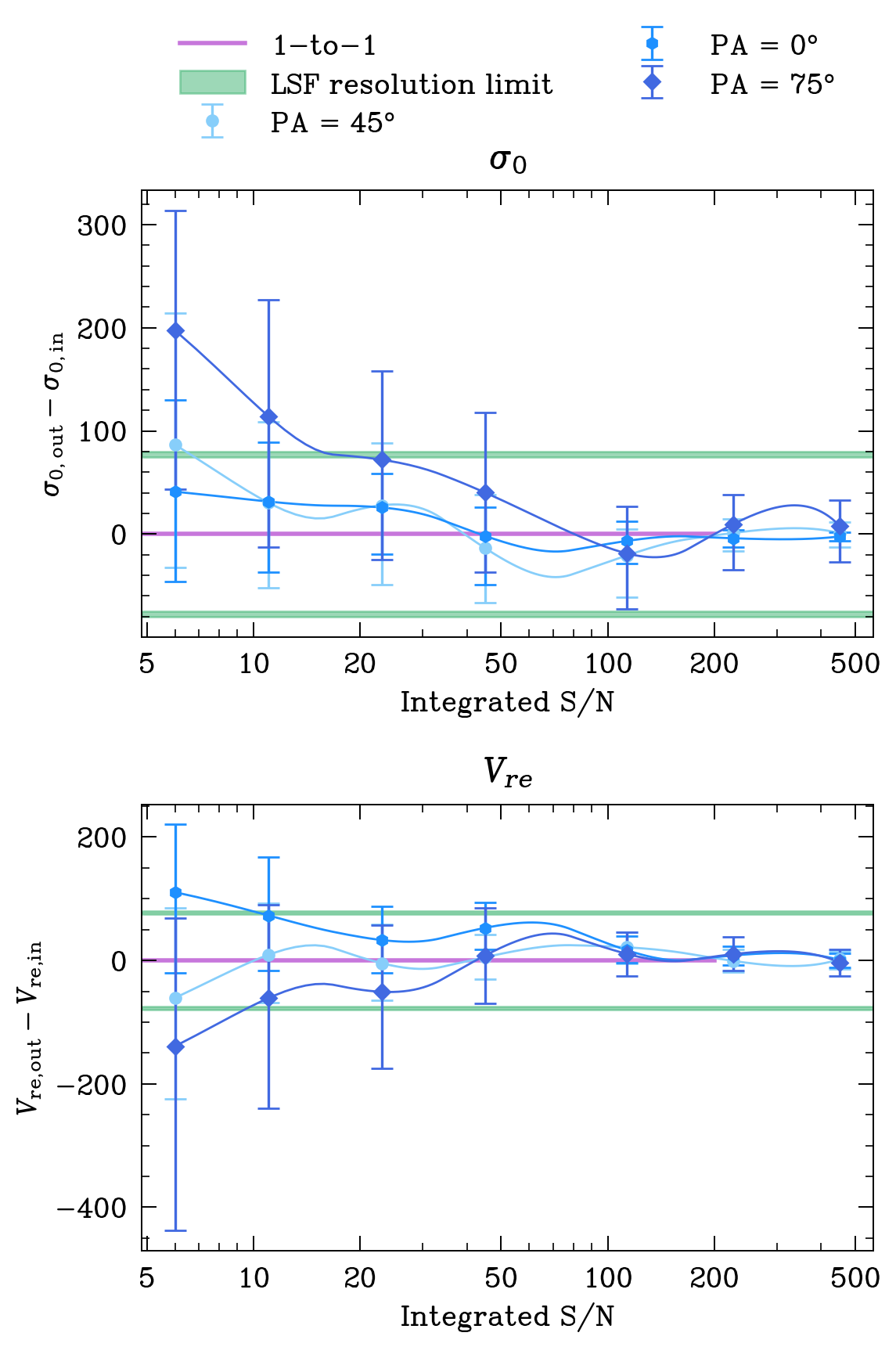}
    \caption{Results of recovery tests for $\disp$ (top panel) and $v_{\rm re}$ (bottom panel) for integrated $\rm S/N = 6-500$. The mock galaxy has $\disp = 100$ km/s and $v_{\rm re} \approx 170$ km/s.}
    \label{fig:fig:geko-test-sn}
\end{figure}

We also test the effect of integrated $\rm S/N$ on our ability to recover kinematic parameters for our mock galaxies. We show the results on Fig. \ref{fig:fig:geko-test-sn}, for $\disp = 100$ km/s and $v_{\rm re} = 170$ km/s. Beyond $\rm S/N = 10$, the recovered value of $\disp$ is consistent with the true input value within the uncertainties. We set $\rm S/N >10$ as our cut for the silver sample, and $\rm S/N >20$ for the gold sample. The uncertainty decreases with increasing $\rm S/N$ and decreasing PA. 

\section{Comparison to $z<4$ measurements}\label{app:z-evol-comp}

In this section, we briefly introduce the studies we compare our measurements to (Figs. \ref{fig:vsigma_z_evol}, \ref{fig:sigma0-z-comp}, and \ref{fig:corr-comp}) and discuss how their methods differ from those used in this work and how that may affect the direct comparisons we make.

The MOSFIRE Deep Evolution Field (MOSDEF) survey data is described in \cite{Price:2020wf}, and is obtained using the MOSFIRE spectrograph on the Keck telescope. This survey is comprised of rest-frame optical 2D moderate-resolution spectra $R\sim 3000-3600$ for galaxies in the CANDELS field at $z\sim 1.4-3.8$. The kinematic analysis is restricted to galaxies whose major axis is roughly aligned with the slit and that are spatially resolved. Their final sample is made up of $\sim 700$ galaxies detected in \Ha, \OIII, and H$\beta$, in order of availability. Their kinematic modelling is similar to the one adopted in this paper. They assume an intrinsic axis ratio (i.e., axis ratio for an edge-on view)  $(b/a)_0 = 0.19$, and model the velocities with an arctangent model and a constant velocity dispersion across the galaxy. Using inferred posterior distributions, they compute $\disp$\ and the rotational support, which they define as $v_{2.2}/ \sigma_0$ where  $v_{2.2} = V(r=2.2r_{\rm s})$. This is important to note as typically  $v_{2.2}>v_{\text{re}}$, which can boost the $v/\sigma_0$ measurements compared to those in this work.

The  K-band multi-object spectrograph (KMOS)$^{\text{3D}}$ survey data described in \cite{Wisnioski:2015vx} is obtained using the KMOS integral field spectrograph on Keck. Galaxies are observed in H$\alpha$ across five deep fields (COSMOS, GOODS-S, GOODS-N, UDS, and AEGIS) and span redshifts $z\sim 0.7-2.7$. For the kinematic fitting, each spaxel is fit with a 1D Gaussian model including the instrument LSF. The velocity dispersions are computed in the outer regions that are less affected by beam smearing. They define the rotational support $v/\sigma_0$ with $v_{\text{rot}} = v_{\text{obs}}/\sin i$, where $v_{\text{obs}}$ is an observed velocity difference from the maximal and minimal velocities of the velocity axis profile. This overall velocity gradient is usually greater than the velocity measured at the effective radius, which could bias the measurements high compared to this work.

The SIGMA survey \citep{Simons:2016aa, Simons:2017aa} is a Keck/MOSFIRE survey, which targeted 97 $z\sim 2$ star-forming galaxies in the GOODS-S, GOODS-N and UDS fields, detected in H$\alpha$ or \OIII. The kinematic modelling assumes an arctangent velocity model and $q_0=0.2$. The velocity used for the rotational support represents the asymptotic velocity measured in the flat outer end of the rotation curve.

The KMOS Deep Survey (KDS) survey is described in \citep{Turner:2017aa}. Galaxies at $z\sim 3.5$ are observed through their \OIII\ emission, where the ionised gas kinematics are computed using an arctangent model. The rotational velocity is computed as the best-fit velocity at $r=2r_{\rm e}$, which is comparable to the velocity at the radius enclosing $80\%$ of the total light $v_{80}$ \citep{Tiley:2016aa}. In the thin disk model, we expect the rotation curve to increase beyond $r_{\rm e}$, where it is measured in this work, so the velocities at $r=2r_{\rm e}$ could be higher. However, the velocities for a thick exponential disk $n=1$ would peak at $r\approx 1.38 r_{\rm e}$, so the relation between velocities at $r=r_{\rm e}$ and $r=2r_{\rm e}$ varies depending on the object.


The KROSS survey \citep{Johnson:2018aa} is also a KMOS survey which observed $\sim 800$ star-forming galaxies at $z\sim1$ using H$\alpha$. The kinematic modelling assumes an intrinsic axis ratio $q_0=0.2$ and computes the velocity for the rotational support as the velocity at $r=2r_{\rm e}$. Both this velocity and the velocity dispersion are computed along the major kinematic axis. However, different from our work, the velocity curve is modelled as an exponential disk. At the small radii probed by the grism data, this curve is similar to the arctangent curve we use, so this should not affect our comparison. 

The Sydney/AAO Multi-object Integral-field spec-
trograph \citep[SAMI; ][]{Croom:2012aa} Galaxy Survey \citep{Allen:2015aa, Green:2018aa} is a census of local Universe galaxies, probing both stellar and gas kinematics.  
The Multi-Unit Spectroscopic Explorer \citep[MUSE; ][]{Bacon:2010aa} samples is presented in \citep{Swinbank:2017aa} and covers galaxies at redshifts $0.28 < z < 1.49$. Both the SAMI and MUSE measurements shown in this work are those presented in \citet{Johnson:2018aa}, and are corrected, when necessary, to match measurements for the KROSS sample and provide an accurate comparison. The kinematics for these two samples are both derived from ionised gas.

The SINS/zC-SINF AO survey \citep{Forster-Schreiber:2018aa} is a survey of 35 galaxies at $z\sim 2$, probing kinematics and star-formation with \Ha and [NII] lines. The rotational velocity is chosen as the maximum observed velicity difference along the kinematic major axis, similar to \citet{Wisnioski:2015vx}. An isotropic velicity dispersion $\disp$ is assumed. \citep{Forster-Schreiber:2018aa} report kinematic measurements consistent with those from the KMOS survey at similar redshifts \citep{Wisnioski:2015vx, Wuyts:2016aa,Ubler:2017aa}, and are hence comparable. 

The Gassendi \Ha\ survey of SPirals \citep[GHASP;][]{Epinat:2008aa, Epinat:2010aa} contains Fabry-Perot observations of \Ha\ in 203 local galaxies. For the kinematic modelling, \citet{Epinat:2010aa} a thin planar disk is assumed, and a variety of rotation curves are used based on how well they fit the observed rotation in each system. However, all remain failry consistent with our arctagent curve within the radii probed in this work. The nature of the data and the modelling approach make a one-to-one comparison difficult, but nonetheless provide an indication of galaxy kinematics in the local Universe.

The the Mapping Nearby Galaxies at Apache Point Observatory \citep[MaNGA;][]{Bundy:2015wp} survey measaurements shown for comparison in this work are from \citet{Yu:2019aa}. The sample contains spatially resolved \Ha\ measurements for 2716 galaxies in the local Universe. The velocity dispersion is inferred by staking the 1D \Ha\ line from each spaxel and correcting for the instrument resolution and beam smearing.

The \cite{de-Graaff:2024ab} measurements are obtained using the JWST NIRSpec MSA 2D spectra. The assumptions adopted in the kinematic modelling are the same as the ones used in this work. However, the NIRSpec slits suffer from slit-losses, which are described in the paper and accounted for in the forward-modelling of the instrument. 

The sample from \citet{Parlanti:2023ab} is selected from the public ALMA data archive and is composed of 22 galaxies at high redshift $(4.2 < z < 7.6)$. The galaxies are observed in the [CII] or [OIII] emission lines, or both. The rotational velocity is modelled using a bulge component and an exponential disk component. The rotational velocity is defined as the maximum observed velocity, which is typically higher than the velocity measured at $r_{\rm e}$, as previously discussed. The cold gas measurements from the [CII] emission line have lower velocity dispersions on average, as shown in \citep{Parlanti:2023ab}, than measurements from ionised gas ([OIII] and \Ha).

We also compare our results with predictions from the TNG simulations \citep{Pillepich:2019aa}, where they compute kinematic properties in the following way. The rotational velocity $v_{\text{rot}}$ is the absolute maximum of a galaxy's rotation curve along its structural major axis, from its edge-on projection and within twice the stellar half-mass radius. The dispersion $\disp$\ is the value of the velocity dispersion from a face-on projection of a galaxy, averaged in pixels of 0.5 ckpc a side that lie along the structural major axis and between one and two the stellar half-mass radius.

We do not discuss the cold gas local Universe measurements as they are only shown in Fig. \ref{fig:sigma0-z-comp} to compare to local Universe measurements of ionised gas \citep[for further details, see ][]{Ubler:2019vg}. 

\section{Gold Sample Fits} \label{app:gold-fits}
Here we present the \textsc{geko} summary plot, which shows the best-fit model compared to the grism data, of all of the objects in the gold sample, whose properties are summarized in Tab. \ref{tab:results-gold}. For a full description of the structure of this plot see Sec. \ref{sec:fitting} and Fig. \ref{fig:geko-summary}. We note that hot or dead pixels are masked in the fitting process, but not in the summary plots.

\begin{figure*}
    \centering
    \includegraphics[width=0.8\linewidth]{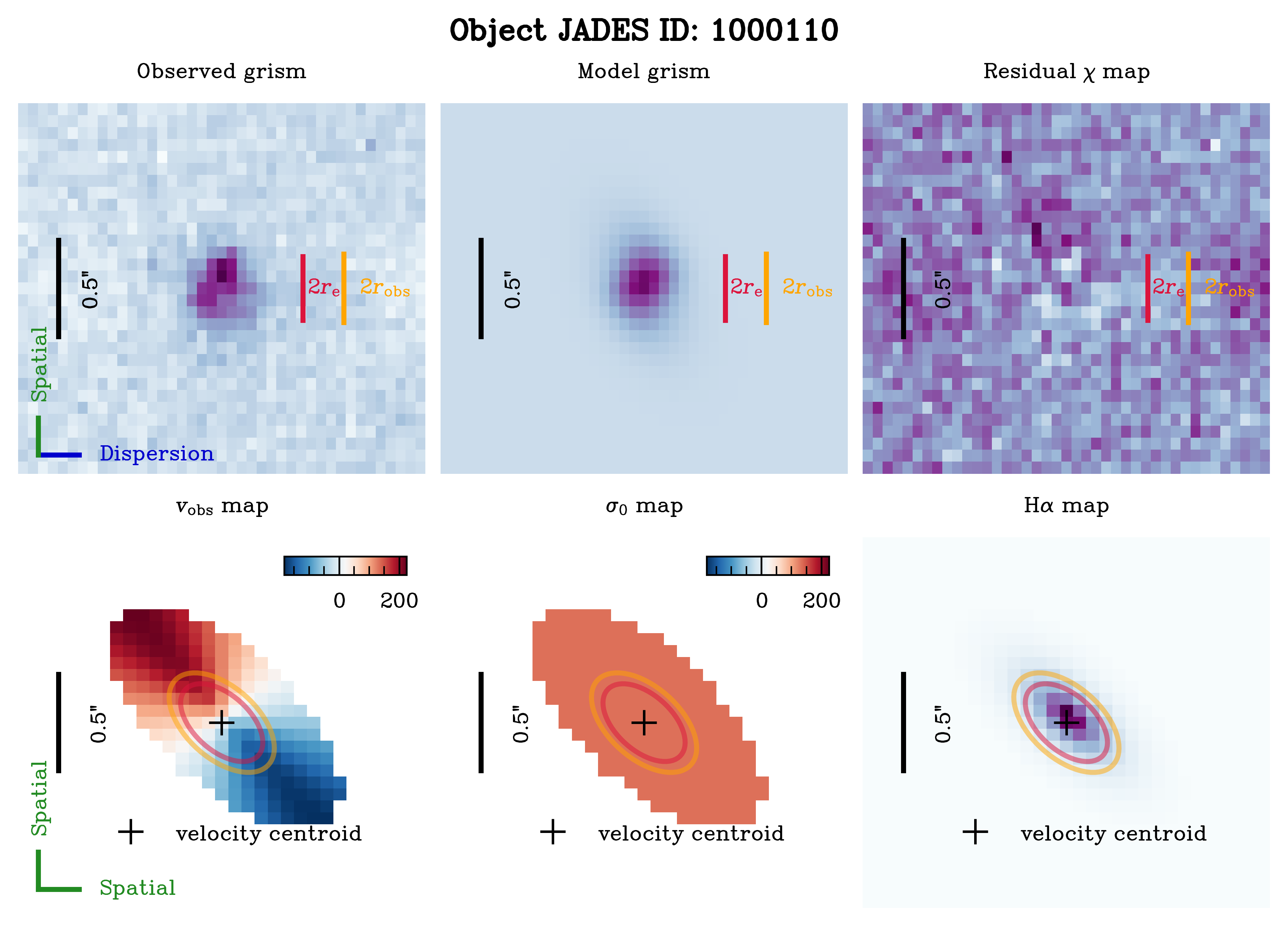}
    \caption{\textsc{geko} summary plots for the objects in the gold sample. The JADES ID for each object is marked at the top. The left-hand panel shows the residuals of the model with respect to the observed grism, along with the best-fit velocity, velocity dispersion, and intrinsic emission maps. For a full description of the structure of this plot see Sec. \ref{sec:fitting} and Figs.\ref{fig:geko-summary}.}
    \label{fig:gold-summaries}
\end{figure*}

\begin{figure*}
    \ContinuedFloat
    \centering
     \includegraphics[width=0.8\linewidth]{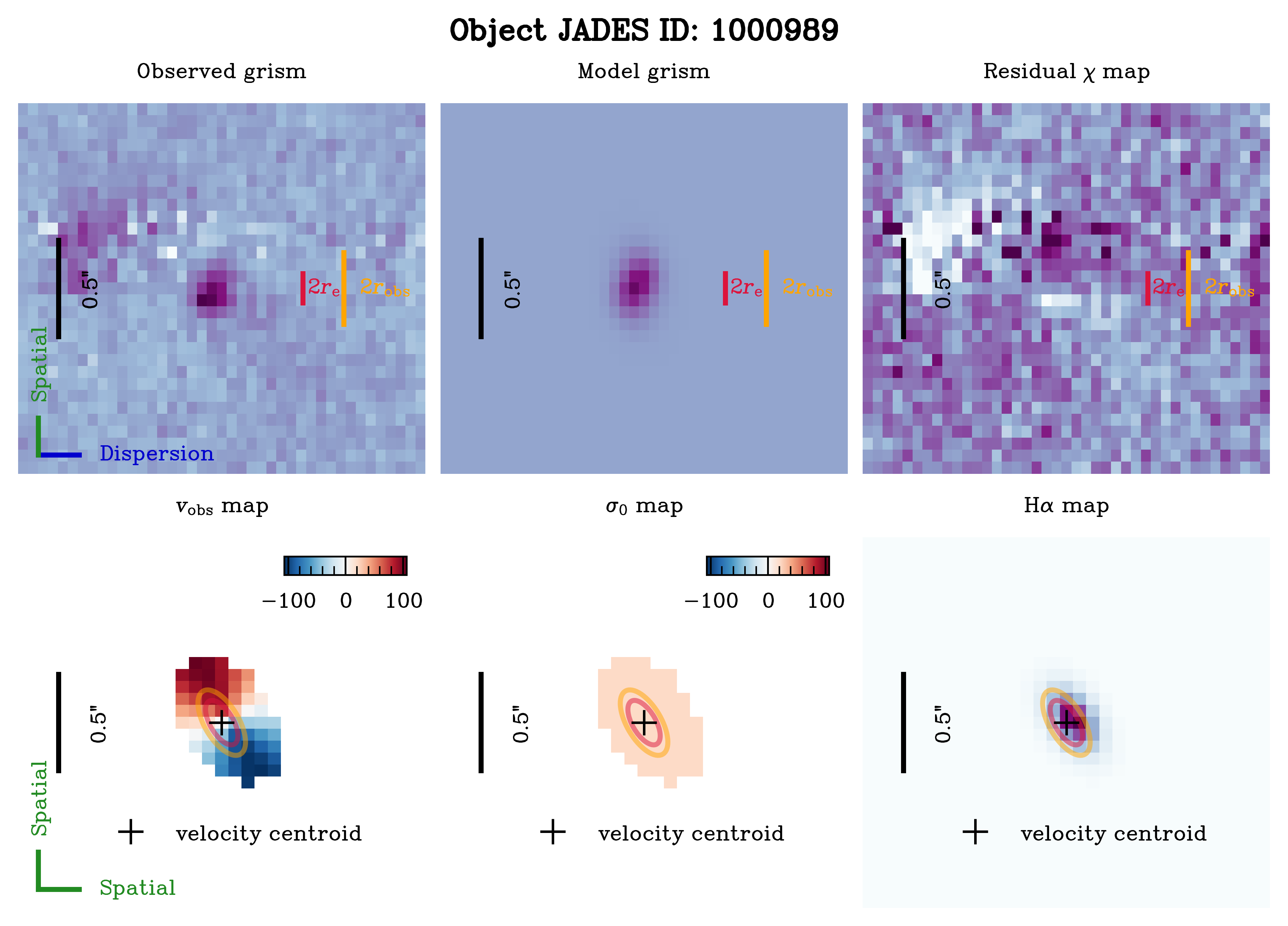}
     \caption{Continued.}
\end{figure*}

\begin{figure*}
    \ContinuedFloat
    \centering
         \includegraphics[width=0.8\linewidth]{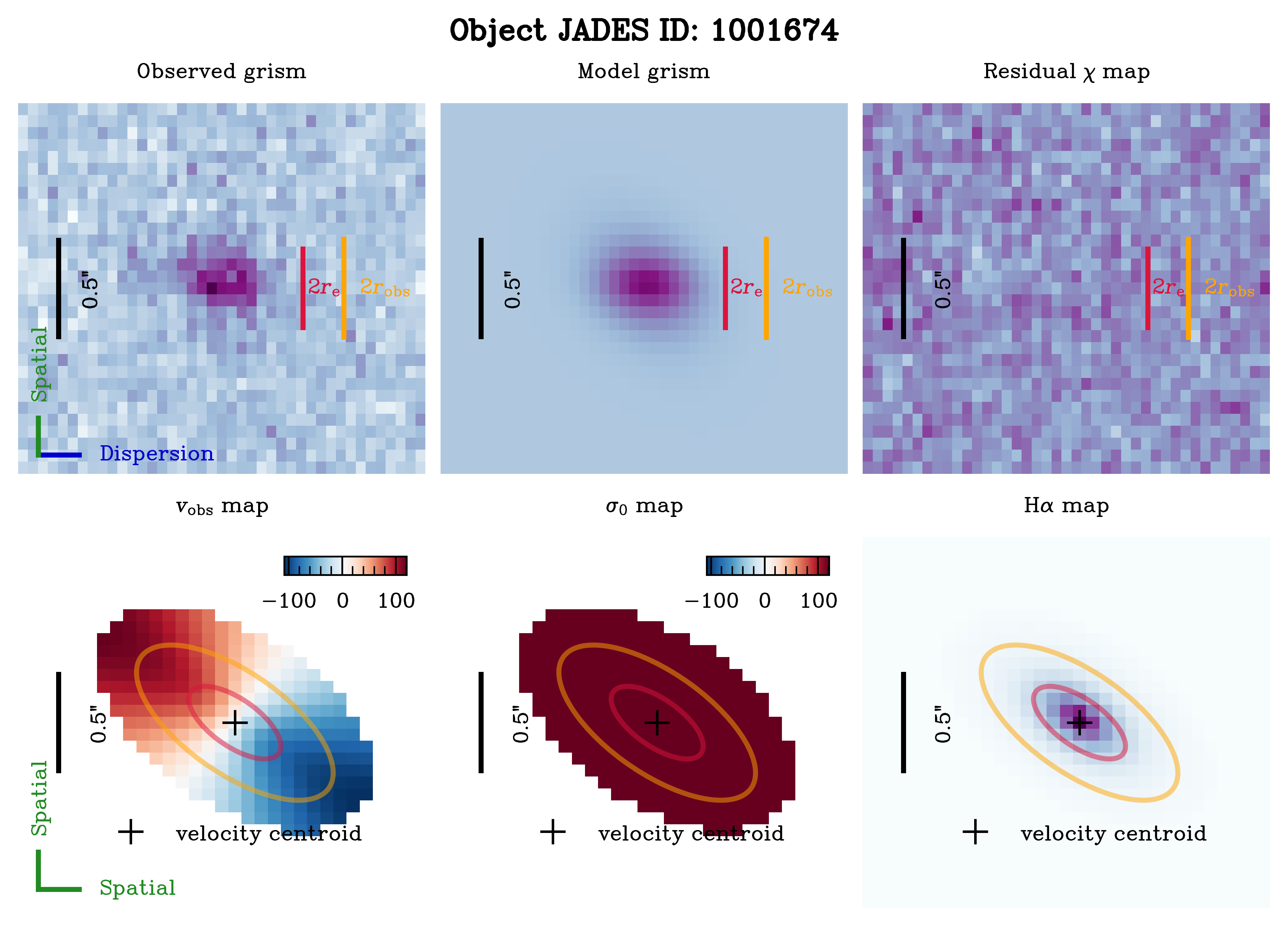}
     \caption{Continued.}
\end{figure*}

\begin{figure*}
    \ContinuedFloat
    \centering
         \includegraphics[width=0.8\linewidth]{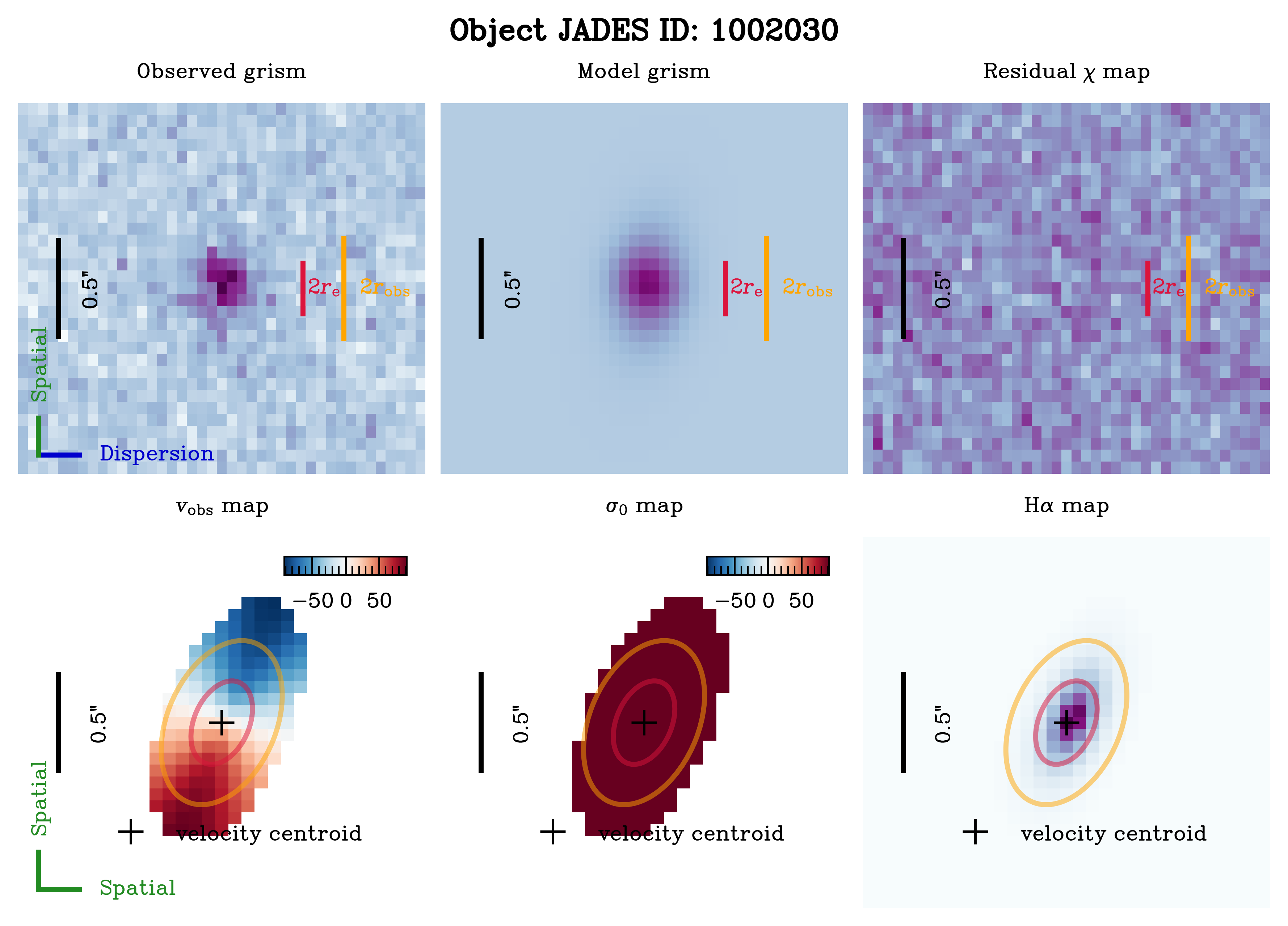}
     \caption{Continued.}
\end{figure*}

\begin{figure*}
    \ContinuedFloat
    \centering
         \includegraphics[width=0.8\linewidth]{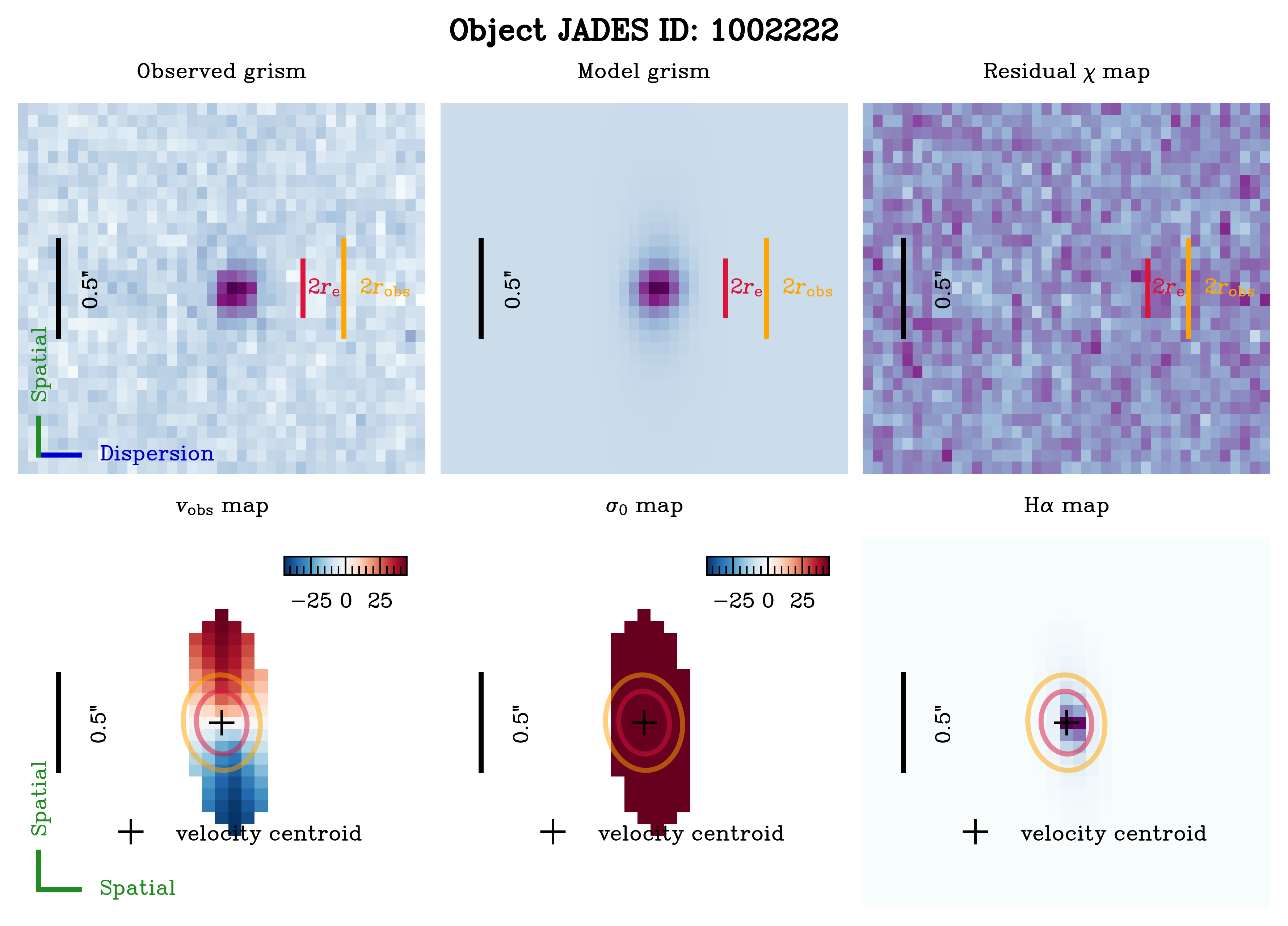}
     \caption{Continued.}
\end{figure*}

\begin{figure*}
    \ContinuedFloat
    \centering
         \includegraphics[width=0.8\linewidth]{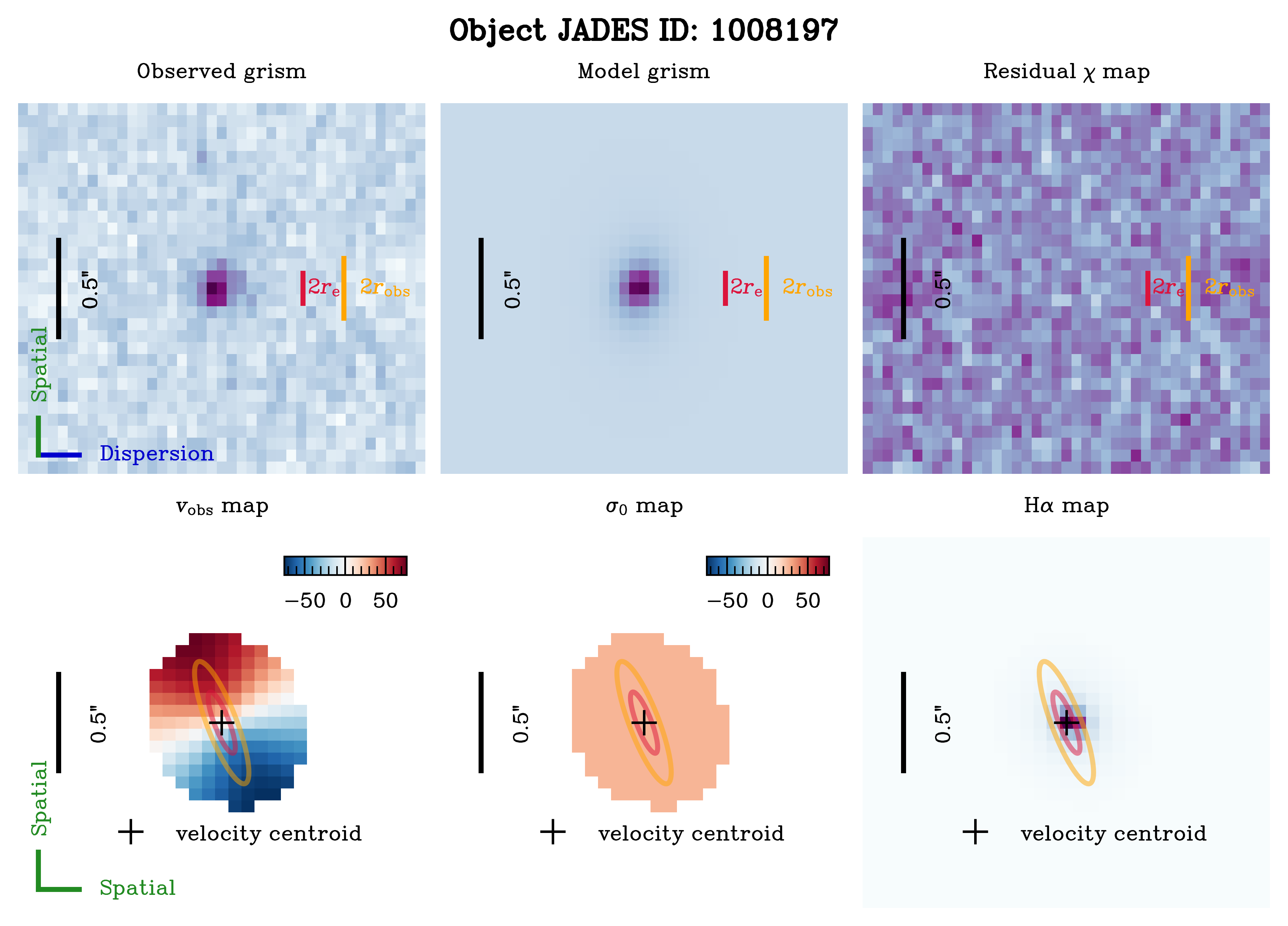}
     \caption{Continued.}
\end{figure*}

\begin{figure*}
    \ContinuedFloat
    \centering
         \includegraphics[width=0.8\linewidth]{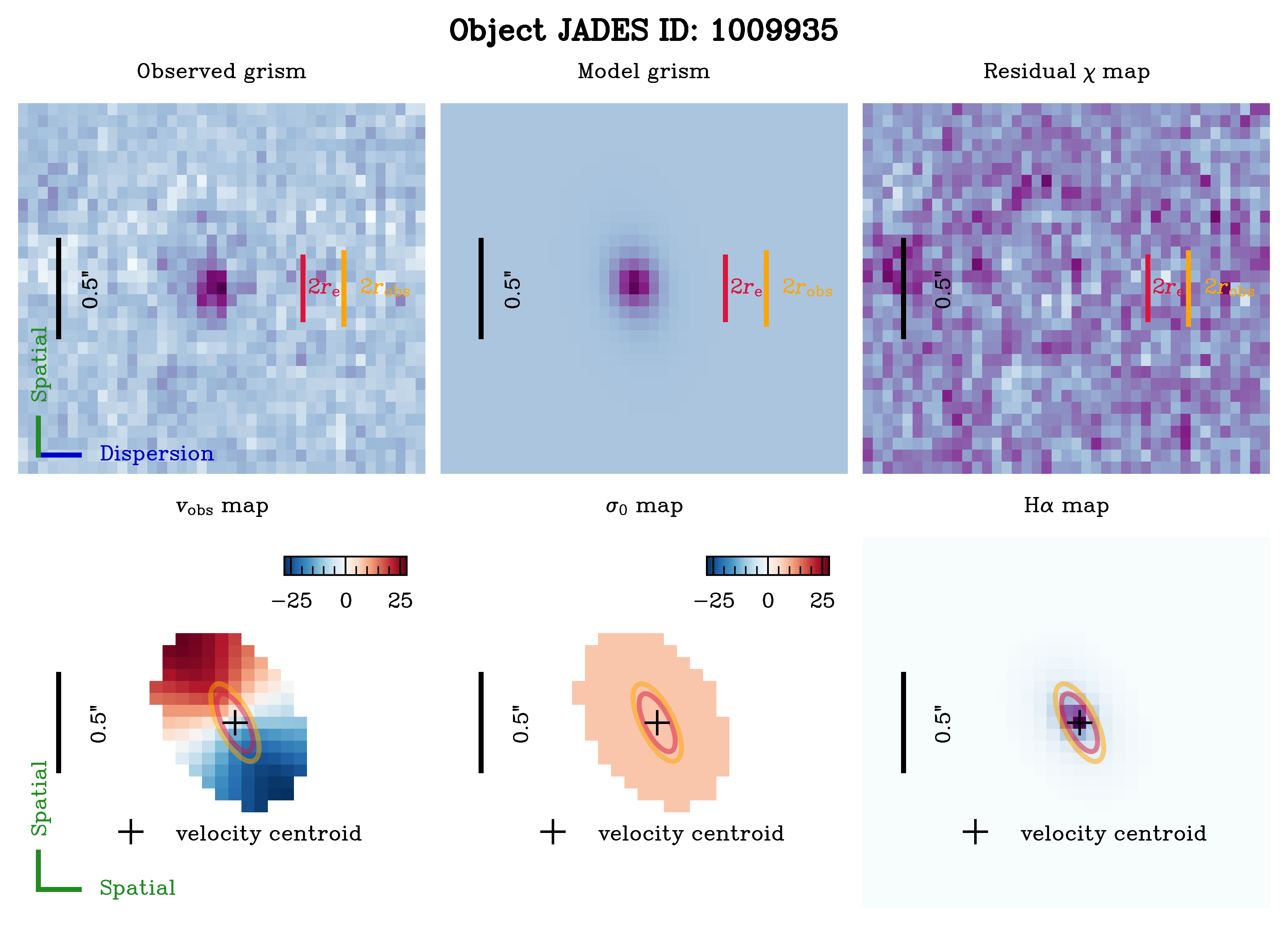}
     \caption{Continued.}
\end{figure*}

\begin{figure*}
    \ContinuedFloat
    \centering
         \includegraphics[width=0.8\linewidth]{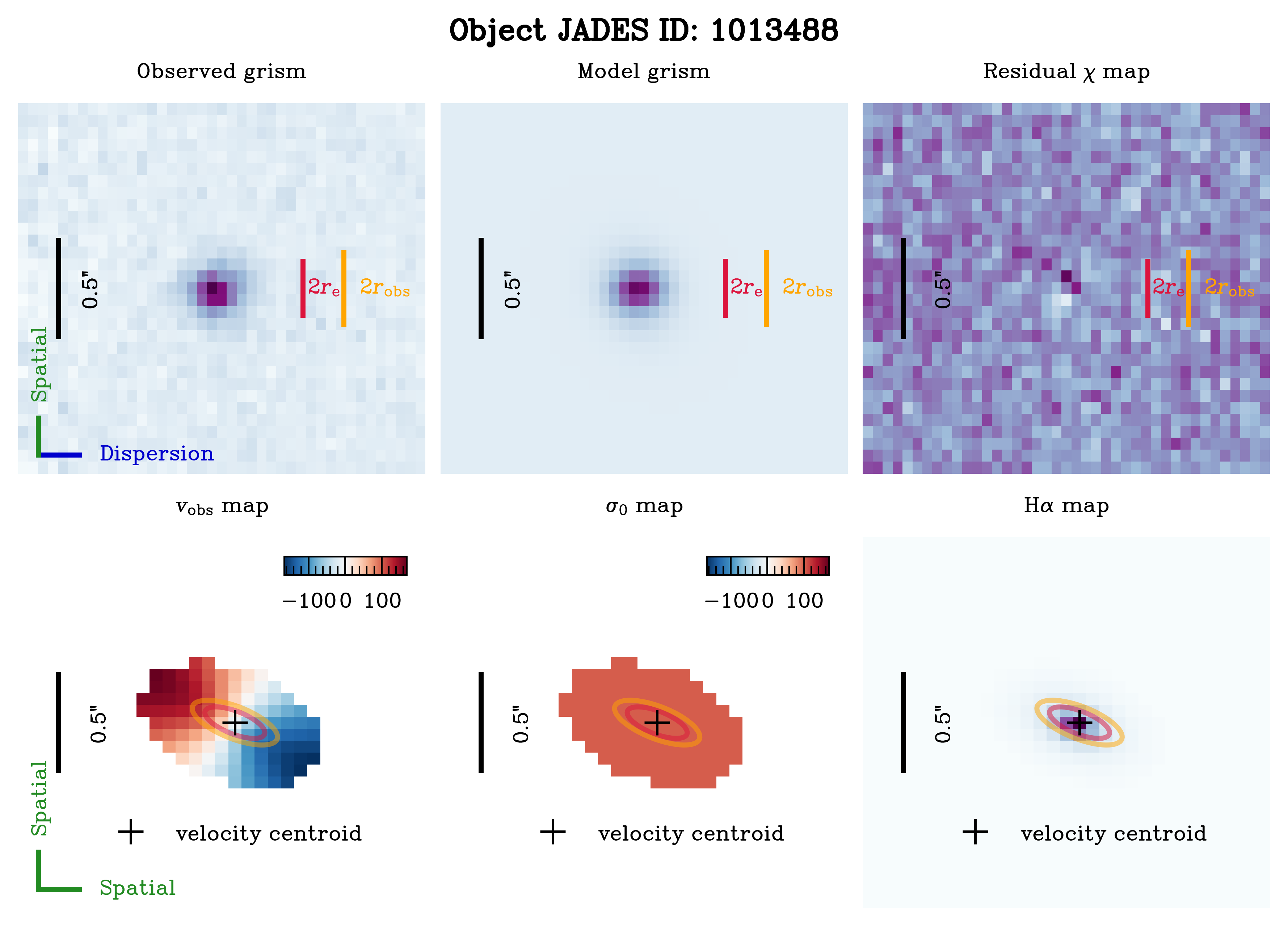}
     \caption{Continued.}
\end{figure*}

\begin{figure*}
    \ContinuedFloat
    \centering
         \includegraphics[width=0.8\linewidth]{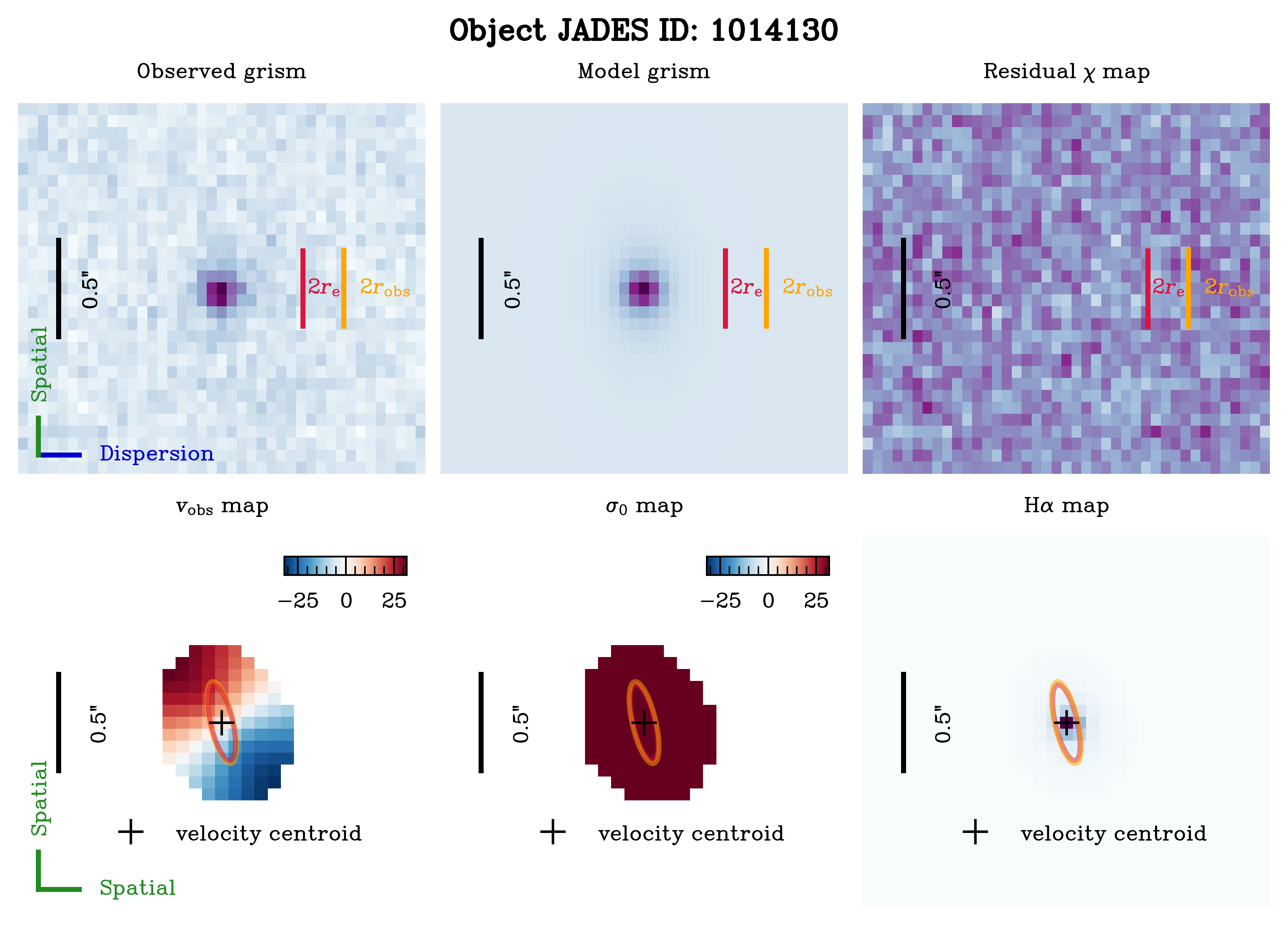}
     \caption{Continued.}
\end{figure*}

\begin{figure*}
    \ContinuedFloat
    \centering
         \includegraphics[width=0.8\linewidth]{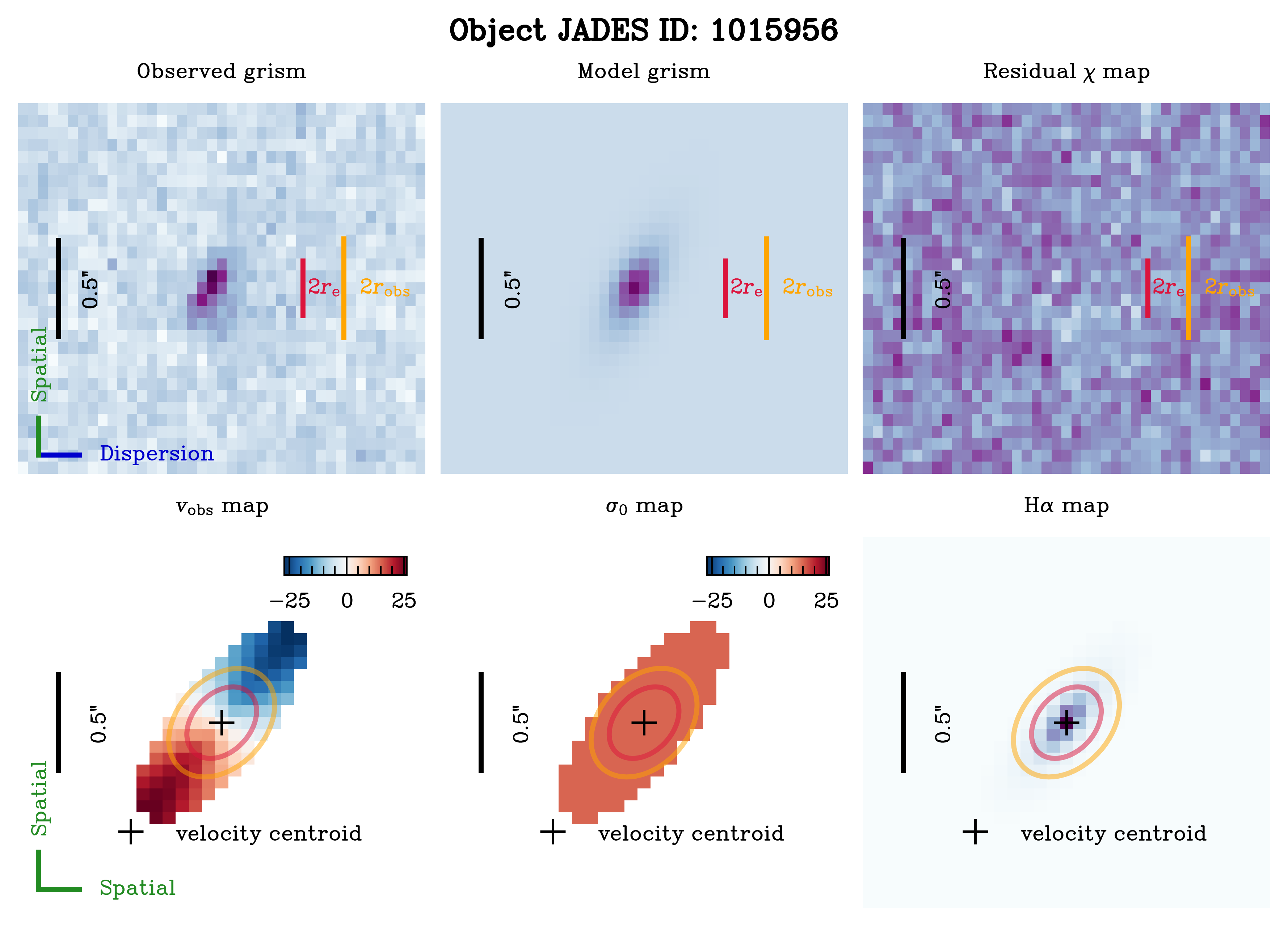}
     \caption{Continued.}
\end{figure*}

\begin{figure*}
    \ContinuedFloat
    \centering
         \includegraphics[width=0.8\linewidth]{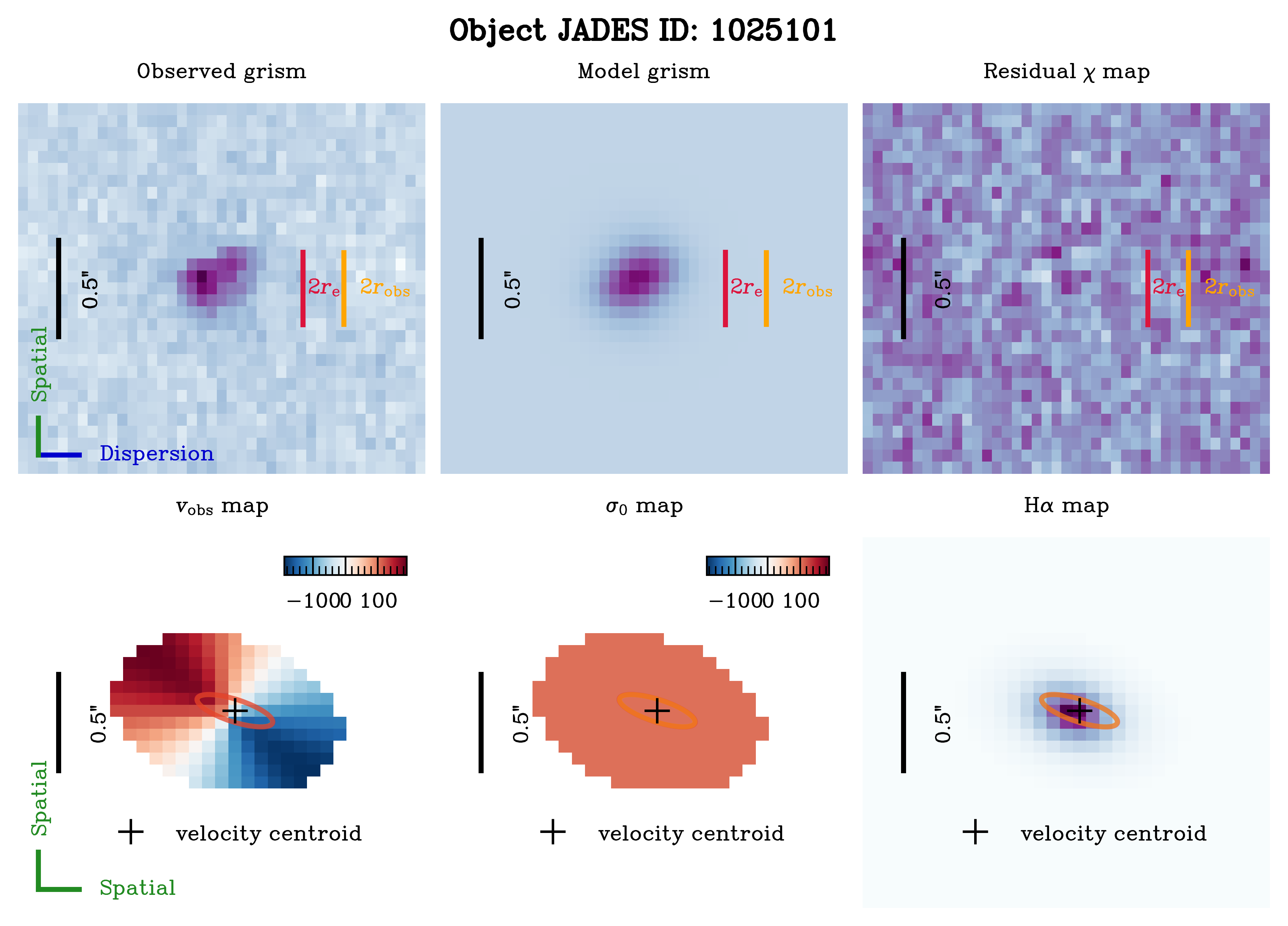}
     \caption{Continued.}
\end{figure*}

\begin{figure*}
    \ContinuedFloat
    \centering
         \includegraphics[width=0.8\linewidth]{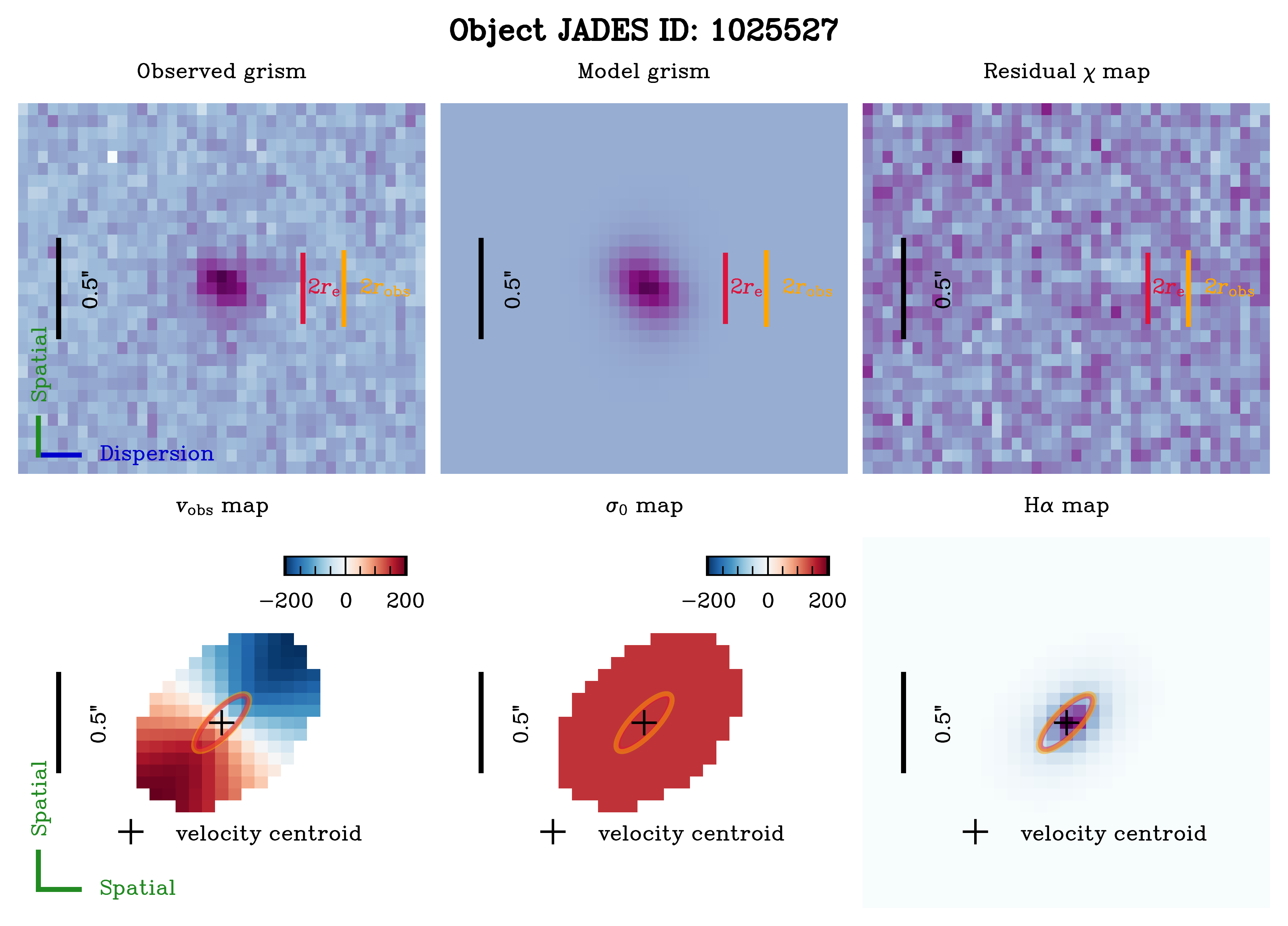}     \caption{Continued.}
\end{figure*}

\begin{figure*}
    \ContinuedFloat
    \centering
         \includegraphics[width=0.8\linewidth]{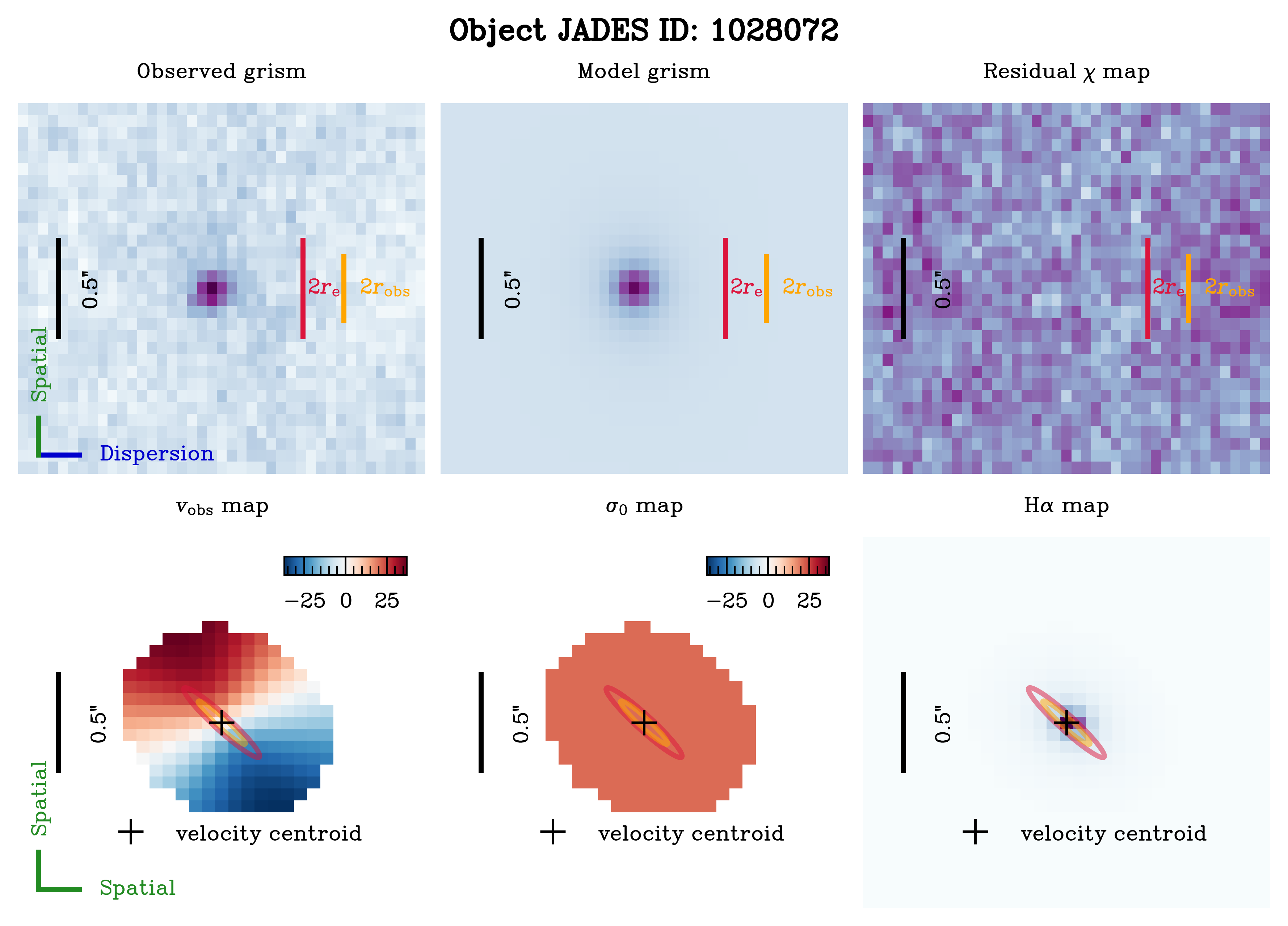}
     \caption{Continued.}
\end{figure*}

\begin{figure*}
    \ContinuedFloat
    \centering
         \includegraphics[width=0.8\linewidth]{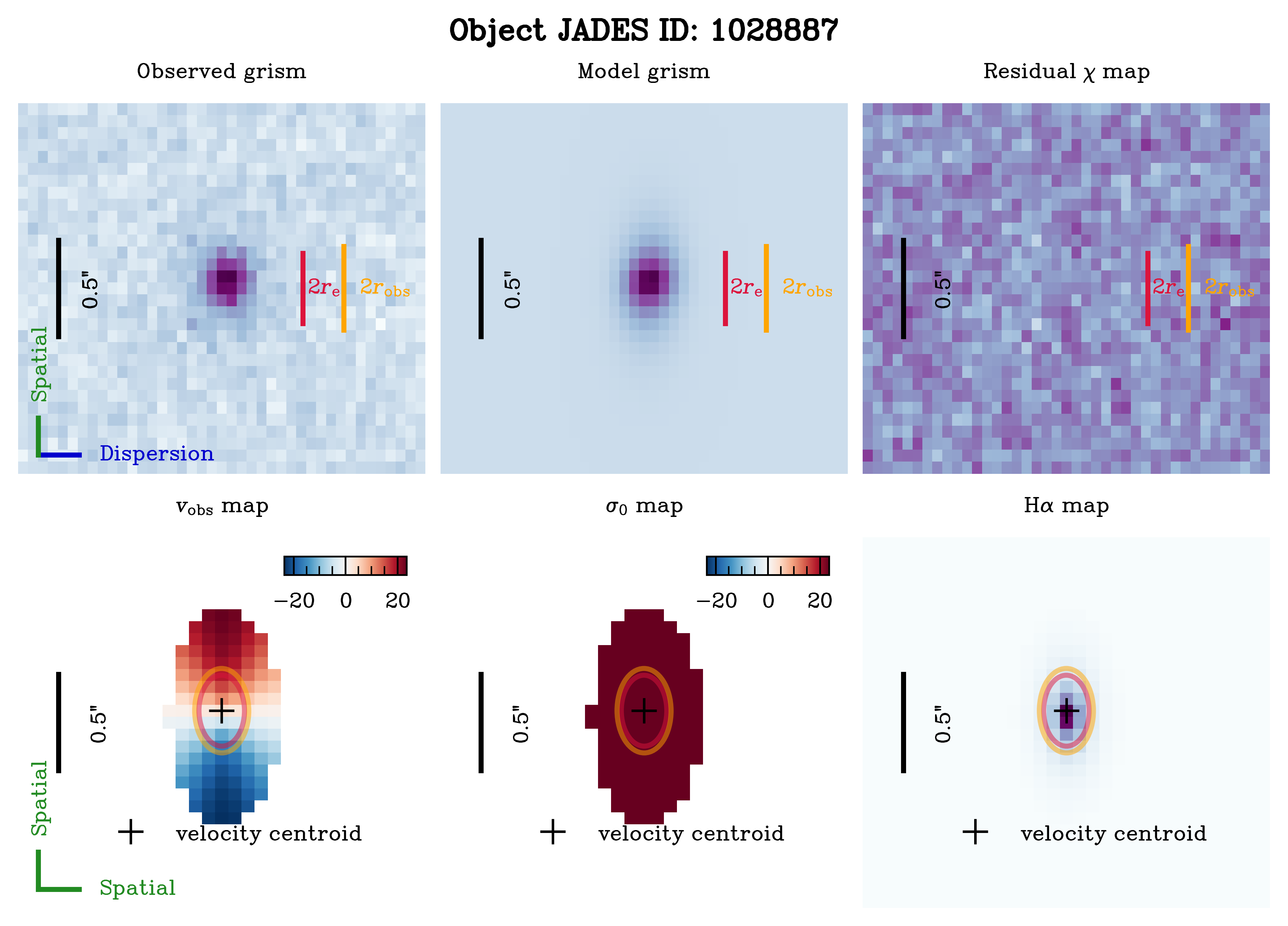}
     \caption{Continued.}
\end{figure*}

\begin{figure*}
    \ContinuedFloat
    \centering
         \includegraphics[width=0.8\linewidth]{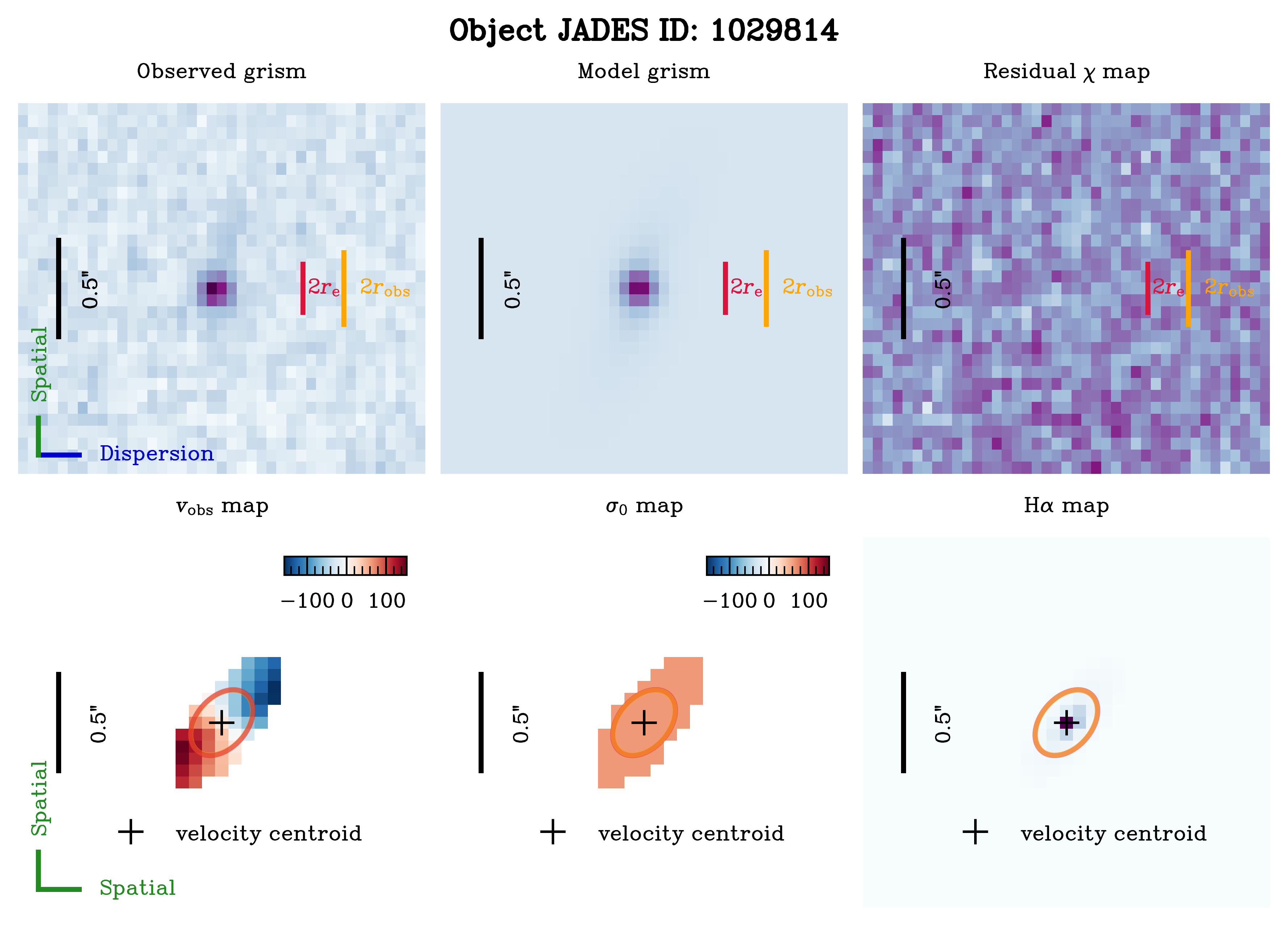}
     \caption{Continued.}
\end{figure*}

\begin{figure*}
    \ContinuedFloat
    \centering
         \includegraphics[width=0.8\linewidth]{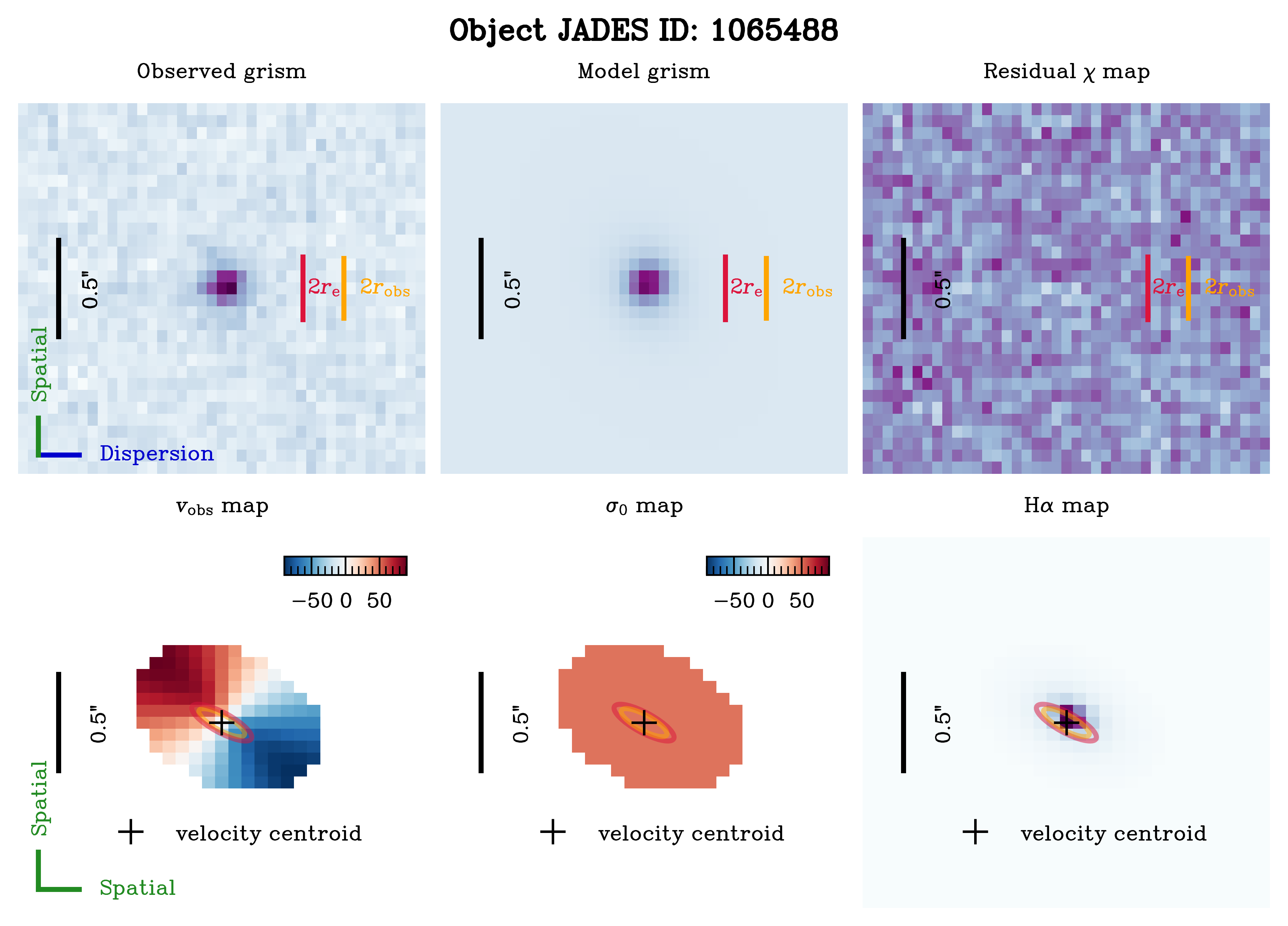}
     \caption{Continued.}
\end{figure*}

\begin{figure*}
    \ContinuedFloat
    \centering
         \includegraphics[width=0.8\linewidth]{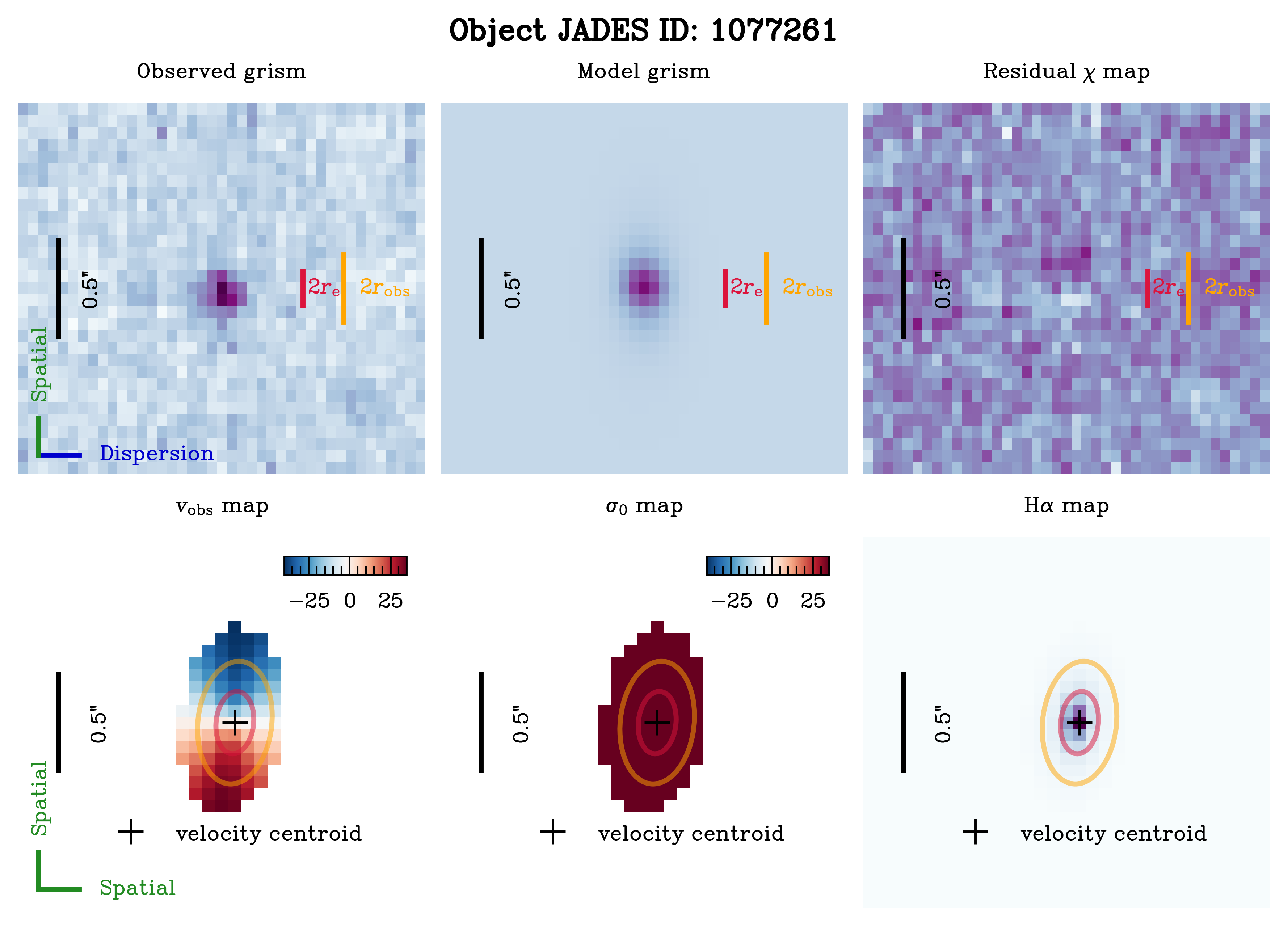}
     \caption{Continued.}
\end{figure*}

\begin{figure*}
    \ContinuedFloat
    \centering
         \includegraphics[width=0.8\linewidth]{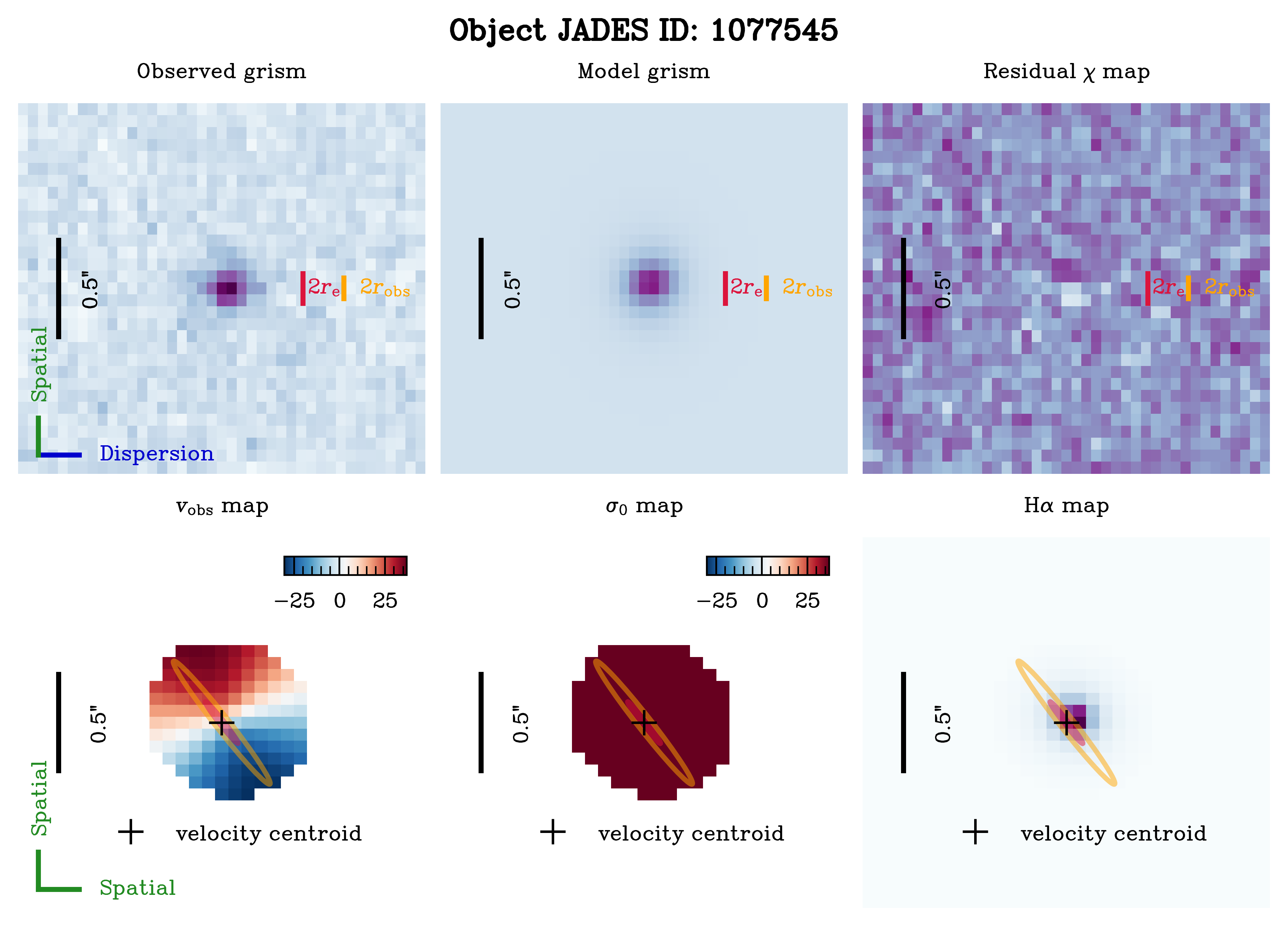}
     \caption{Continued.}
\end{figure*}

\begin{figure*}
    \ContinuedFloat
    \centering
         \includegraphics[width=0.8\linewidth]{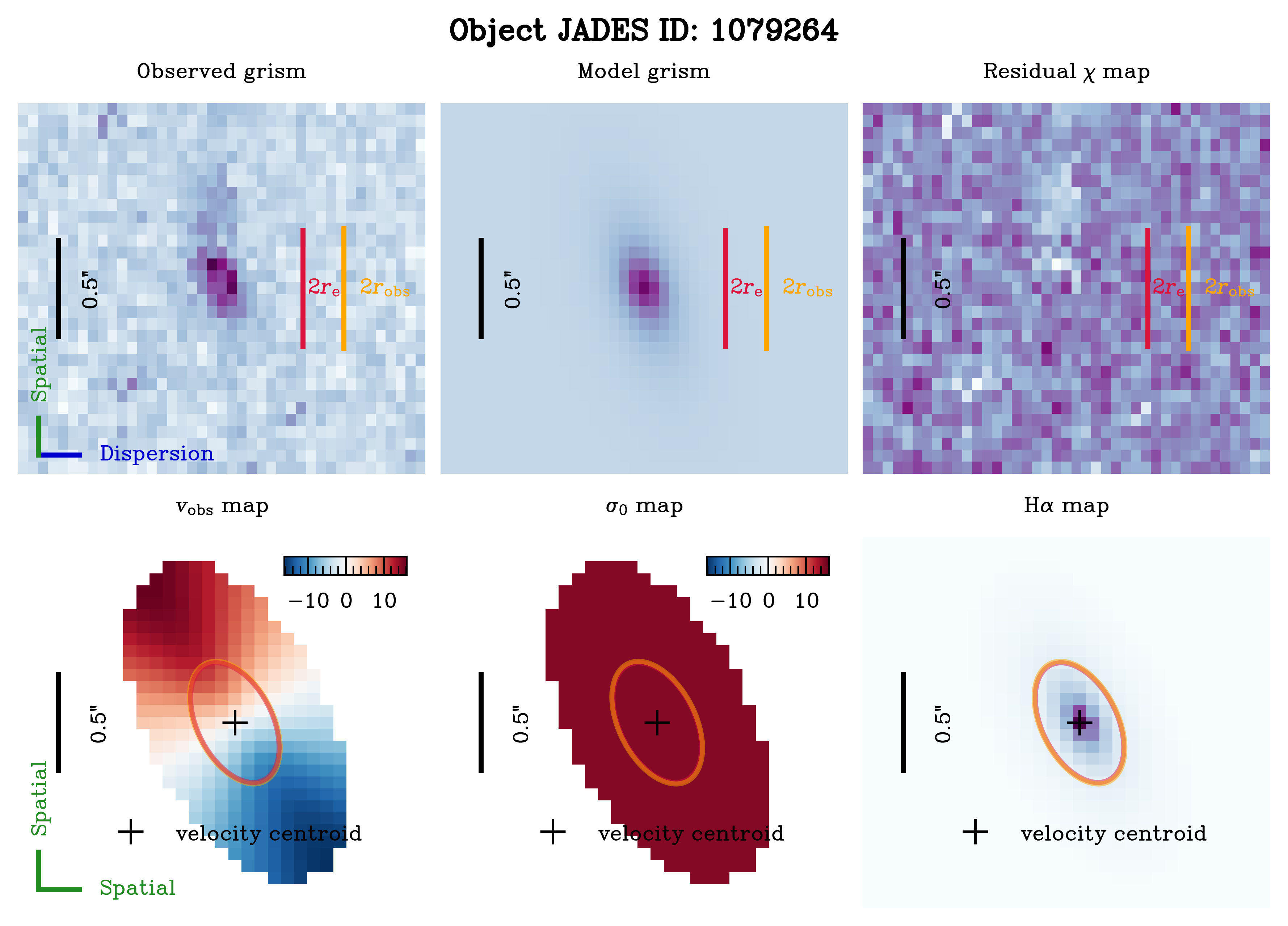}
     \caption{Continued.}
\end{figure*}

\begin{figure*}
    \ContinuedFloat
    \centering
         \includegraphics[width=0.8\linewidth]{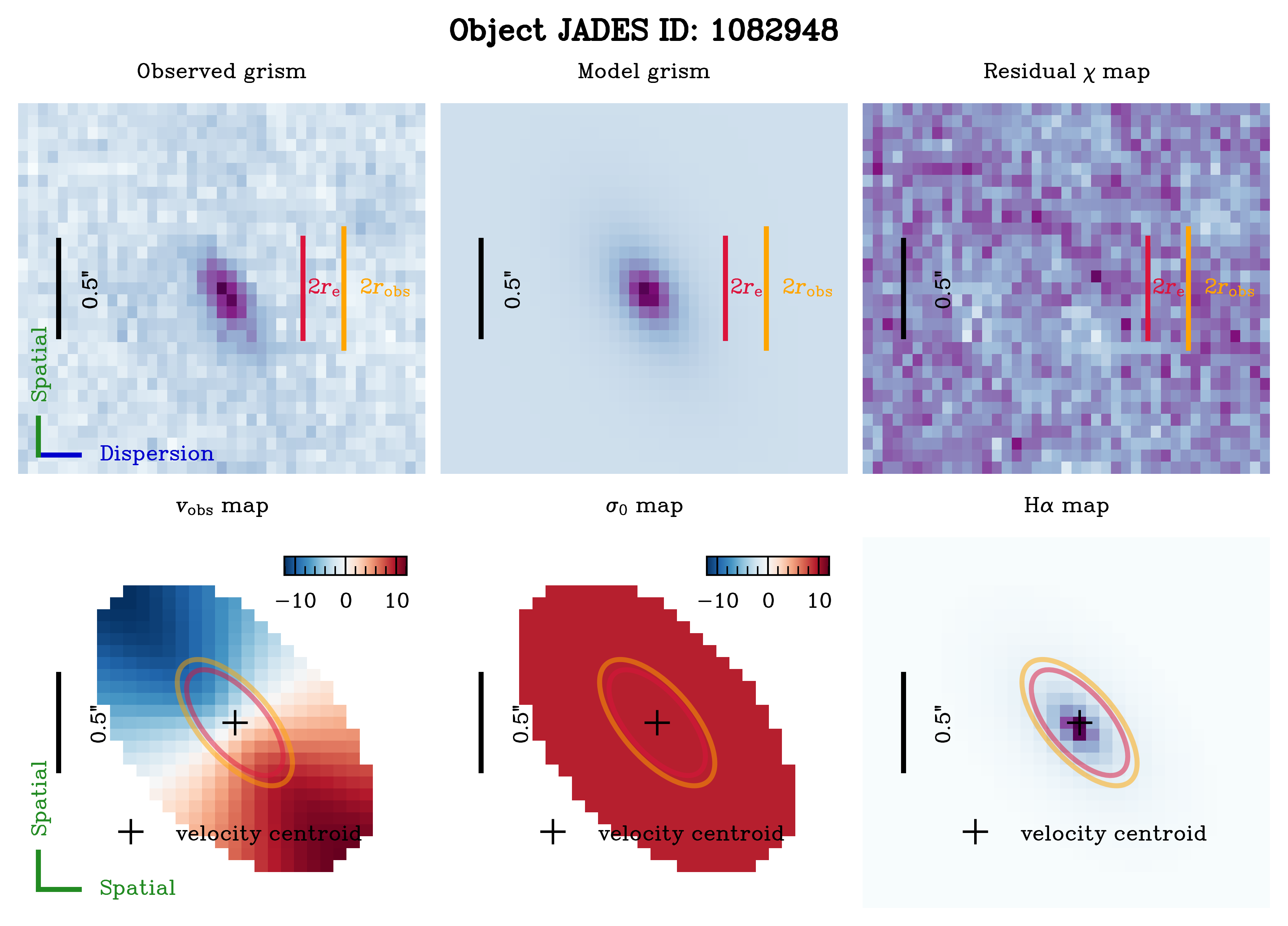}
     \caption{Continued.}
\end{figure*}

\begin{figure*}
    \ContinuedFloat
    \centering
         \includegraphics[width=0.8\linewidth]{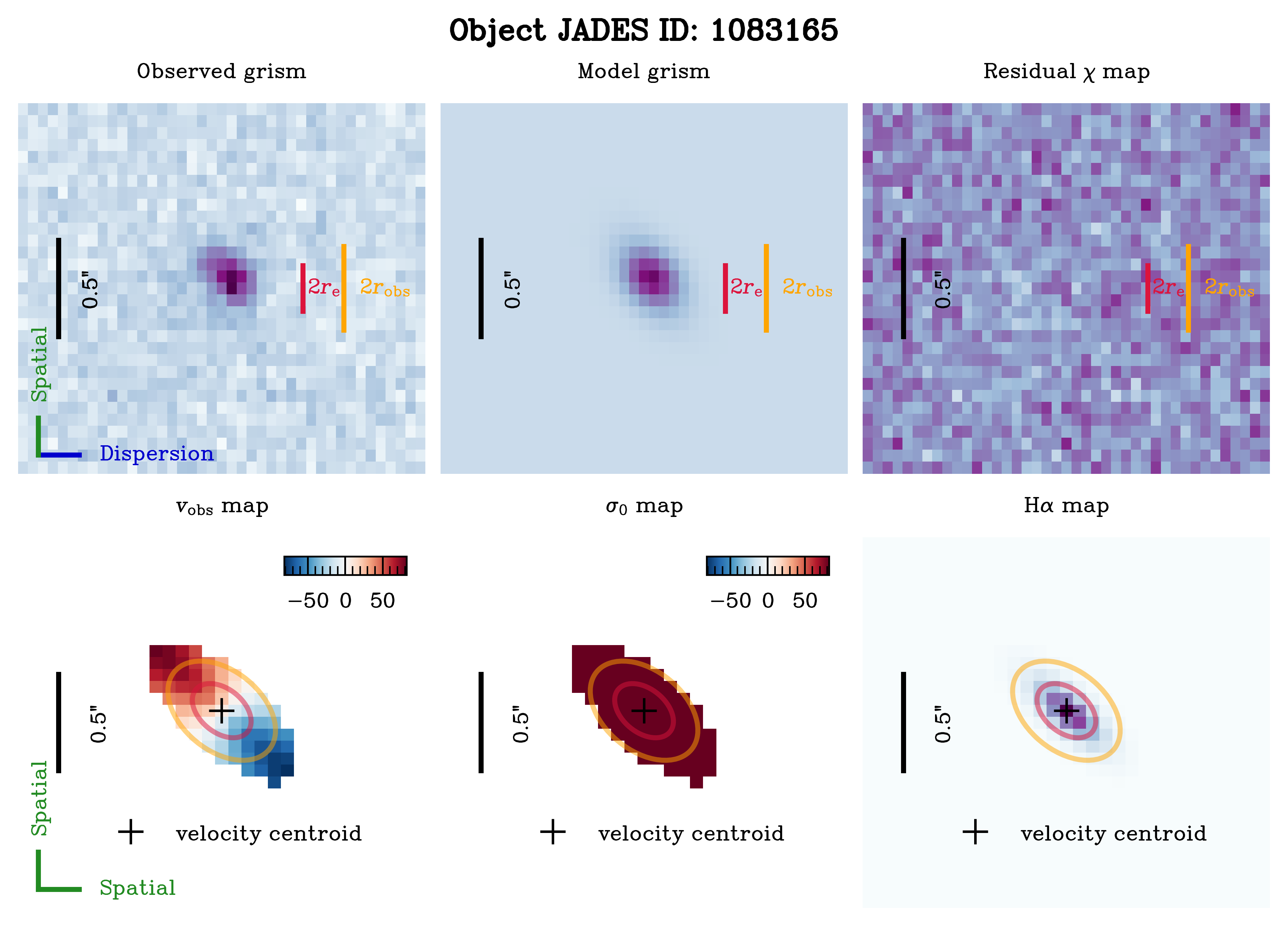}
     \caption{Continued.}
\end{figure*}

\begin{figure*}
    \ContinuedFloat
    \centering
         \includegraphics[width=0.8\linewidth]{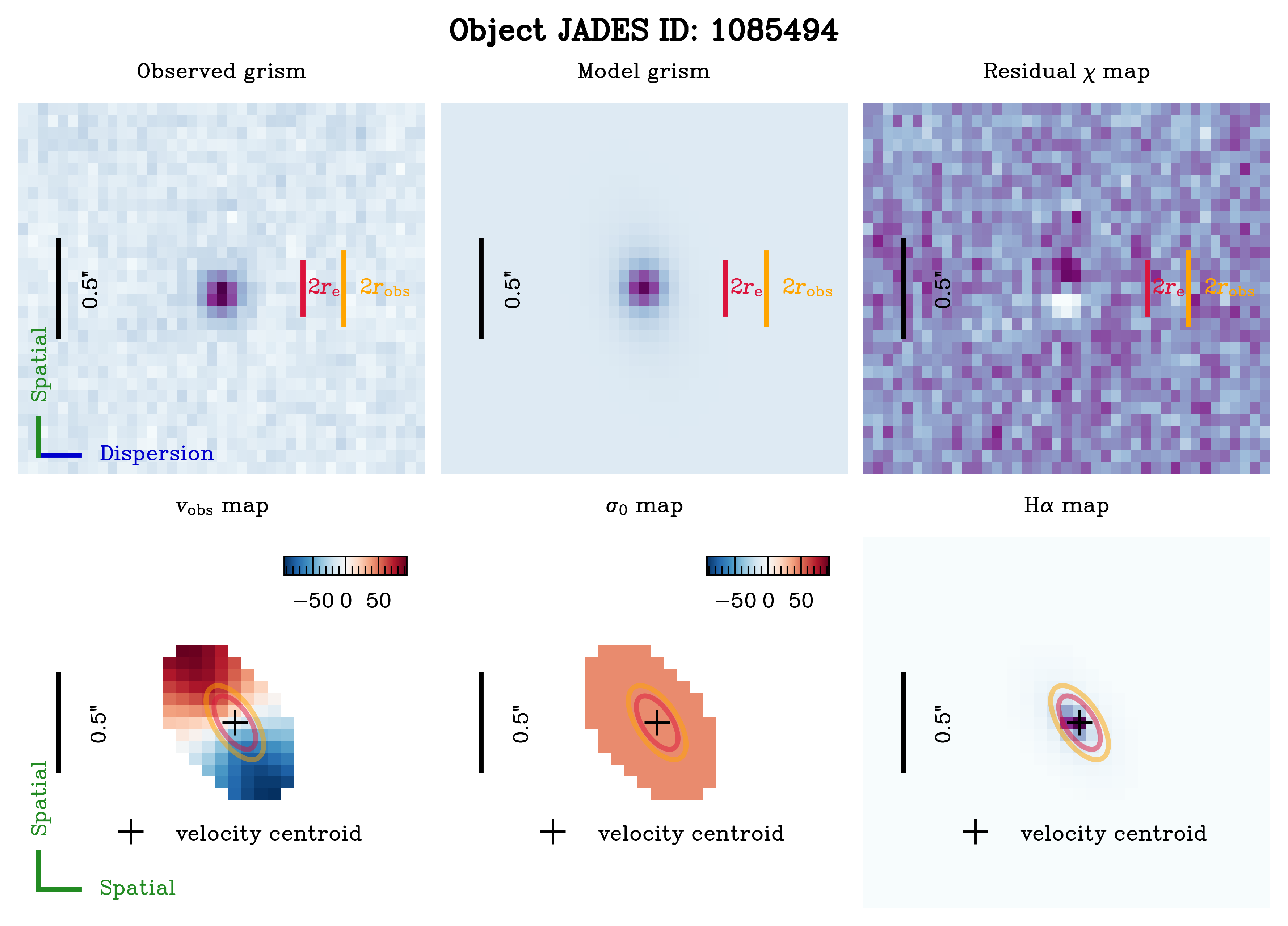}
     \caption{Continued.}
\end{figure*}

\begin{figure*}
    \ContinuedFloat
    \centering
         \includegraphics[width=0.8\linewidth]{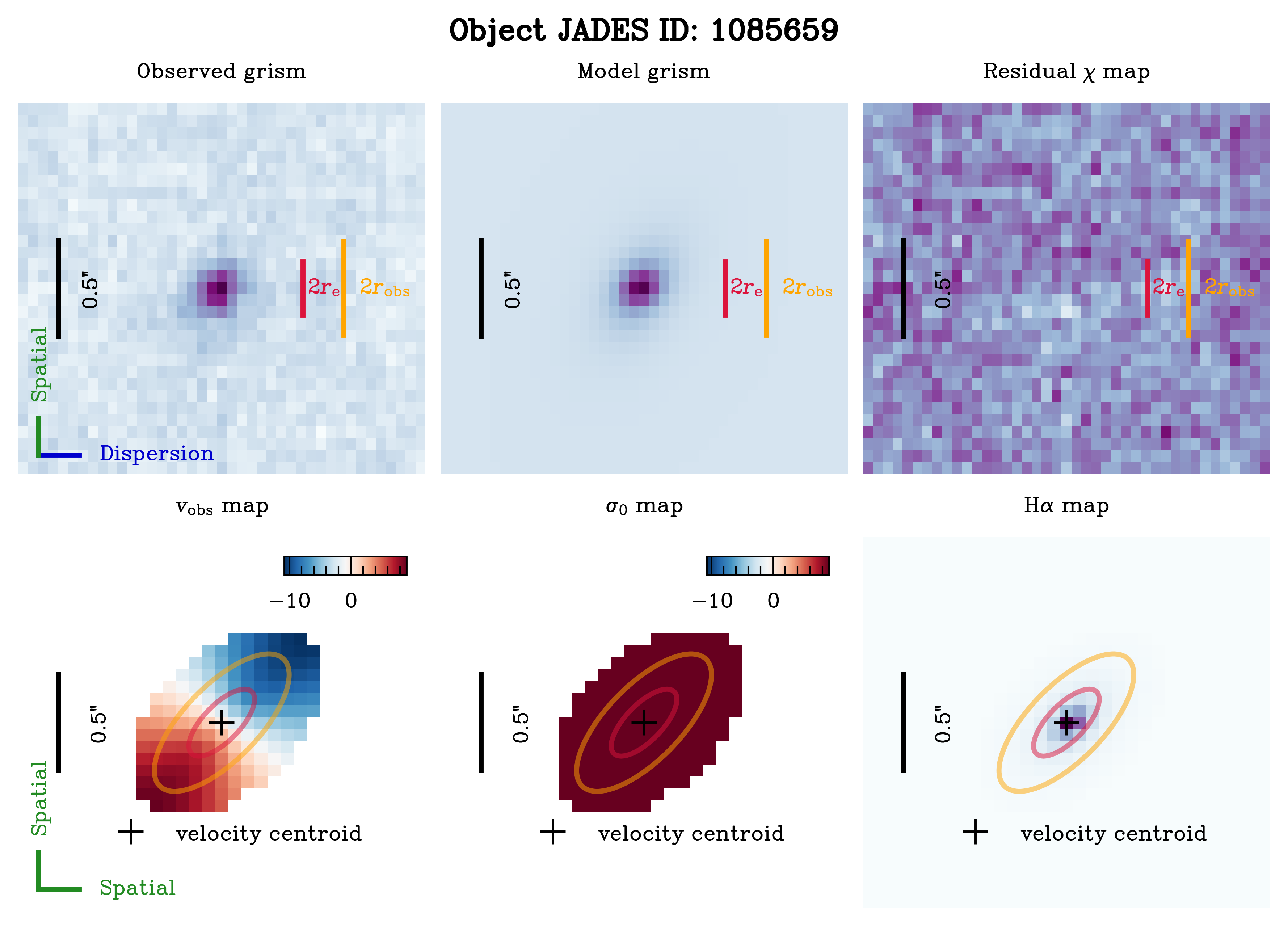}
     \caption{Continued.}
\end{figure*}

\begin{figure*}
    \ContinuedFloat
    \centering
         \includegraphics[width=0.8\linewidth]{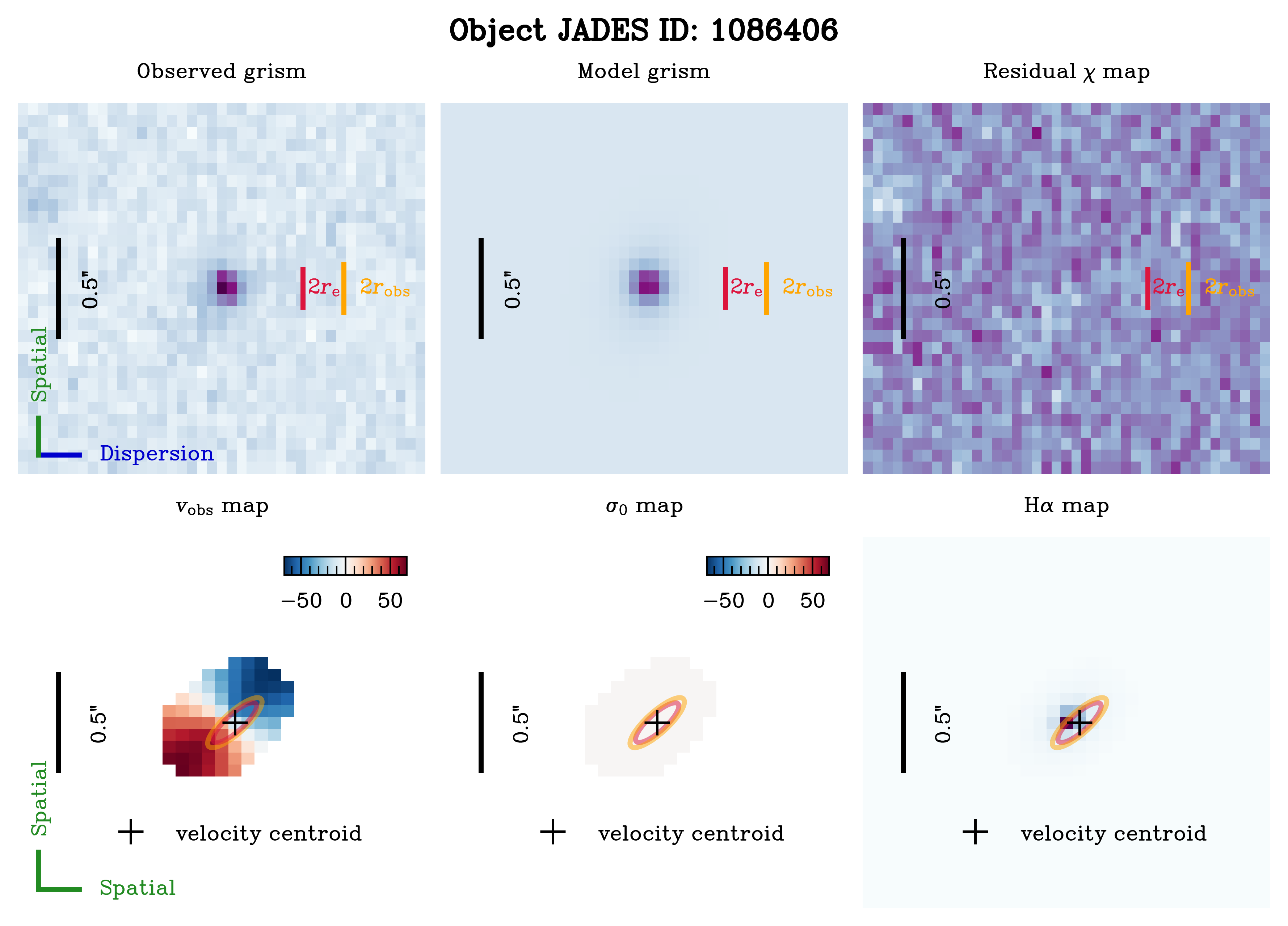}
     \caption{Continued.}
\end{figure*}

\begin{figure*}
    \ContinuedFloat
    \centering
         \includegraphics[width=0.8\linewidth]{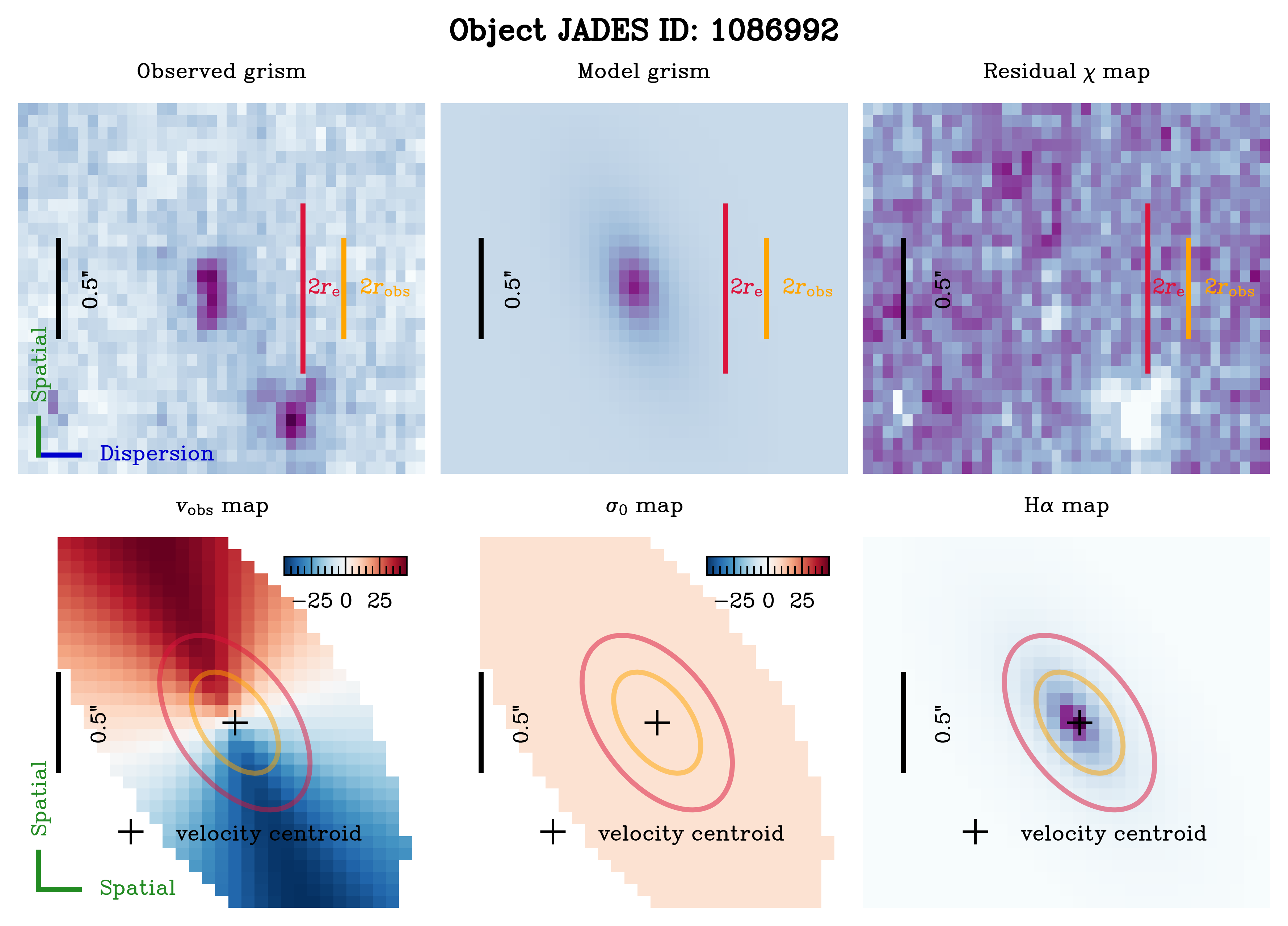}
     \caption{Continued.}
\end{figure*}

\begin{figure*}
    \ContinuedFloat
    \centering
         \includegraphics[width=0.8\linewidth]{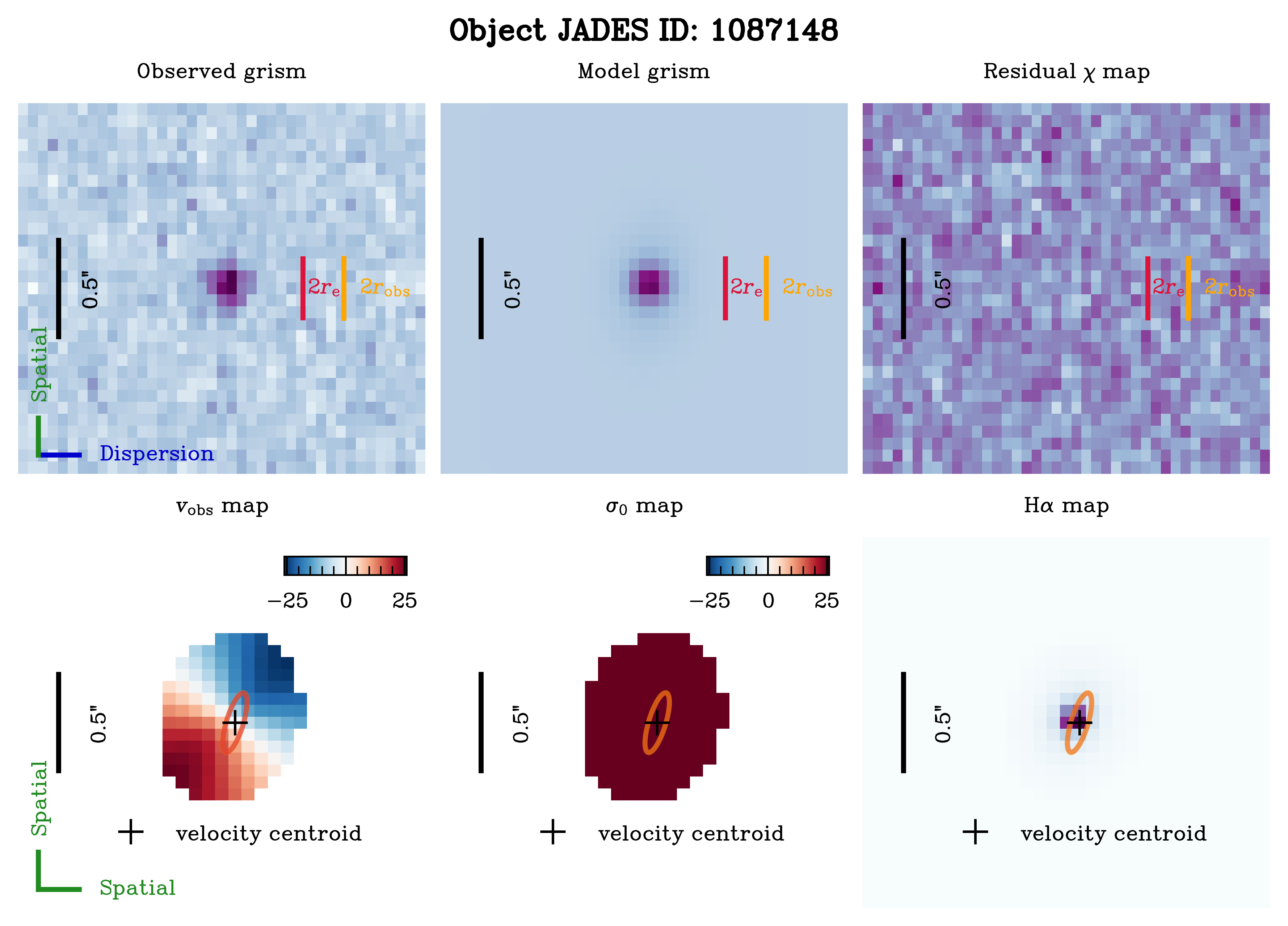}
     \caption{Continued.}
\end{figure*}

\begin{figure*}
    \ContinuedFloat
    \centering
         \includegraphics[width=0.8\linewidth]{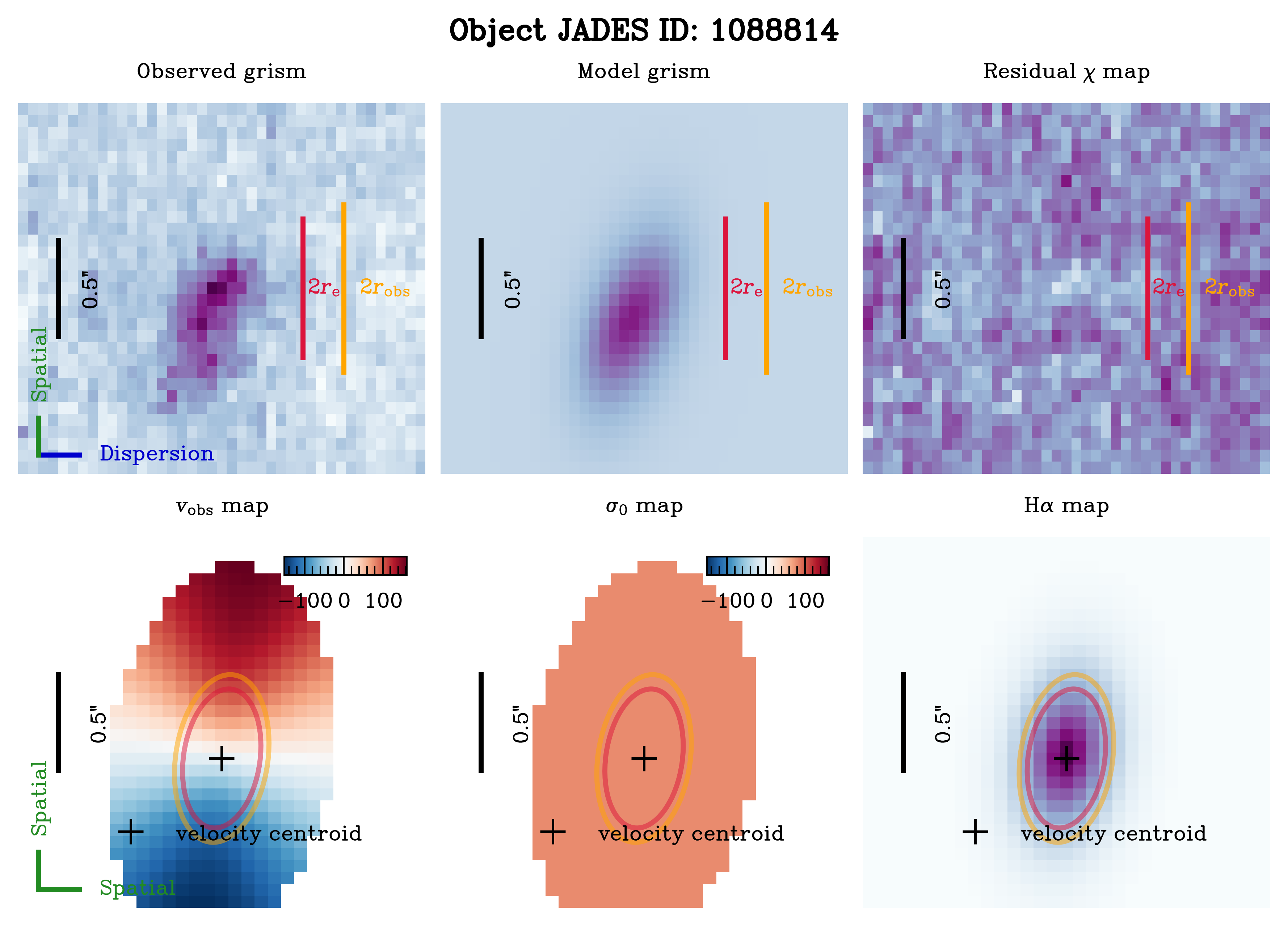}
     \caption{Continued.}
\end{figure*}

\begin{figure*}
    \ContinuedFloat
    \centering
         \includegraphics[width=0.8\linewidth]{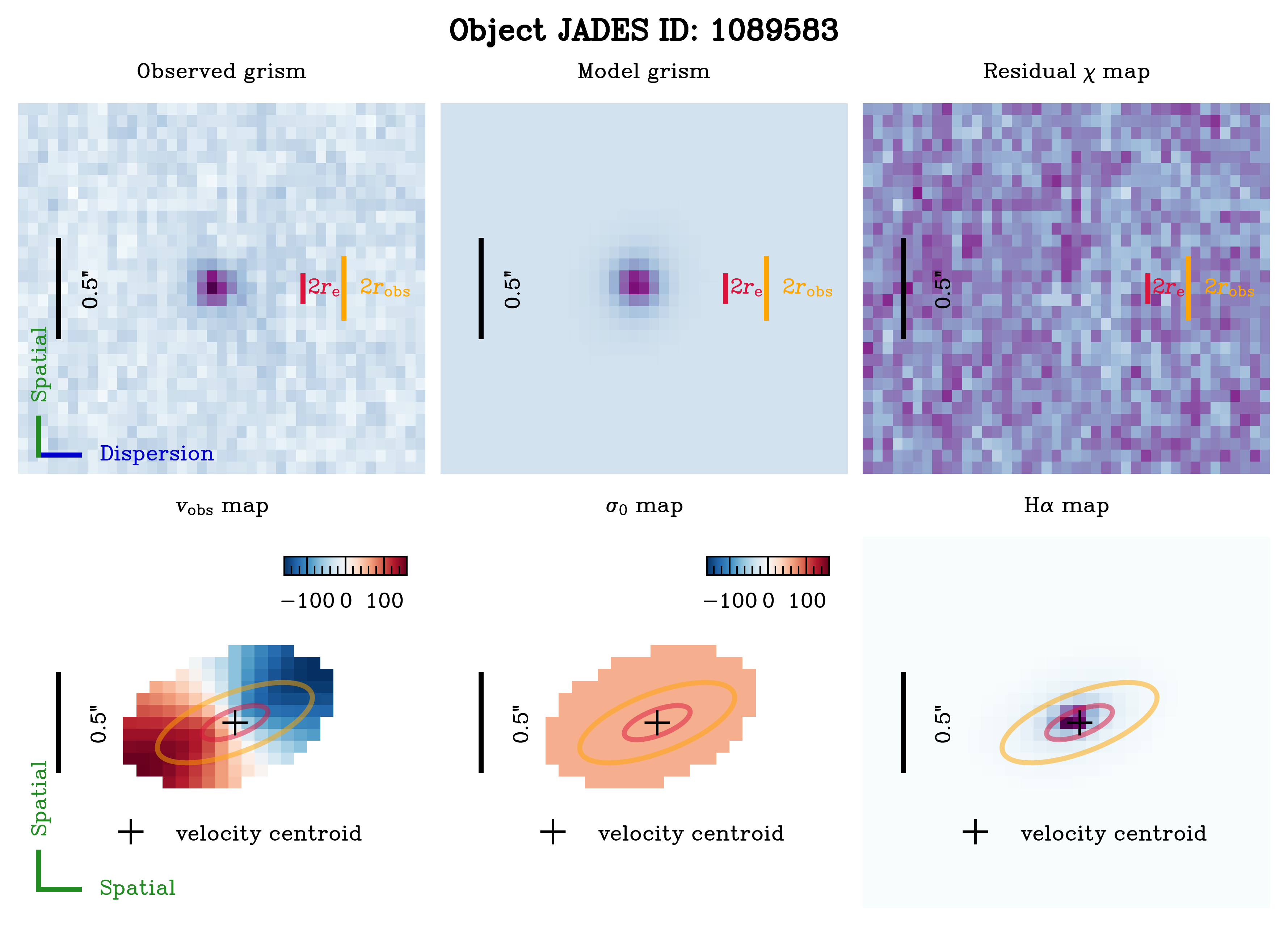}
     \caption{Continued.}
\end{figure*}

\begin{figure*}
    \ContinuedFloat
    \centering
         \includegraphics[width=0.8\linewidth]{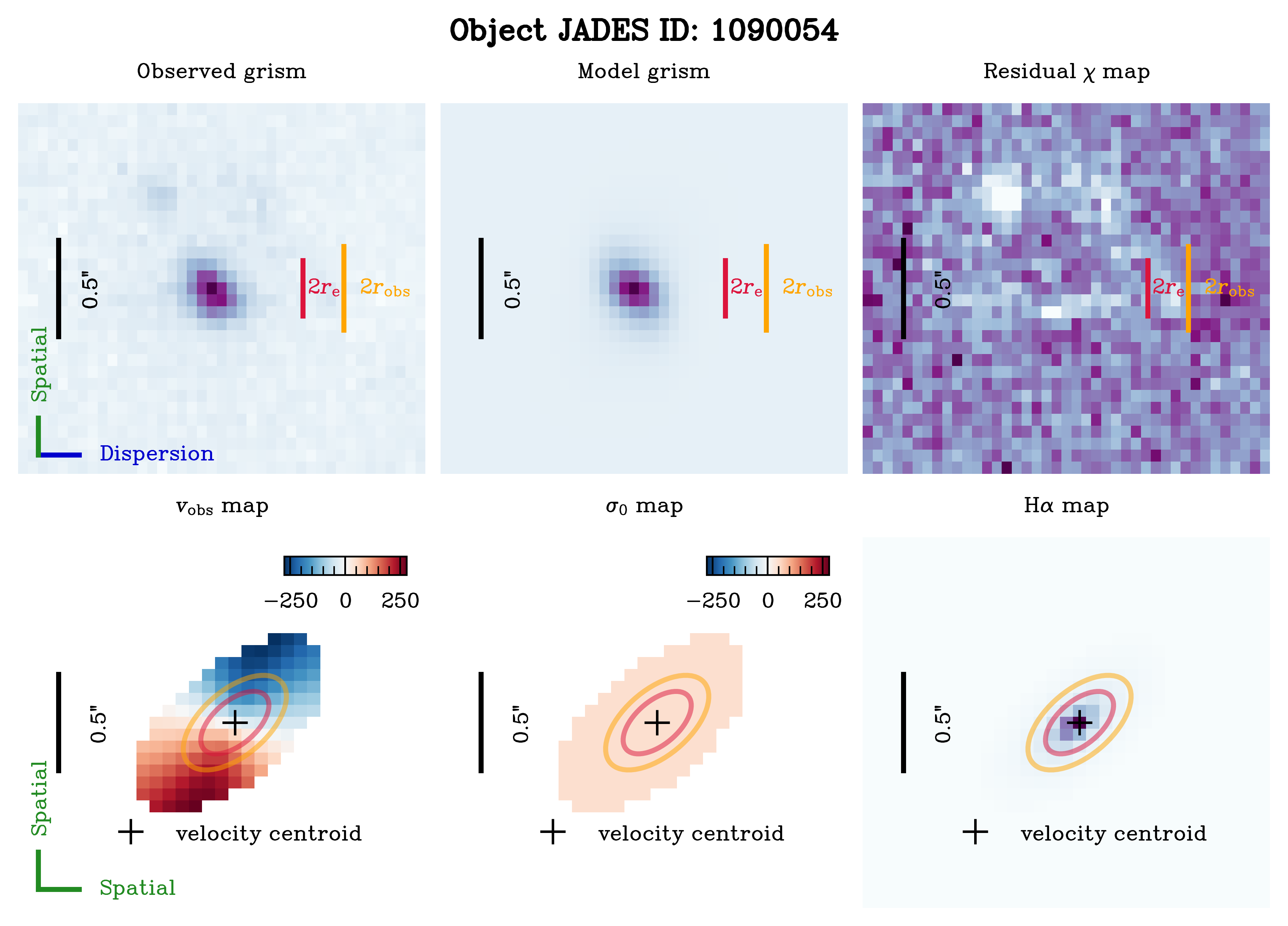}
     \caption{Continued.}
\end{figure*}

\begin{figure*}
    \ContinuedFloat
    \centering
         \includegraphics[width=0.8\linewidth]{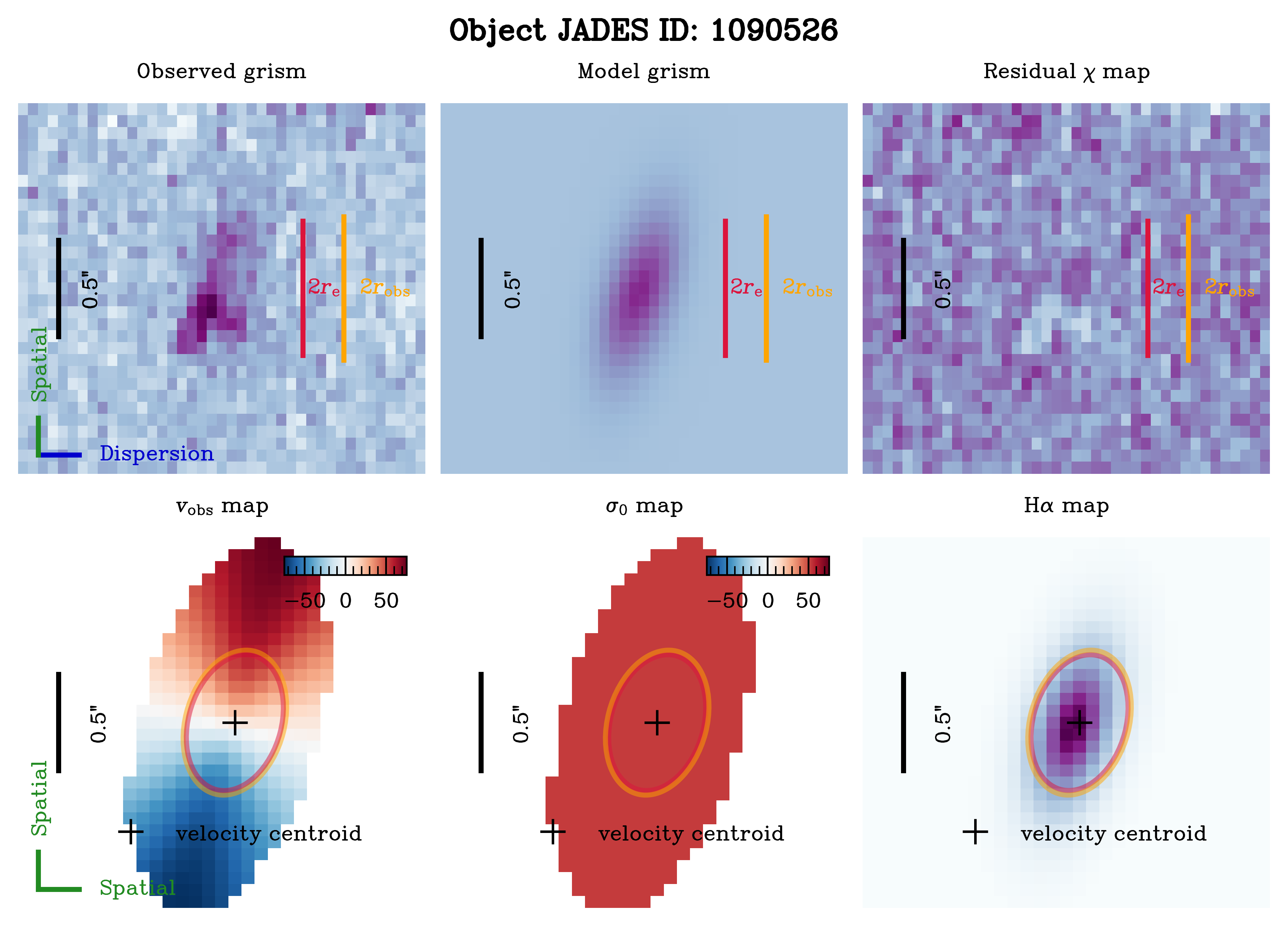}
     \caption{Continued.}
\end{figure*}

\begin{figure*}
    \ContinuedFloat
    \centering
         \includegraphics[width=0.8\linewidth]{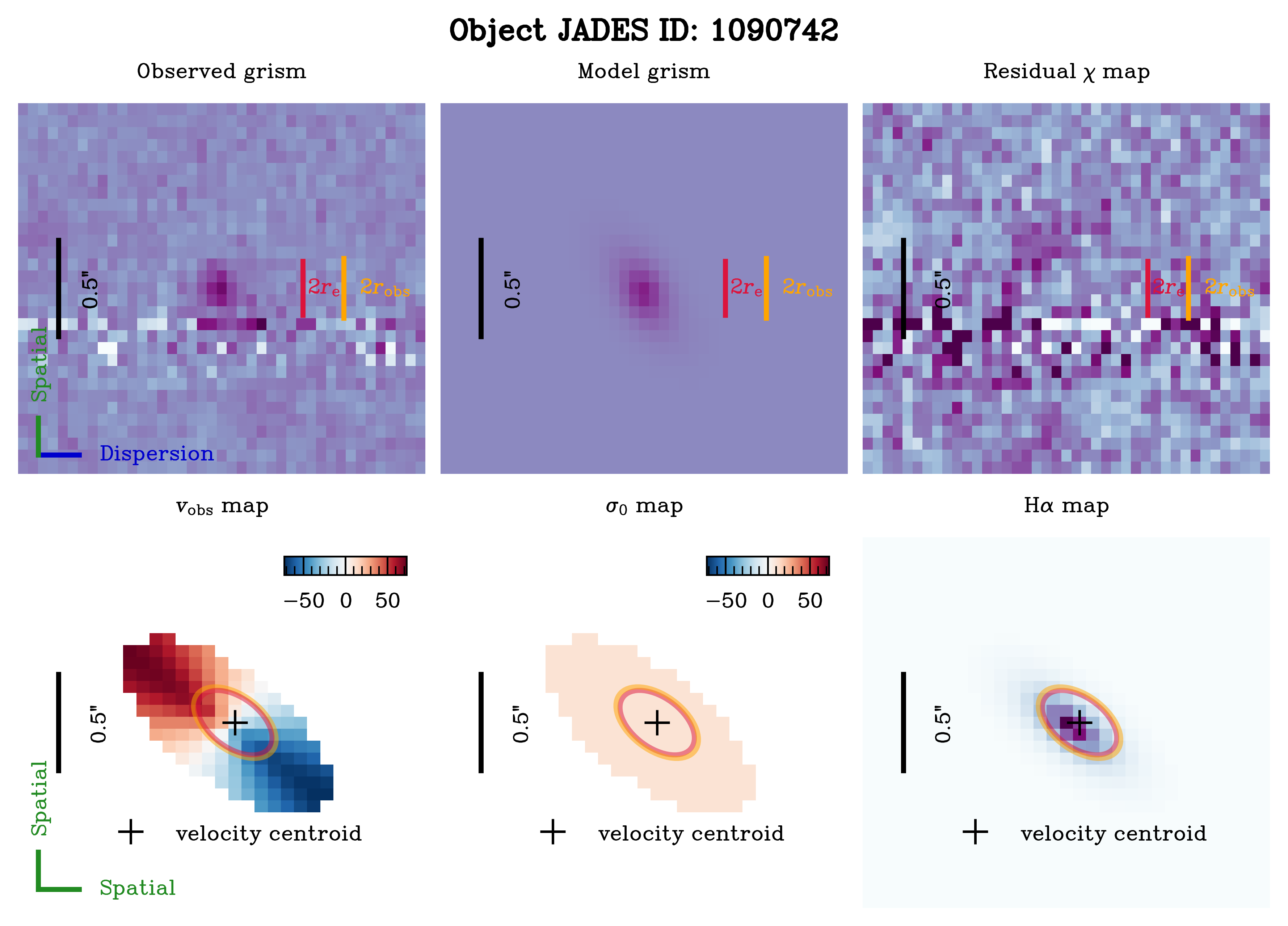}
     \caption{Continued.}
\end{figure*}

\begin{figure*}
    \ContinuedFloat
    \centering
         \includegraphics[width=0.8\linewidth]{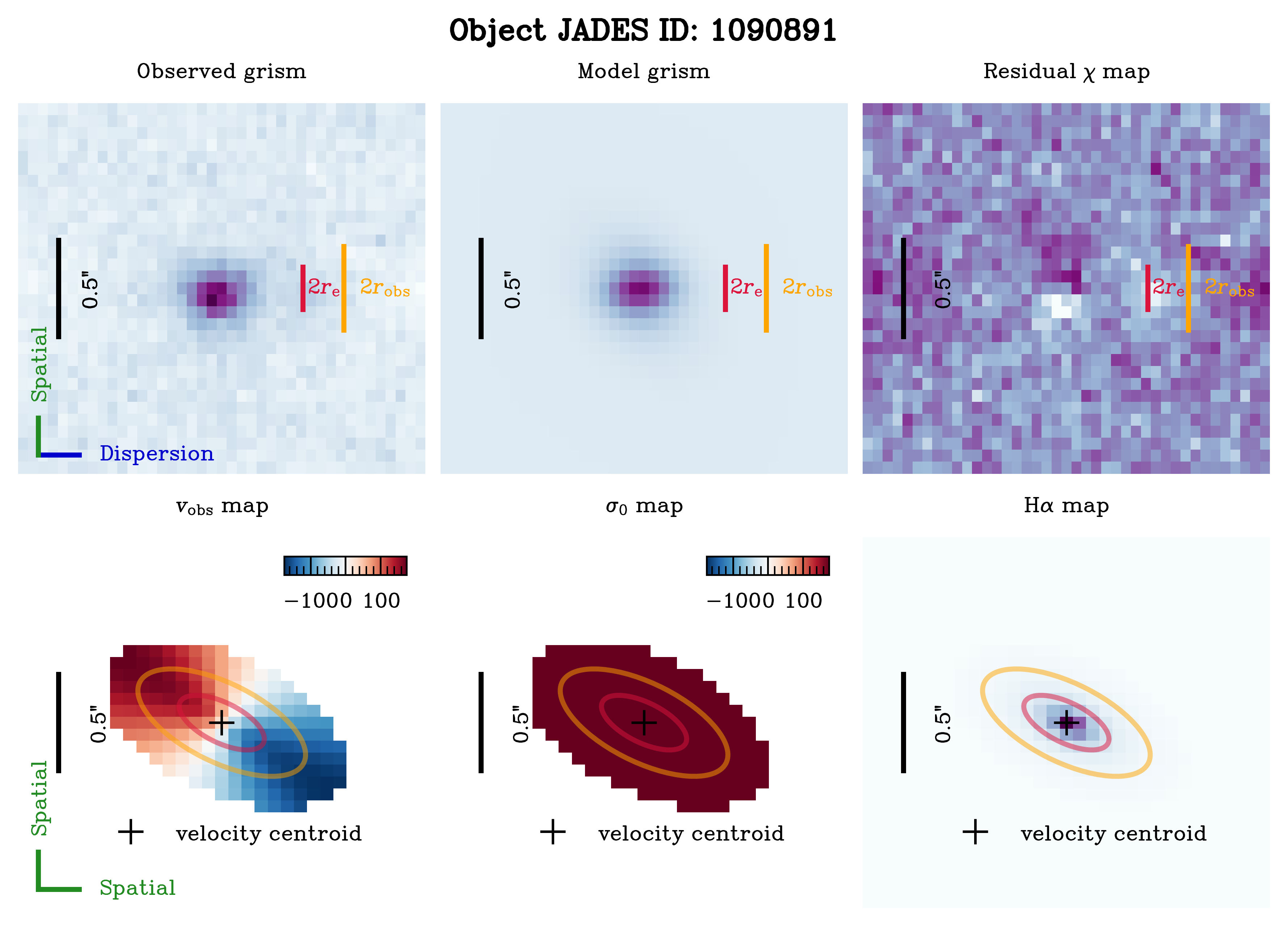}
     \caption{Continued.}
\end{figure*}

\begin{figure*}
    \ContinuedFloat
    \centering
         \includegraphics[width=0.8\linewidth]{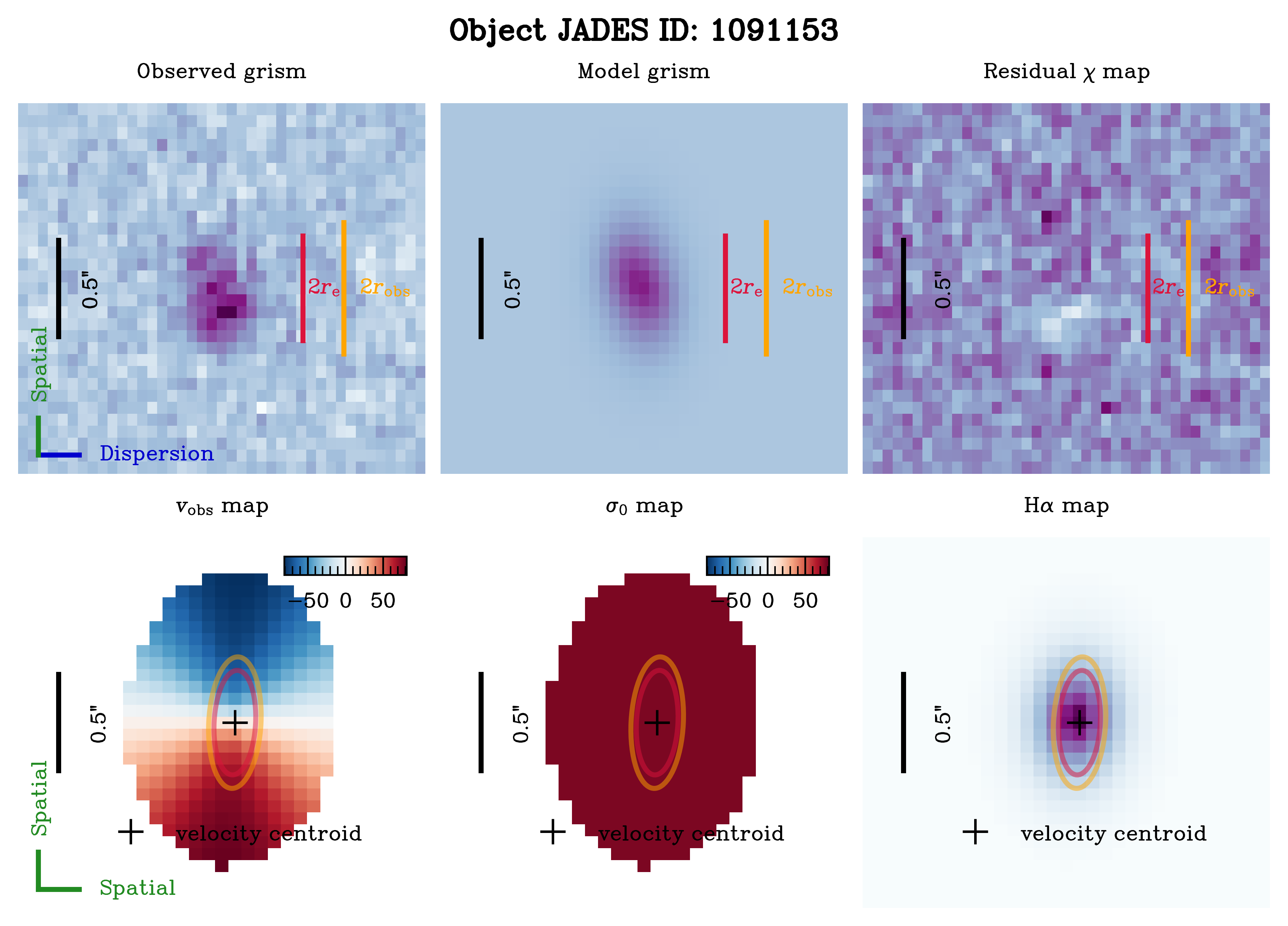}
     \caption{Continued.}
\end{figure*}

\begin{figure*}
    \ContinuedFloat
    \centering
         \includegraphics[width=0.8\linewidth]{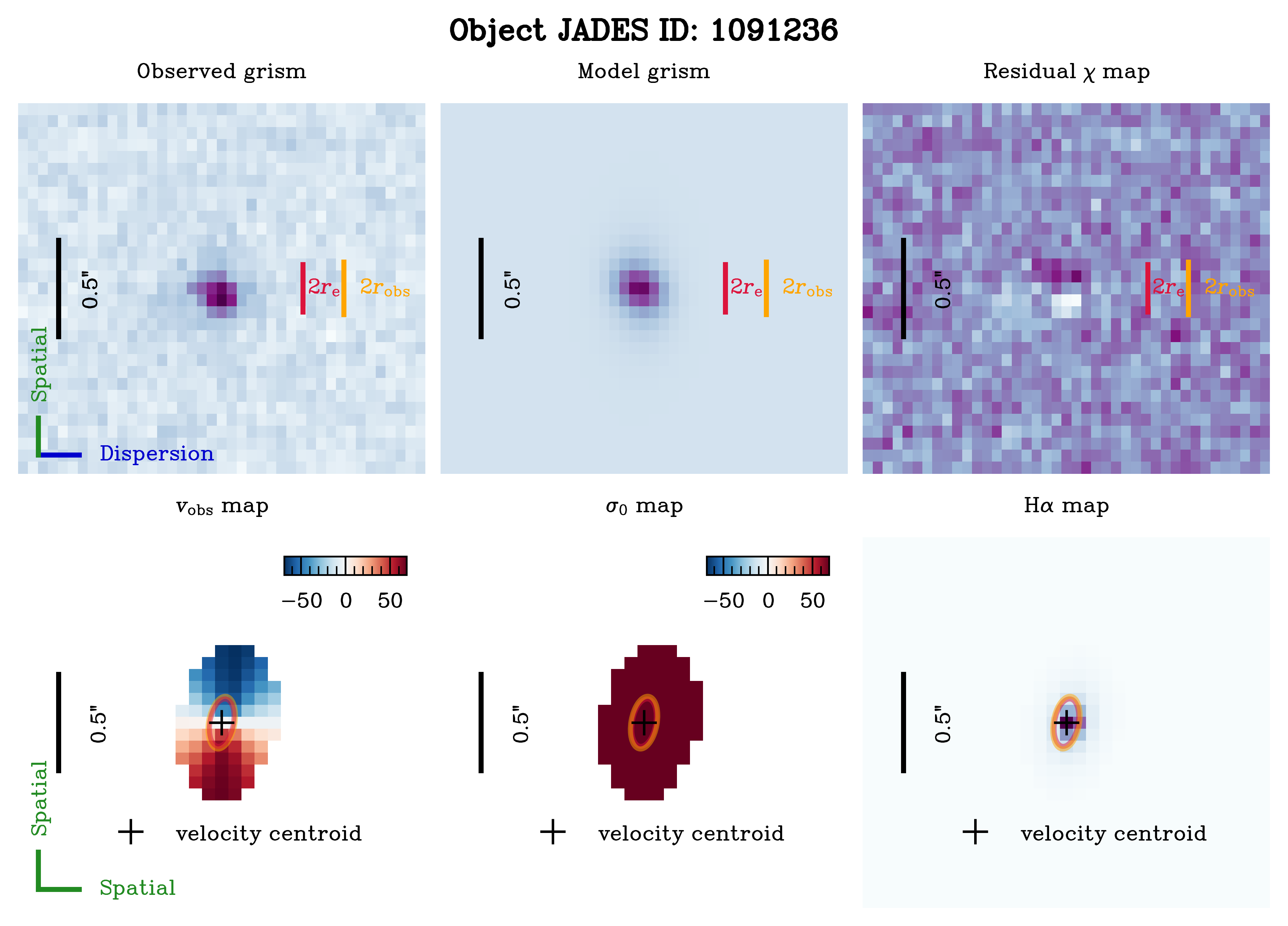}
     \caption{Continued.}
\end{figure*}

\begin{figure*}
    \ContinuedFloat
    \centering
         \includegraphics[width=0.8\linewidth]{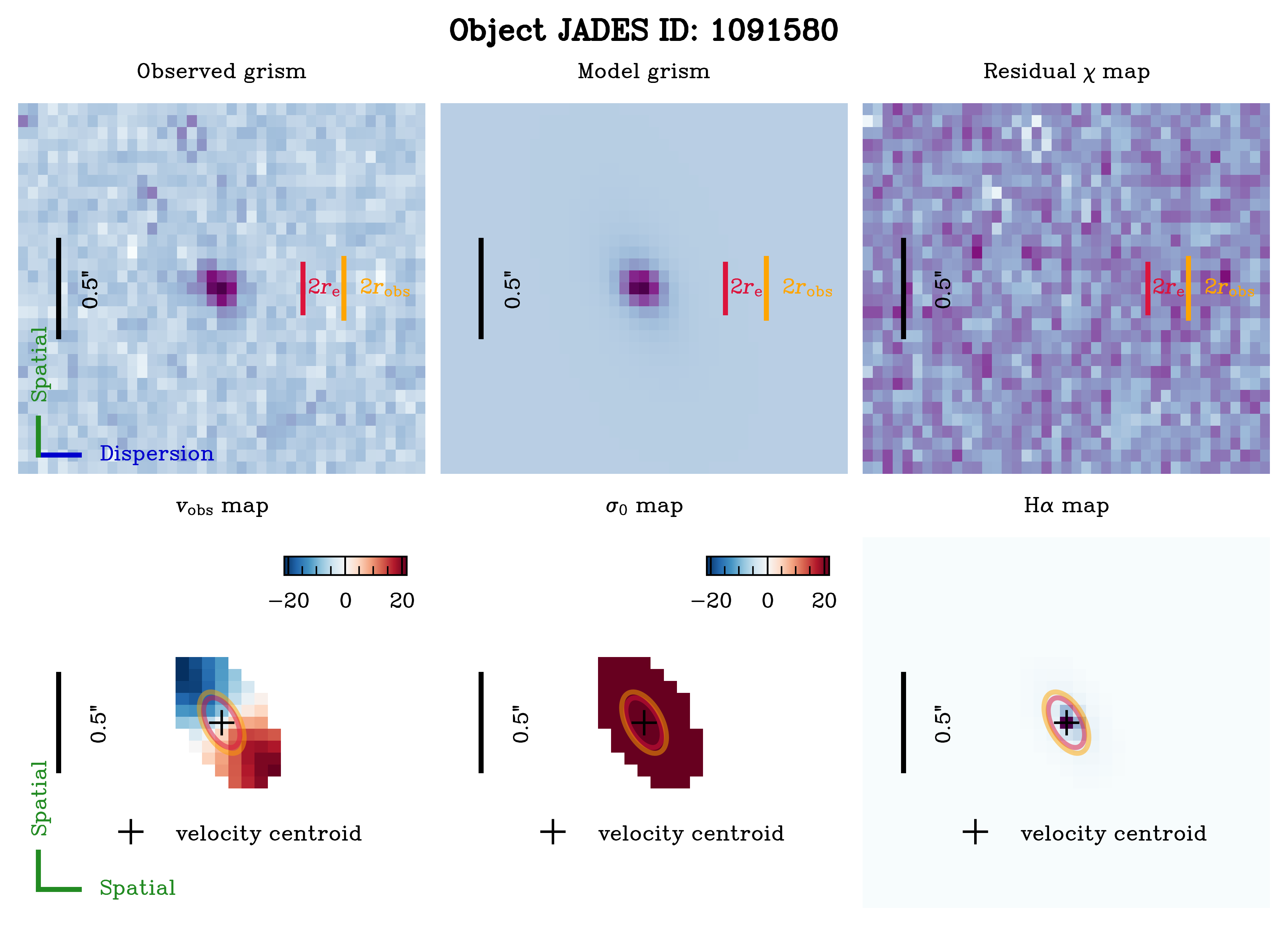}
     \caption{Continued.}
\end{figure*}

\begin{figure*}
    \ContinuedFloat
    \centering
         \includegraphics[width=0.8\linewidth]{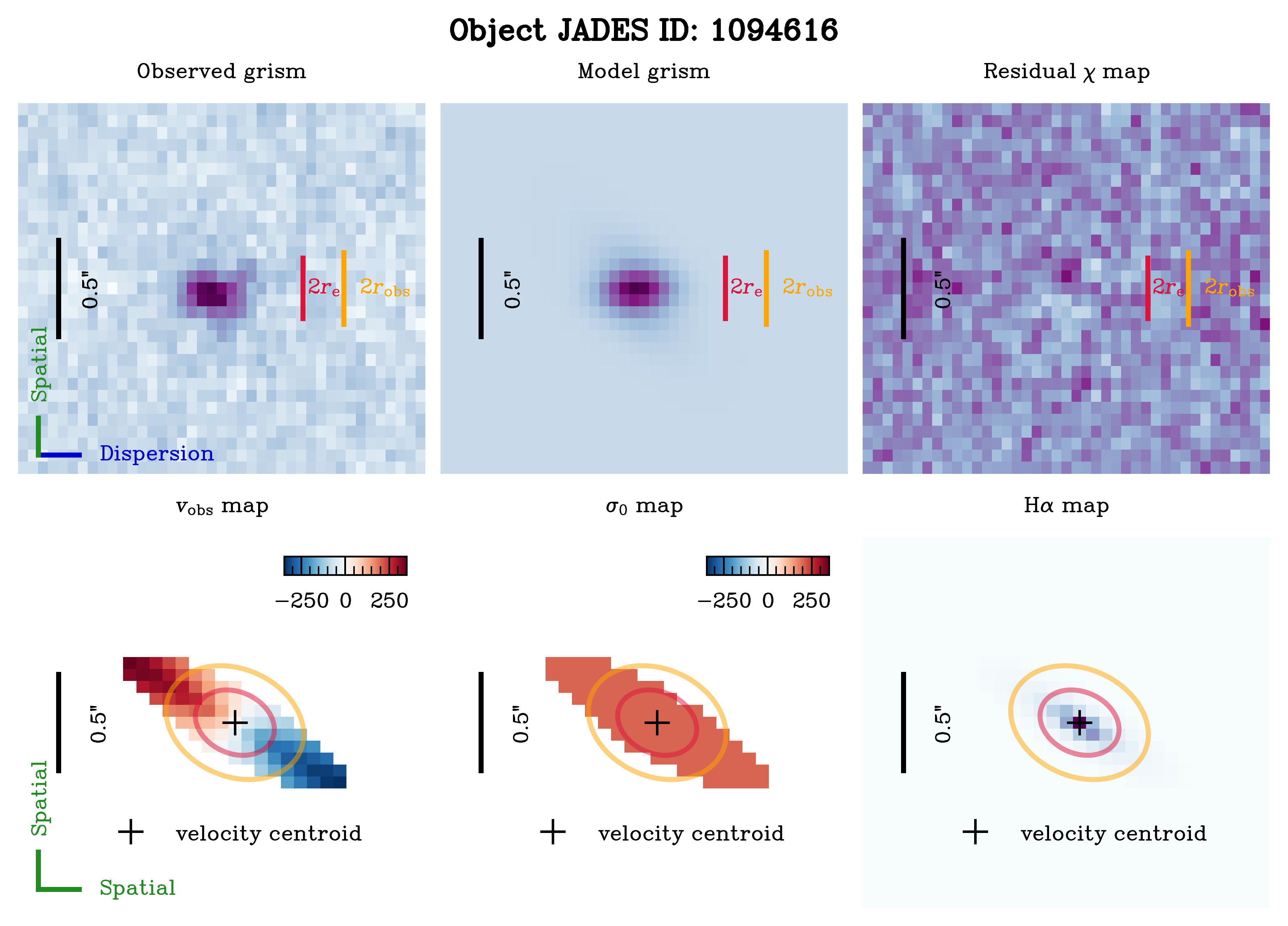}
     \caption{Continued.}
\end{figure*}

\begin{figure*}
    \ContinuedFloat
    \centering
         \includegraphics[width=0.8\linewidth]{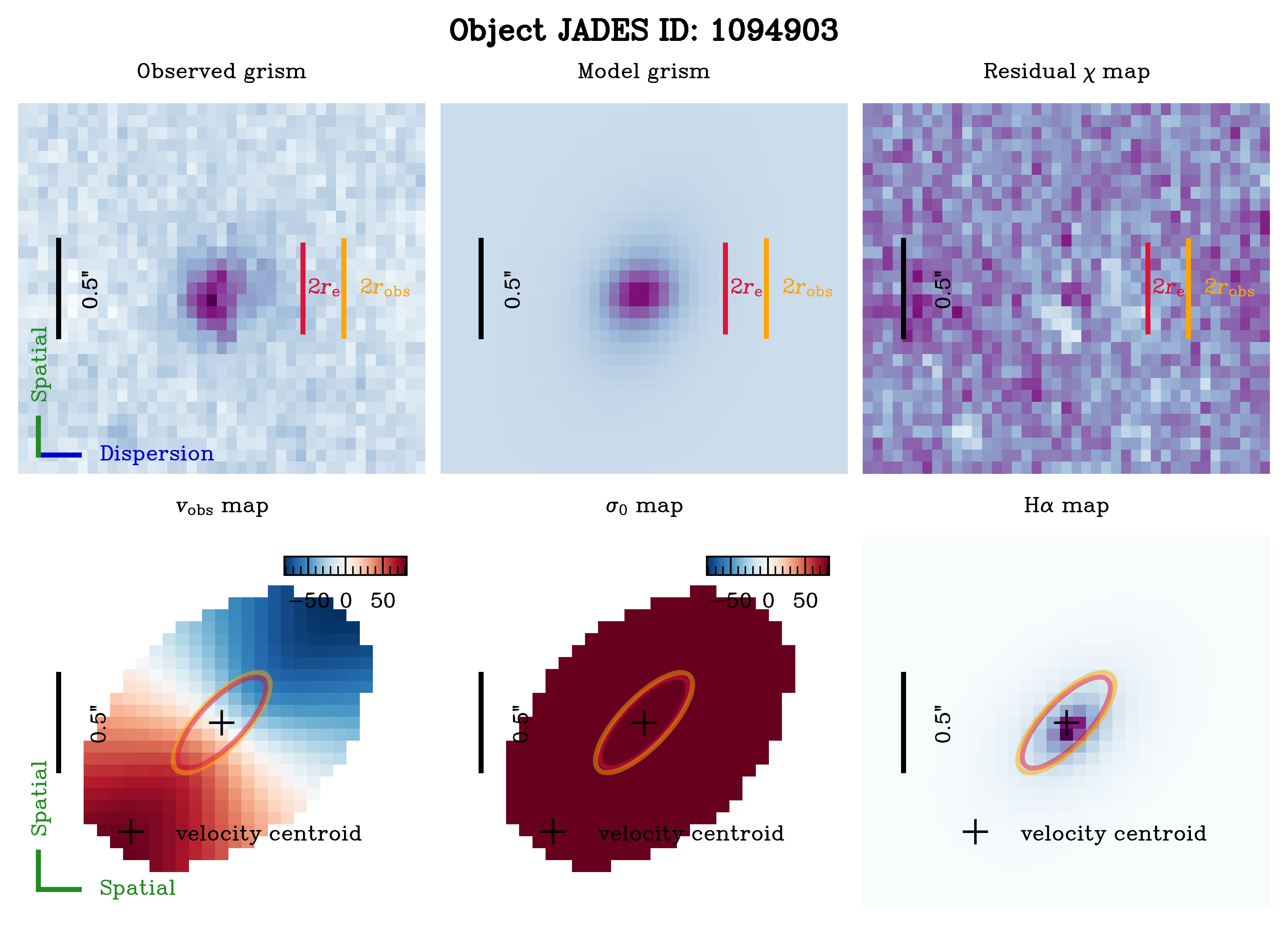}
     \caption{Continued.}
\end{figure*}

\begin{figure*}
    \ContinuedFloat
    \centering
         \includegraphics[width=0.8\linewidth]{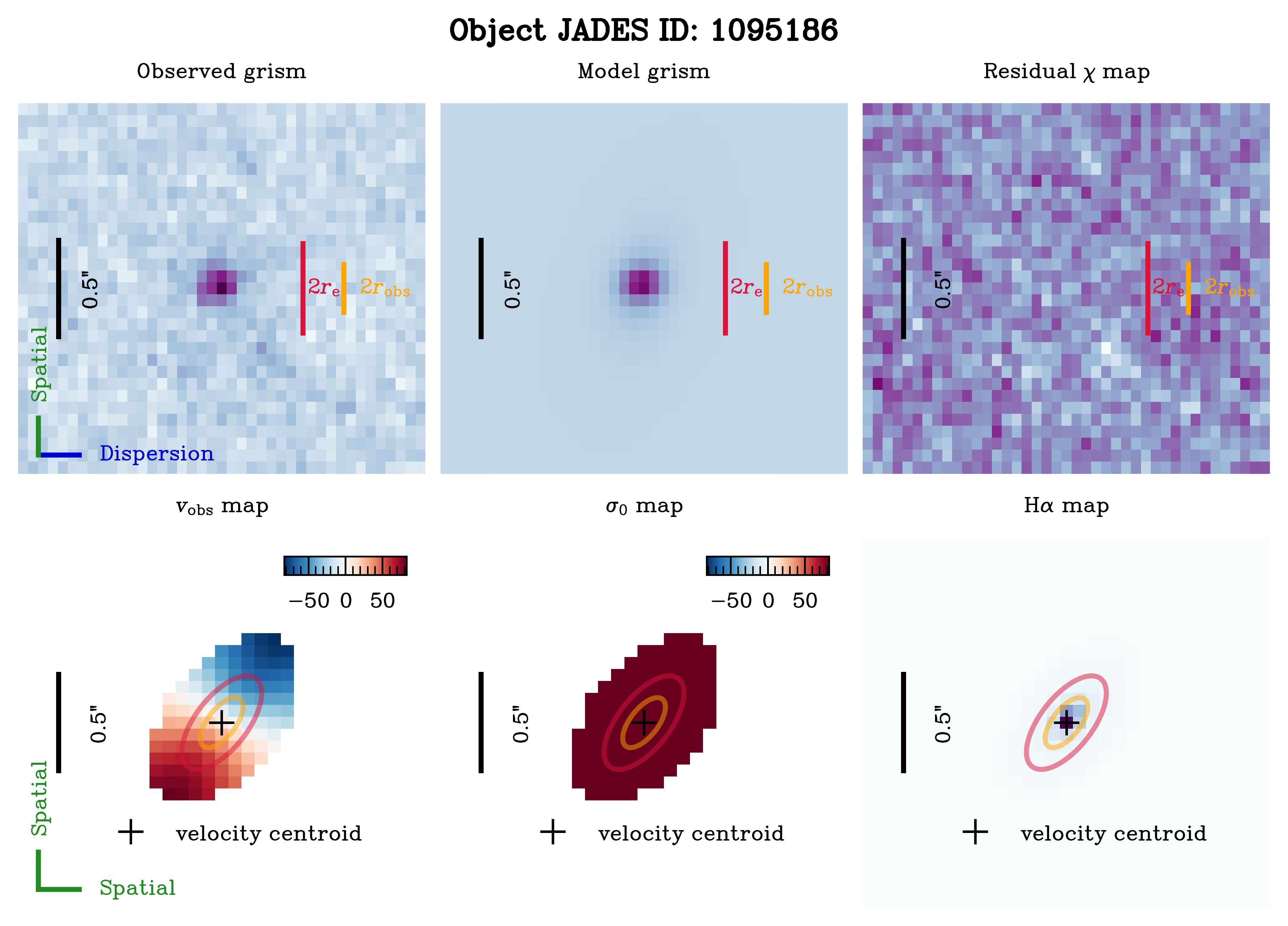}
     \caption{Continued.}
\end{figure*}

\begin{figure*}
    \ContinuedFloat
    \centering
         \includegraphics[width=0.8\linewidth]{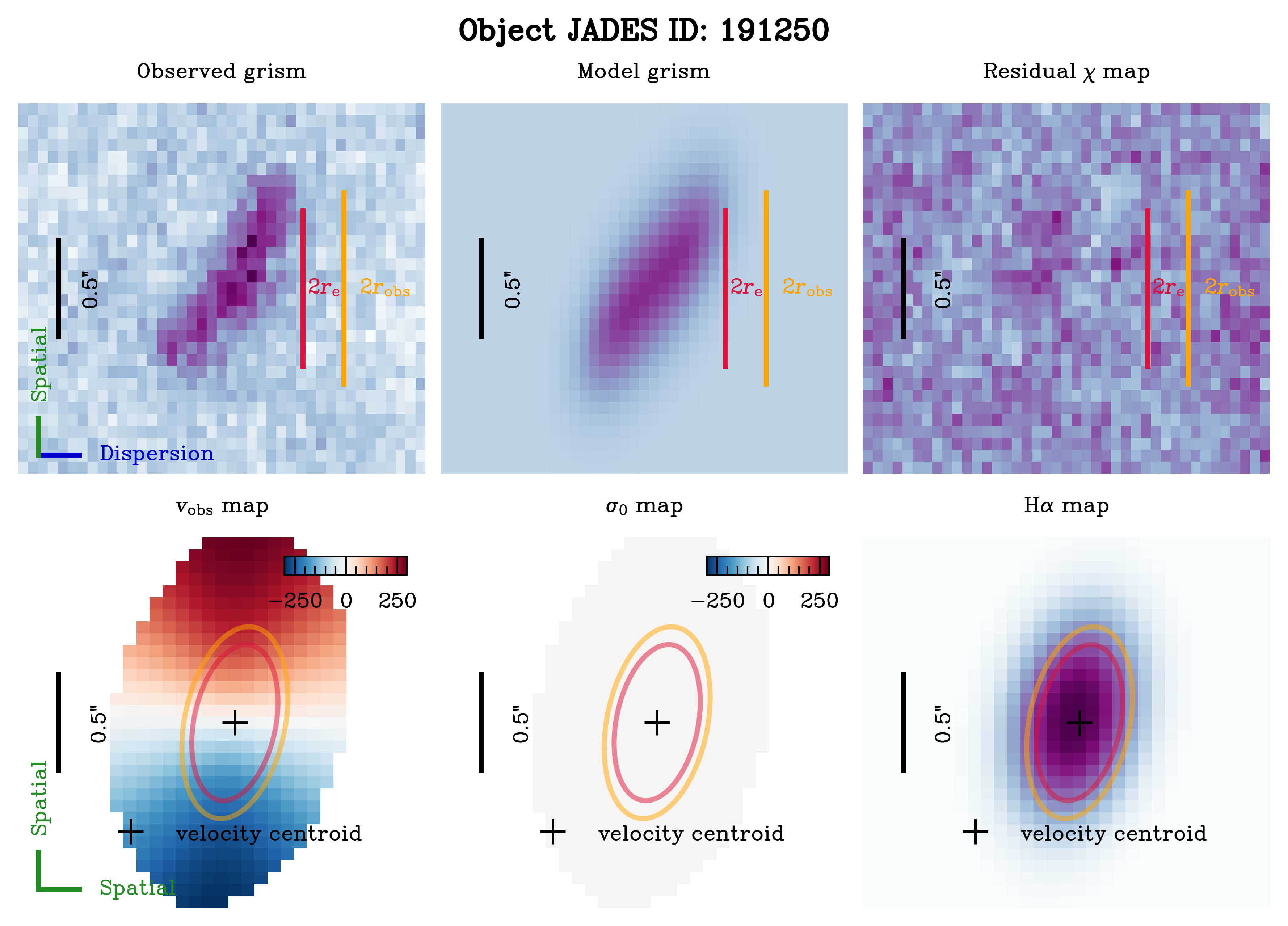}
     \caption{Continued.}
\end{figure*}

\begin{figure*}
    \ContinuedFloat
    \centering
         \includegraphics[width=0.8\linewidth]{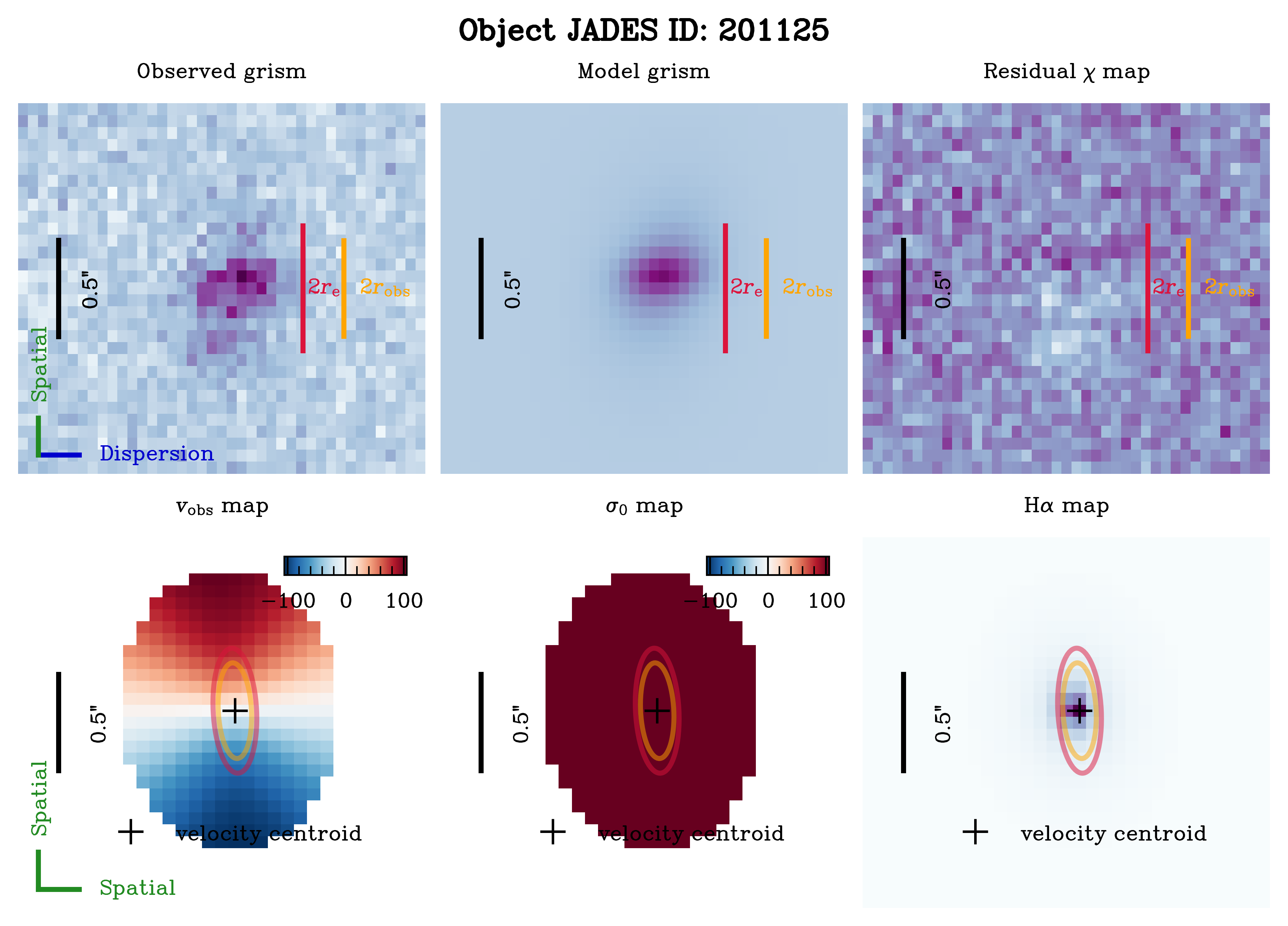}
     \caption{Continued.}
\end{figure*}

\begin{figure*}
    \ContinuedFloat
    \centering
         \includegraphics[width=0.8\linewidth]{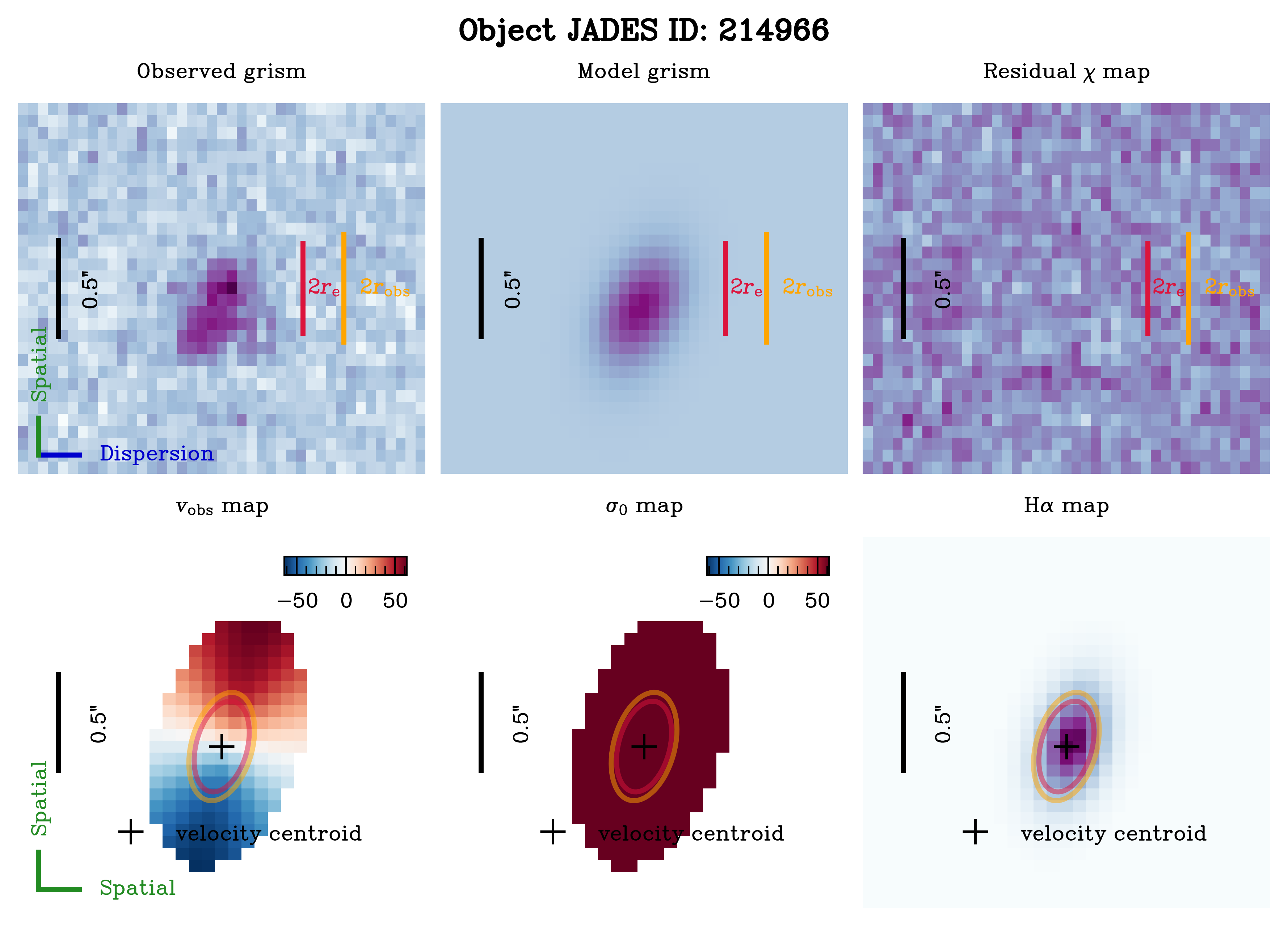}
     \caption{Continued.}
\end{figure*}


\bsp	
\label{lastpage}
\end{document}